%% file: modular_tomography_I.tex
\pdfoutput=1
\documentclass[11pt,a4paper]{article}
\input{modular_tomography_preamble.tex}

\title{Modular tomography of perturbative quantum causal shadows I: Averaged positivity}

\author[a]{Liangyu Chen} 
\affiliation[a]{Yau Mathematical Sciences Center, Tsinghua University, Beijing 100084, China} 

\emailAdd{liangyu-chen@mail.tsinghua.edu.cn}

\hypersetup{pdfauthor={Liangyu Chen},pdftitle={Modular tomography of perturbative quantum causal shadows I: Averaged positivity}}
\abstract{Correlations between disconnected bulk regions can lift a classical degeneracy between the extremal and causal surfaces, producing a perturbative quantum causal shadow, as shown by Chen and Wang.
However, it remained unclear how the positivity of this displacement is encoded in the boundary theory. Here we address this question in vacuum $\text{AdS}_3/\text{CFT}_2$ by constructing a positive full-modular response $\mathcal{P}[f]\geq 0$ from relative-entropy monotonicity, where $|f\rangle$ is a normalized pure Euclidean boundary probe state prepared by a single scalar insertion. In Einstein gravity with two minimally coupled scalar fields, its growing cross-channel contribution measures a radial average of the quantum extremal surface (QES) displacement weighted by the horizon null-energy flux of the dual bulk wavepacket. We establish this relation non-circularly by matching the transverse scattering source against an independent generalized-entropy calculation of the QES displacement, verified via a two-variable Selberg integral. To transfer the boundary sign constraint to this geometric response, we control both the complete same-side ($AA$) contribution and the integrated cross-channel large-boost remainder at tree level in the specified bulk theory. For every fixed single-insertion preparation, the complete $AA$ response is $O(e^{-\beta})$ at large Rindler boost rapidity $\beta$, and the rescaled cross-channel remainder $e^{-\beta}r_\times(\beta)$ decays as $O(e^{-\beta})$ via an exact Mellin discontinuity estimate. A positive Laplace projection therefore isolates the growing cross response and yields an all-scale averaged boundary inequality for the correlation-induced QES displacement.}

\begin{document}
\maketitle
\flushbottom

\section{Introduction}
\label{sec:intro}

In semiclassical holographic duality, causal propagation and entanglement associate two generally distinct bulk regions with a boundary subregion $R$. The causal wedge consists of bulk points that can both receive signals from and send signals to the boundary domain of dependence $D[R]$; its future and past null boundaries meet at the causal surface~\cite{DensityMatrixDual,CausalInfo}. The entanglement wedge is the bulk domain of dependence of a spatial region bounded by $R$ and the quantum extremal surface (QES), which extremizes the generalized entropy~\cite{RT,HRT,FLM,EW}. Consistency of entanglement-wedge reconstruction~\cite{HKLL,ADH,JLMS,DHW} requires causal wedge inclusion---under standard energy and quantum focusing assumptions, the entanglement wedge must contain the causal wedge, leaving a bulk \emph{causal shadow} between them~\cite{GaoWald,Maximin,CausalityHEE,EW,QFC}. What boundary inequality expresses the local sign of this separation? Addressing this question brings together two complementary uses of entanglement in holography: determining which bulk region a boundary density matrix describes, and constraining the geometry of that region through information-theoretic positivity.

When the classical Ryu--Takayanagi (RT) or Hubeny--Rangamani--Takayanagi (HRT) surface is already separated from the causal surface, small quantum corrections perturb a pre-existing classical gap. In that case, positivity of the total separation does not by itself constrain the sign of the quantum correction, since a small shift in either direction leaves the gap positive. A useful setting for isolating the quantum contribution is therefore a degenerate configuration in which the classical extremal and causal surfaces coincide. Correlations between disconnected bulk wedges can then displace the QES even when the background metric is held fixed. This mechanism played a central role in the analysis of Levine, Shahbazi-Moghaddam and Soni, who used cocycle flow and its gravitational backreaction to bring a causally inaccessible portion of the entanglement wedge into causal contact with the boundary~\cite{SeeingEW}. Subsequently, Chen and Wang developed a perturbative description of this correlation-induced causal shadow, computing the QES displacement from the shape dependence of bulk mutual information~\cite{Shape,CW}.

We continue that analysis in vacuum $\text{AdS}_3/\text{CFT}_2$, using a conformal frame in which $R=A\cup B$ consists of a half-line $A$ and a distant small interval $B$ of dimensionless size $\lambda\ll 1$ in the disconnected entanglement-wedge phase. On the $t=0$ slice, the classical RT curve coincides with the causal surface along $x_{\rm P}=0$ on the branch near $A$, where $x_{\rm P}$ and $z$ denote the Poincar\'e spatial and radial coordinates. The $O(\lambda^{2\Delta})$ bulk mutual information between the disconnected wedges~\cite{AgonFaulkner} lifts this degeneracy perturbatively and shifts the QES towards $B$ by a leading displacement of order $\GN\lambda^{2\Delta}$~\cite{Shape,CW},
\begin{equation}
 \eta(z)=G_N\lambda^{2\Delta}zH(\log z)+\cdots,
 \label{eq:introshape}
\end{equation}
where $\GN=3/(2c)$ is Newton's constant, $c$ is the boundary central charge, $\Delta\in(1,2)$ is the dimension of the lightest nonidentity scalar primary $\OO$, and $x=\log z$ is the proper arclength coordinate along the reference curve $x_{\rm P}=0$. Writing $\eta(z)=G_N\lambda^{2\Delta}\eta_*(z)+\cdots$, the dimensionless profile $H(x)=\eta_*(z)/z$ directly measures the leading proper normal thickness of the resulting \emph{quantum causal shadow}. Because there is no classical offset to subtract, the sign of $H(x)$ is the direct bulk witness of quantum causal wedge inclusion on this branch.

The bulk calculation of ref.~\cite{CW} explained this sign through the monotonicity of mutual information and the positivity of the geometric response kernel. Sections 5.1--5.2 of ref.~\cite{CW} then asked how the same constraint could be expressed in the boundary theory. This is the question we take up here: \emph{which positive boundary observable constrains the correlation-induced QES displacement?}

Modular flow provides a natural boundary language for this question because it retains fine-grained dynamical information about a reduced state beyond its von Neumann entropy. While the Bisognano--Wichmann theorem identifies vacuum wedge modular flow with Lorentz boosts~\cite{BW}, conformal symmetry yields a local geometric flow for vacuum balls~\cite{CHM}, and null-plane deformations still generate local Markovian flows on the horizon~\cite{CTTNull}, multi-component and general deformed regions exhibit rich nonlocal modular dynamics even in free theories~\cite{CHNonlocal,NullCuts}. At the same time, the algebraic positivity and monotonicity of relative entropy~\cite{ArakiRE} impose stringent constraints on energy and geometry in quantum field theory and holography, ranging from the Bekenstein and modular-energy bounds~\cite{CasiniRE,BCFull,ModEnergy} to the holographic entanglement first law and canonical energy~\cite{BCHM,LVFisher} as well as the averaged and quantum null energy conditions~\cite{FLPW,HKT,BFKLW,QNEC,CFQNEC}. In bulk reconstruction, modular flow and distinguishability metrics have likewise been used to localize bulk operators and extract extremal-surface geometry~\cite{FL,KLIntersect,FLWToolkit,InfoMetricEW,ShadowsEW}.

The proposed connection between causal-shadow positivity and the nonlocal part of the boundary modular Hamiltonian~\cite{CW} motivated a study of entropy changes under modular flow~\cite{MES}. Evolving the product of the vacuum marginals with the joint modular Hamiltonian, that work established nonnegative entropy shifts for two disjoint intervals in a two-dimensional CFT vacuum.

To connect boundary positivity directly to the QES displacement, we use the modular scattering framework of Chandrasekaran, Faulkner and Levine (CFL), who detected QES positions through high-boost modular-flowed correlators~\cite{CFL}. Using an operator product expansion (OPE) of twist operators~\cite{AgonFaulkner} and replica analytic continuation, CFL derived the leading small-region modular interaction---decomposed into cross terms containing one light operator in each region and same-region terms containing both in a single region---and related its high-boost growth to bulk eikonal gravitational scattering~\cite{EikonalAdS,Regge,SS,RS} and modular chaos~\cite{MSS,ModChaos}. To obtain a boundary inequality for this response, we combine relative-entropy monotonicity under the inclusions $A\subset R$ and $R^c\subset A^c$. For any normalized pure global probe state $|f\rangle$, this gives the exact sign constraint~\cite{CasiniRE,BCFull,ModEnergy,FLPW}:
\begin{equation}
 \PP[f]=(D_R-D_A)+(D_{A^c}-D_{R^c})\ge0,
 \qquad R=A\cup B,
 \label{eq:introP}
\end{equation}
where $D_X$ abbreviates the relative entropy of the reduced probe state $\rho_X^f$ with respect to the reduced vacuum state $\rho_X^\Omega$:
\[
 D_X\equiv D(\rho_X^f\Vert\rho_X^\Omega)
 =\operatorname{Tr}_X\!\left[\rho_X^f
 \left(\log\rho_X^f-\log\rho_X^\Omega\right)\right].
\]
Let $K_X=-\log\rho_X^\Omega$ be the vacuum modular Hamiltonian, and write $\widehat K_X=K_X-K_{X^c}$ for its full modular Hamiltonian. Global purity cancels the entropy variations between complementary regions, leaving $\PP[f]=\langle f|\widehat{K}_R-\widehat{K}_A|f\rangle$. Thus $\PP[f]\ge 0$ is a positive semidefinite expectation value in the probe state, with the reference half-line contribution subtracted exactly. In this paper, we evaluate this observable on normalized single-insertion Euclidean probe states across all boundary scales, all probing the modular operators of the same reference vacuum.

Turning this boundary inequality into a constraint on the QES displacement requires identifying the complete modular response at the relevant perturbative order. In the small-interval expansion, microcausality removes the term with both light operators in $B$, but leaves a cross term with one operator in each region and a same-side term with both operators in $A$. The cross term has a direct scattering interpretation. A Euclidean preparation of a distinct scalar primary creates a normalizable bulk wavepacket, and its horizon null flux samples the QES displacement. To fix the normalization non-circularly, we compare the transverse scattering source with an independent calculation of the generalized-entropy source. Their agreement establishes the quantitative relation between the cross-channel response and the QES displacement.

The same-side $AA$ contribution enters the same boundary observable and must therefore be controlled before the positivity of $\PP$ can constrain the cross response. We therefore evaluate the complete tree-level graviton exchange in the $AA$ channel, including the double-trace families built from the external scalar operators~\cite{WittenBlocks,SpinningPropagators}. Integrating this graph against the modular kernel explores moving endpoints and relative-time tails on the Lorentzian sheets fixed by the operator order. We use Euclidean vacuum preparation and the ordered real-time holographic prescription~\cite{SkenderisVanRees,SkenderisVanReesLong} to evaluate this complete contribution. After transferring the external boost $\beta$ to the modular kernel, a uniform bound on the unboosted graph controls the full integration domain and proves that the complete $AA$ response decays as $O(e^{-\beta})$ for every fixed single-insertion preparation.

The QES displacement is encoded in the coefficient of the growing $e^\beta$ cross response. To extract this coefficient and preserve boundary positivity, we use a positive Laplace projection: a boost integral with a positive weight that selects the $e^\beta$ coefficient and removes all subleading modes. In Einstein gravity with minimally coupled scalars, we first extract the leading small-interval coefficient and then isolate its term linear in $\GN$, before taking the large-boost limit of this gravitational contribution. Using the exact Mellin representation of the tree-level graviton exchange, we also prove directly that the integrated cross-channel remainder beyond the leading $e^\beta$ eikonal term is $O(1)$ at large boost (so its boost-rescaled value decays as $O(e^{-\beta})$). Together with complete $AA$ suppression, the projected boundary inequality becomes positivity of the radial flux average of the QES displacement across all boundary preparation scales $s>0$.

This all-scale averaged inequality supplies the dynamical boundary constraint sought here, but a single Euclidean preparation creates a wavepacket of fixed radial width; positive averages across all scales need not by themselves determine the local pointwise sign of $H(x)$. Overcoming this fixed-width limitation without introducing heavier bulk fields requires retaining quantum interference between distinct preparations. That next step is developed in Paper II~\cite{CompanionTomography}, which generalizes the single-insertion response and tree-level remainder bounds established here to finite coherent superpositions, constructs a complete hierarchy of local positivity criteria via finite preparation derivatives, and applies the resulting tomography to derive exact benchmarks and sharp pointwise bounds for the scalar QES. The present paper establishes the boundary observable, its gravitational displacement dictionary, and its averaged positivity inequality; the second determines how much local information about the causal shadow can be extracted from it.

This paper is organized as follows. Section~\ref{sec:setup} specifies the geometry, bulk theory and boundary inequality. Section~\ref{sec:probe} constructs the single-insertion Euclidean probes and their horizon flux. Section~\ref{sec:dictionary} derives the cross-channel displacement dictionary and its independent source normalization. Section~\ref{sec:abel} establishes complete $AA$ suppression, the cross-channel remainder bound, the positive Laplace projection yielding the all-scale averaged boundary inequality, and the perturbative boost regime, and Section~\ref{sec:discussion} summarizes our conclusions and outlook. The appendices contain the probe normalization, entropy-source Selberg verification, complete $AA$ contour and endpoint analysis, the Mellin proof of the cross-channel Regge remainder bound, and a stress-family check.

\section{Geometry and the boundary inequality}
\label{sec:setup}

To turn causal wedge inclusion into a boundary modular constraint, we first establish the reference bulk geometry, specify the bulk effective theory and real-time holographic prescription, and formulate the exact boundary relative-entropy inequality whose perturbative modular expansion drives our analysis.

\subsection{The selected quantum causal shadow}
We work in a vacuum boundary geometry where the causal surface and the classical extremal surface coincide on the branch of interest. This degeneracy eliminates any classical background offset and fixes a canonical reference surface from which the quantum extremal displacement and its local sign are measured.

Specifically, in the vacuum state $|\Omega\rangle$ of a two-dimensional conformal field theory ($\text{CFT}_2$), we choose the spatial subregions on the time-symmetric slice $t=0$ to be
\begin{equation}
 A=(-\infty,0),\qquad
 B=[1-\lambda/2,1+\lambda/2],\qquad
 R=A\cup B,\qquad r=R^c,
 \quad 0<\lambda\ll1,
 \label{eq:regions}
\end{equation}
where we have set the separation scale between $A$ and the center of $B$ to unity. Throughout this paper, the operator algebra $\mathcal{A}(X)$ of a spatial region $X$ refers to the algebra of its causal domain of dependence. Because the causal developments of $A$ and $B$ are spacelike separated, their local algebras commute. For $\lambda\ll 1$, the classical entanglement wedge of $R$ lies in the disconnected phase, consisting of two separate bulk components anchored to $A$ and $B$, respectively; in what follows, we focus on the branch near the half-line $A$.

The dual reference spacetime is unit-radius Poincar\'e $\text{AdS}_3$, with metric
\begin{equation}
 \dd s^2=\frac{-\dd t^2+\dd x_{\rm P}^2+\dd z^2}{z^2},
 \qquad z>0,\qquad \GN=\frac{3}{2c}.
 \label{eq:metric}
\end{equation}
Here $\GN=3/(2c)$ follows from the Brown--Henneaux relation~\cite{BrownHenneaux} for unit AdS radius, where $\GN$ is the three-dimensional Newton constant and $c$ is the boundary central charge.

\begin{figure}[t]
\centering
\includegraphics[width=\linewidth]{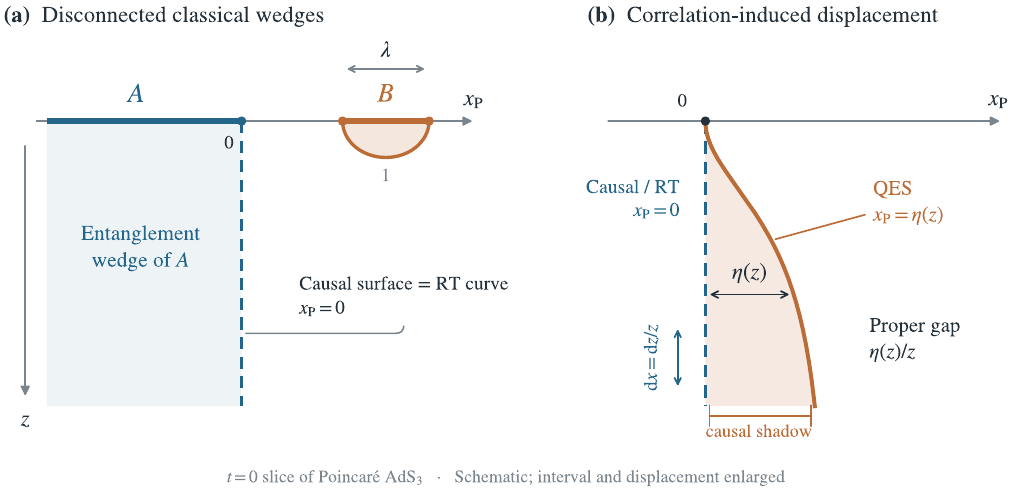}
\caption{Reference geometry and the displacement being reconstructed. (a) A half-line $A$ and a small interval $B$ have disconnected classical entanglement wedges. On the selected branch, the causal surface and Ryu--Takayanagi (RT) curve coincide at $x_{\rm P}=0$. (b) Bulk correlations move the QES and create the shaded causal-shadow strip. The coordinate and leading proper normal widths are $\eta(z)$ and $\eta(z)/z$. The drawings are schematic: the interval and quantum displacement are enlarged, and the second quantum branch is omitted.}
\label{fig:geometry}
\end{figure}

On the $t=0$ slice, the classical Ryu--Takayanagi (RT) curve~\cite{RT} of the component adjacent to $A$ coincides with its causal surface along the radial line $x_{\rm P}=0$, so this branch has vanishing classical causal shadow thickness. When bulk quantum effects are included, nonlocal correlations between the bulk wedges of $A$ and $B$ modify the bulk generalized entropy and displace the quantum extremal surface (QES) towards $B$~\cite{CW}. To probe this vacuum QES, we will evaluate matrix elements of the fixed vacuum modular operator in the boundary states introduced in Section~\ref{sec:probe}. We parameterize the reference and displaced surfaces on the $t=0$ slice by
\begin{equation}
 X_{\rm causal}=X_{\rm RT}:\ x_{\rm P}=0,
 \qquad
 X_{\rm QES}:\ x_{\rm P}=\eta(z),
 \label{eq:surfaces}
\end{equation}
where $\eta(z)>0$ corresponds to a displacement towards $B$ (into the exterior of the causal wedge of $A$), as illustrated schematically in Figure~\ref{fig:geometry}.

The proper normal displacement in the metric~\eqref{eq:metric} is $\eta(z)/z$ to first order in the deformation. We describe its profile along the reference curve using the proper arclength coordinate $x=\log z$, distinct from the Poincar\'e spatial coordinate $x_{\rm P}$. Factoring out the leading perturbative scale $\GN\lambda^{2\Delta}$, we define
\begin{equation}
 \eta(z)=\GN\lambda^{2\Delta}\eta_*(z)+\cdots,
 \qquad x=\log z,
 \qquad H(x)=e^{-x}\eta_*(e^x)=\frac{\eta_*(z)}z.
 \label{eq:H}
\end{equation}
By construction, $H(x)$ is the leading dimensionless coefficient of the proper normal gap, so that the physical proper thickness is $\GN\lambda^{2\Delta}H(x)+\cdots$. Working in the proper coordinate $x=\log z$ is natural both geometrically and analytically, as radial boundary dilations $z\mapsto s z$ act simply as translations $x\mapsto x+\log s$. The profile $H(x)$ is the primary unknown function that we aim to probe through boundary modular matrix elements.

We emphasize that \eqref{eq:H} is thus far only a kinematic definition and does not by itself imply that $H(x)\to 0$ at the asymptotic boundary or in the deep bulk. Near the boundary ($z\to 0$, equivalently $x=\log z\to -\infty$), decay of $H(x)$ requires $\eta_*(z)=o(z)$, which is stronger than simply anchoring the coordinate position $\eta_*(0)=0$. In Section~\ref{sec:source}, we derive the QES Jacobi equation from the generalized entropy and show that the physical solution obeys $\eta_*(z)\sim z^2$ as $z\to 0$ and approaches a finite constant as $z\to\infty$, which in turn guarantees exponential decay $H(x)\sim C_2 e^{-|x|}$ as $x\to\pm\infty$.

\subsection{The bulk theory and vacuum prescription}
We assume that the leading correlation correction between $A$ and $B$ is controlled by a unique lightest nonidentity primary $\OO$ of any spin, which is a scalar of scaling dimension $1<\Delta<2$ dual to a bulk scalar field $\varphi$ (in particular, there are no lower-dimension spinning primaries with scaling dimension $\le\Delta$). To probe the resulting geometry without mixing the probe and background sectors at generalized-free order, we prepare boundary excitations using a second, distinct scalar primary $\OO_\phi$ of dimension $d>2$, dual to a bulk scalar field $\phi$. (Throughout, the boundary spacetime dimension is two; neither $d$ nor $\Delta$ denotes a spacetime dimension.) Choosing $1<\Delta<2$ selects standard $\text{AdS}_3$ quantization, places the scalar double-trace dimensions $2\Delta+2n$ strictly above two, and cleanly separates the leading scalar-pair small-interval scaling $\lambda^{2\Delta}$ from the vacuum stress-tensor contribution at order $\lambda^4$.

Concretely, we work with the minimally coupled two-derivative Lorentzian effective action
\begin{align}
 I={}&\frac1{16\pi G_N}\int\sqrt{-g}\,(\mathcal R+2)\dd^3X
 -\frac12\int\sqrt{-g}\,
 \big[(\nabla\varphi)^2+m_\varphi^2\varphi^2
      +(\nabla\phi)^2+m_\phi^2\phi^2\big]\dd^3X,
 \nonumber\\[-2pt]
 &m_\varphi^2=\Delta(\Delta-2),\qquad m_\phi^2=d(d-2),
 \label{eq:selectedaction}
\end{align}
where $\mathcal R$ is the bulk Ricci scalar and standard Gibbons--Hawking and holographic renormalization counterterms are implicit. Although $m_\varphi^2<0$, it lies above the $\text{AdS}_3$ Breitenlohner--Freedman bound $m_{\mathrm{BF}}^2=-1$~\cite{BreitenlohnerFreedman}. Both real scalar fields are canonically normalized and minimally coupled to Einstein gravity, with fixed renormalized external dimensions and unit-normalized boundary two-point functions. The dimensions $d$ and $\Delta$ are held fixed as $c\to\infty$. The $\mathbb Z_2$ symmetry $\varphi\mapsto-\varphi$ of \eqref{eq:selectedaction} forbids the cubic vertices $\varphi\phi^2$ and $\varphi^3$. At the connected tree-level order $\GN$ retained below, we omit all other direct cubic matter couplings and all quartic matter couplings. The latter include $\varphi^2\phi^2$ and higher-derivative spin-2 contact vertices such as $(\nabla_\mu\varphi\nabla_\nu\varphi)(\nabla^\mu\phi\nabla^\nu\phi)$. We also omit additional light exchange fields and higher-derivative curvature interactions.

It is important to be precise about what ``Einstein tree order'' means in our analysis. This phrase specifies two independent restrictions: two-derivative Einstein gravitational dynamics and tree-level Witten diagrams without bulk loops. At order $G_N$, the single-stress contribution consists of the two chiral global conformal families of $T$ and $\bar T$. The complete exchange diagram also contains scalar double-trace families~\cite{WittenBlocks}; its crossed-channel Mellin expansion used in Appendix~\ref{app:crossremainder} has both regular and logarithmic terms. The QES displacement is a quantum entanglement correction even though this connected probe four-point coefficient is represented by a tree-level graviton exchange. While a large central charge $c$, a sparse light spectrum, and a large higher-spin gap $\Delta_{\rm gap}\to\infty$ motivate a local two-derivative bulk effective action~\cite{HPPS,AdSEFT,CEMZ}, they do not automatically forbid independent matter contact interactions; in particular, a spin-2 contact term $(\nabla_\mu\varphi\nabla_\nu\varphi)(\nabla^\mu\phi\nabla^\nu\phi)$ would contribute at order $\GN e^\beta$ to the Regge cross response without shifting the generalized-entropy QES profile $\eta(z)$, while in Regge-behaved theories where a finite higher-spin gap $\Delta_{\rm gap}<\infty$ lowers the leading Regge intercept to $J_0<2$~\cite{Regge,SS}, an $O(e^{(J_0-1)\beta})$ cross response is projected to zero by the strict $\epsilon_A\downarrow 0$ Laplace filter \eqref{eq:Abel} (defined in Section~\ref{sec:Abelresidue}). Setting such contact terms to zero and taking the strict Einstein limit in \eqref{eq:selectedaction} is therefore an explicit specification of our bulk model.

We evaluate real-time modular-flowed correlators using the holographic prescription of Skenderis and van Rees~\cite{SkenderisVanRees,SkenderisVanReesLong}. The Euclidean segments prepare the states used in the Lorentzian calculation, with the bulk fields and their canonical momenta matched at the Euclidean--Lorentzian interfaces. Infinitesimal imaginary-time shifts specify the Wightman operator ordering; the corresponding imaginary-time positions and integration contours for the two $AA$ orderings are given in Appendix~\ref{app:AAfull} (eqs.~\eqref{eq:AA_slots_full} and \eqref{eq:AA_cycles_full}), and those for the cross channel in Appendix~\ref{app:crossremainder}. We then continue the Witten diagram along these contours, retaining the branch of the double-trace logarithm in \eqref{eq:AA_full_normal_sum} (Appendix~\ref{app:crossremainder}) reached along the continuation path. Vacuum modular flow in the half-line wedge $A$ is a Rindler boost, with a thermal circle of period $2\pi$ in imaginary Rindler time. The Kubo--Martin--Schwinger (KMS) condition constrains the imaginary-time insertion positions to respect the cyclic operator ordering on this circle. In Appendices~\ref{app:AAfull} and~\ref{app:crossremainder}, we track the integration contours for both channels explicitly to prove that the complete $AA$ exchange decays as $O(e^{-\beta})$ and that the integrated cross-channel remainder beyond the leading $e^\beta$ eikonal term is $O(1)$.

Finally, we emphasize the separation of roles between the vacuum state and the probe state. The reference vacuum state $|\Omega\rangle$, its causal horizon, and the light-pair modular correlation correction are held fixed throughout. Varying the boundary probe state $|f\rangle$ simply selects different expectation values of this fixed operator, with their gravitational interaction computed at order $\GN$ via the connected four-point Witten diagram; changing the probe does not alter the reference vacuum profile $H(x)$ in \eqref{eq:introshape}. At finite $\GN$, the corresponding probe regime requires weak scattering, as discussed in Section~\ref{sec:finitewindow}.

\subsection{Relative entropy and the full modular difference}
Before introducing any bulk or perturbative approximation, relative-entropy monotonicity provides an exact, nonperturbative positivity constraint on the boundary. When evaluated in pure global probe states, complementary von Neumann entropy variations cancel identically, reducing the relative-entropy combination to a positive semidefinite expectation value of the full modular Hamiltonian difference.

For any normalized pure global state $|f\rangle$ and spatial region $X$, we write the relative entropy, one-sided vacuum modular Hamiltonian, and full vacuum modular Hamiltonian as
\begin{equation}
 D_X(f\Vert\Omega)=\operatorname{Tr}\rho_X^f
   (\log\rho_X^f-\log\rho_X^\Omega),\qquad
 K_X=-\log\rho_X^\Omega,\qquad
 \widehat K_X=K_X-K_{X^c}.
 \label{eq:modulardefs}
\end{equation}
In continuum quantum field theory, reduced density matrices and one-sided modular Hamiltonians serve as regulated shorthand for the algebraic Tomita--Takesaki modular operator and Araki relative entropy~\cite{ArakiRE,WittenEntanglement}. Throughout our operator derivations, we assume the standard vacuum representation together with a common factorizing ultraviolet regulator whose quadratic-form continuum limit yields the full modular generators of the corresponding local algebras, and we assume Haag duality when identifying algebraic commutants with geometric complements $X^c$. The single-insertion Euclidean preparations introduced in Section~\ref{sec:probe} (as well as their coherent superpositions analyzed in Paper II~\cite{CompanionTomography}) lie in the common quadratic-form domain of these generators.

Because $A\subset R$ and $r=R^c\subset A^c$, the monotonicity of relative entropy under subsystem inclusion immediately yields the exact double-difference inequality~\cite{CasiniRE,BCFull,ModEnergy,FLPW}
\begin{equation}
 \PP[f]=(D_R-D_A)+(D_{A^c}-D_r)\ge0.
 \label{eq:P}
\end{equation}
Note that $\PP[f]$ is a sum of two monotonic relative-entropy differences---each comparing $|f\rangle$ against $|\Omega\rangle$ on nested regions---rather than a relative entropy between two regions. Using the standard identity
\begin{equation}
 D_X=\Delta\langle K_X\rangle-\Delta S_X,
 \label{eq:Didentity}
\end{equation}
together with the purity of both $|f\rangle$ and $|\Omega\rangle$ on the full Cauchy slice ($\Delta S_R=\Delta S_r$ and $\Delta S_A=\Delta S_{A^c}$), all four von Neumann entropy variations cancel out of \eqref{eq:P}. Furthermore, because the vacuum is invariant under full modular flow, $(K_X-K_{X^c})|\Omega\rangle=0$, the vacuum expectation values vanish as well, leaving the pure quadratic form
\begin{equation}
 \PP[f]=\langle f|\widehat\PP|f\rangle,\qquad
 \widehat\PP=\widehat K_R-\widehat K_A.
 \label{eq:Poperator}
\end{equation}
(If one instead used a mixed global probe state, the monotonicity inequality \eqref{eq:P} would still hold, but the entropy cancellation leading to the linear operator form \eqref{eq:Poperator} would no longer apply.) Because \eqref{eq:Poperator} holds for any pure state in the common form domain, it will also retain all off-diagonal interference terms when applied to coherent superpositions in Paper II~\cite{CompanionTomography}.

It is instructive to compare $\PP[f]$ with the full modular energy studied by CFL~\cite{CFL}, which in our notation reads $D_r-D_R=-\langle f|\widehat K_R|f\rangle$. Reorganizing \eqref{eq:P} as
\begin{equation}
 \PP=(D_{A^c}-D_A)-(D_r-D_R)
 \label{eq:CFLcomparison}
\end{equation}
shows that $\PP[f]$ subtracts the known half-line Rindler boost $\widehat K_A$ from $\widehat K_R$. This subtraction accomplishes two things simultaneously: it isolates the nonlocal modular interaction induced by the distant interval $B$, and it upgrades the sign-indefinite expectation value $\langle f|\widehat K_R|f\rangle$ into a manifestly nonnegative observable $\PP[f]\ge 0$.

\subsection{The modular correlation correction}
In the commutator representation developed in Section~\ref{sec:dictionary}, the complementary generators drop out by locality, so evaluating $\PP[f]$ requires only the one-sided difference $K_R-K_A$. We therefore expand $K_R$ for $R=A\cup B$ at small interval size $\lambda\ll1$. Adopting the modular flow convention
\begin{equation}
 \OO_X(\sigma)=(\rho_X^\Omega)^{i\sigma/(2\pi)}\OO_X
              (\rho_X^\Omega)^{-i\sigma/(2\pi)}
 =e^{-i\sigma K_X/(2\pi)}\OO_Xe^{i\sigma K_X/(2\pi)},
 \label{eq:flow}
\end{equation}
so that $\partial_\sigma\OO_X(\sigma)=-\frac{i}{2\pi}[K_X,\OO_X(\sigma)]$, the leading light-scalar contribution to the small-interval modular OPE takes the form~\cite{CFL,CW}
\begin{equation}
 K_R-K_A=K_B+\delta K_\times+\delta K_{AA}+\delta K_{BB}+\cdots,
 \label{eq:modularOPE}
\end{equation}
where the subscripts $\times$, $AA$, and $BB$ label the regional supports of the two light scalar insertions $\OO$ appearing in each bilocal modular integral:
\begin{center}
\begingroup
\renewcommand{\arraystretch}{1.25}
\begin{tabularx}{\linewidth}{@{}>{\centering\arraybackslash}p{0.28\linewidth}Y@{}}
\toprule
\textbf{Term} & \textbf{Operator content}\\
\midrule
$\delta K_\times$ & one $\OO_A$ and one $\OO_B$\\
\addlinespace[2pt]
$\delta K_{AA}$ & two $\OO_A$ insertions\\
\addlinespace[2pt]
$\delta K_{BB}$ & two $\OO_B$ insertions\\
\bottomrule
\end{tabularx}
\endgroup
\end{center}
All three bilinear terms arise at order $\lambda^{2\Delta}$ from nonlocal correlations between $A$ and $B$. Although supported on $A$, $\delta K_{AA}$ arises from correlations with $B$ and therefore survives the subtraction of $K_A$.

Before computing any four-point correlator, bulk modular geometry offers a helpful pre-calculation intuition for why taking a large Rindler boost $\beta\gg 1$ isolates the QES displacement $\eta(z)$. Under the Jafferis--Lewkowycz--Maldacena--Suh (JLMS) holographic modular-Hamiltonian correspondence~\cite{JLMS}, $\widehat K_A$ generates the exact $\text{AdS}_3$ Killing boost $-u\partial_u+v\partial_v$ around the degenerate reference surface $u=v=0$ (in the lightcone coordinates $u=t+x_{\rm P}$, $v=t-x_{\rm P}$ introduced in \eqref{eq:nullcoords}). Displacing the entanglement-wedge boundary at $t=0$ to the QES at $x_{\rm P}=\eta(z)$, i.e.\ $(u,v)=(\eta(z),-\eta(z))$, shifts the modular generator on the $u=0$ horizon from $v\partial_v$ to $(v+\eta(z))\partial_v$ (plus terms of boost weight $\le 0$), yielding a positive boost-weight $+1$ null translation:
\[
 \widehat K_R-\widehat K_A
 \;\sim\; \bigl(v+\eta(z)\bigr)\partial_v - v\partial_v
 \;=\; \eta(z)\partial_v
 \;\sim\; 2\pi\int\frac{\dd z}{z}\,\dd v\,\eta(z)\,T_{vv}.
\]
Acting on a probe state with the unitary vacuum Rindler boost $U(\beta)$ preserves the exact inequality $\PP_\beta[f]:=\PP[U(\beta)f]\ge 0$ while amplifying the null momentum $P_v$ by $e^\beta$, thereby enhancing the $\eta(z)$ signal as $e^\beta$ and driving the bulk gravitational exchange into the Regge shockwave regime. The same heuristic suggests why the growing contribution should originate from the horizon-straddling cross term $\delta K_\times$ (connecting $\OO_A$ and $\OO_B$) rather than the single-sided term $\delta K_{AA}$. We emphasize, however, that this JLMS picture is strictly pre-calculation motivation: our quantitative derivation in Sections~\ref{sec:dictionary} and~\ref{sec:abel} does not assume it, and instead proves the QES dictionary non-circularly by matching the cross-channel scattering source against an independent generalized-entropy calculation while directly bounding both $\delta K_{AA}$ and the cross-channel remainder.

In Section~\ref{sec:dictionary}, we show how rewriting $\PP[f]$ as an analytically continued Lorentzian commutator eliminates $K_B$ and $\delta K_{BB}$ by microcausality, and we compute the cross-channel response $\delta K_\times$ to establish the QES displacement dictionary. Because the exact boundary inequality $\PP[f]\ge 0$ constrains the sum of all surviving channels, transferring this sign constraint to the geometric cross readout requires proving that both $\delta K_{AA}$ and the subleading cross-channel remainder do not contaminate the maximal $e^\beta$ growth rate---both bounds are established in Section~\ref{sec:abel} and proved in Appendices~\ref{app:AAfull} and~\ref{app:crossremainder}. The next section derives the bulk horizon wavefunctions and null-energy fluxes needed to evaluate these modular responses.

\section{Euclidean probes and horizon flux}
\label{sec:probe}

In the high-boost Regge regime, graviton exchange with the light scalar pair admits an effective shockwave description~\cite{CFL,EikonalAdS,SS}. The probe couples to this shock through the null component of its bulk stress tensor on the reference Rindler horizon $u=0$. A Euclidean boundary insertion of the probe scalar primary $\OO_\phi$ prepares a normalizable one-particle bulk wavepacket. We compute its wavefunction on the horizon and its null-momentum flux integrated along the horizon generators in closed form. The total flux is fixed by the probe's mean CFT energy, and its radial distribution determines how the probe weights the QES displacement along the reference RT curve.

\subsection{Euclidean preparation, boundary norm, and mean energy}
We prepare a single-insertion probe state by acting on the vacuum with $\OO_\phi$ at spatial position $\xi>0$ (inside $A^c$) and damping with the CFT Hamiltonian $H_{\rm CFT}$ over a Euclidean preparation time $\tau>0$:
\begin{equation}
 |\widetilde f_{\tau,\xi}\rangle
 =e^{-\tau H_{\rm CFT}}\OO_\phi(0,\xi)|\Omega\rangle
 =\OO_\phi(i\tau,\xi)|\Omega\rangle,
 \label{eq:unnormalizedstate}
\end{equation}
where we used $H_{\rm CFT}|\Omega\rangle=0$ and Heisenberg evolution $\OO_\phi(t,\xi)=e^{itH_{\rm CFT}}\OO_\phi(0,\xi)e^{-itH_{\rm CFT}}$. In complexified Lorentzian coordinates $(t,x_{\rm P})$, the ket insertion therefore sits at $(i\tau,\xi)$ in the upper half-plane, while its Hermitian conjugate bra insertion sits at $(-i\tau,\xi)$ in the lower half-plane.

Adopting the unit-normalized boundary two-point convention on the Euclidean-preparation sheet,
\begin{equation}
 \langle\OO_\phi(t_1,x_1)\OO_\phi(t_2,x_2)\rangle_\Omega
 =\bigl[(x_1-x_2)^2-(t_1-t_2)^2\bigr]^{-d},
 \label{eq:boundary2pt}
\end{equation}
the norm of the unnormalized state \eqref{eq:unnormalizedstate} at equal spatial position ($x_1=x_2=\xi$, $t_1-t_2=-2i\tau$) is simply
\begin{equation}
 Z(\tau):=\langle\widetilde f|\widetilde f\rangle=(2\tau)^{-2d}.
 \label{eq:Ztau}
\end{equation}
Dividing by $\sqrt{Z(\tau)}$ yields the normalized single-insertion probe state
\begin{equation}
 |f_{\tau,\xi}\rangle=(2\tau)^d
 e^{-\tau H_{\rm CFT}}\OO_\phi(0,\xi)|\Omega\rangle.
 \label{eq:state}
\end{equation}
Throughout this paper we work with single-insertion probes of the form \eqref{eq:state}; coherent superpositions of distinct single-insertion states are developed in Paper II~\cite{CompanionTomography}.

The mean CFT energy of the normalizable wavepacket $|f_{\tau,\xi}\rangle$ is obtained by differentiating the Euclidean norm $Z(\tau)$:
\begin{equation}
 E_f:=\langle f|H_{\rm CFT}|f\rangle
 =-\frac12\partial_\tau\log Z(\tau)
 =\frac d\tau.
 \label{eq:Ef}
\end{equation}
As we will see below, $E_f$ sets the overall signal amplitude and governs the weak-scattering window at finite $G_N$.

To distinguish the bulk sampling depth from the probe energy, we introduce polar coordinates $(s,\delta)$ in the Euclidean preparation plane:
\begin{equation}
 s=\sqrt{\tau^2+\xi^2},\qquad
 \tau=s\sin\delta,\qquad \xi=s\cos\delta,
 \qquad0<\delta<\frac\pi2.
 \label{eq:prep}
\end{equation}
As shown below, $s$ sets the characteristic bulk sampling depth $z\sim s$. At fixed $s$, varying $\delta$ adjusts the mean energy $E_f=d/(s\sin\delta)$ without changing the normalized radial sampling profile.

\subsection{From the boundary insertion to a horizon wavefunction}
In the linearized bulk theory \eqref{eq:selectedaction}, the excitation $|f_{\tau,\xi}\rangle$ is dual to a one-particle state of the canonically normalized scalar field $\phi$ of mass $m_\phi^2=d(d-2)$, characterized by the bulk wavefunction
\begin{equation}
 \psi_f(t,x_{\rm P},z):=\langle\Omega|\phi(t,x_{\rm P},z)|f\rangle.
 \label{eq:wavedef}
\end{equation}
As derived in Appendix~\ref{app:probenorm}, matching the canonical bulk action normalization to the unit boundary two-point convention \eqref{eq:boundary2pt} fixes the mixed bulk--boundary propagator coefficient in $\text{AdS}_3$ to be $1/\sqrt{2\pi}$, yielding
\begin{equation}
 \psi_f(t,x_{\rm P},z)
 =\frac{(2\tau)^d}{\sqrt{2\pi}}
 \frac{z^d}{[z^2+(x_{\rm P}-\xi)^2-(t-i\tau)^2]^d},
 \label{eq:fullwave}
\end{equation}
where the complex power is evaluated on the analytic sheet continuable from positive Euclidean time $\tau>0$.

Introducing bulk lightcone coordinates adapted to the Rindler horizon of $A$,
\begin{equation}
 u=t+x_{\rm P},\qquad v=t-x_{\rm P},
 \qquad -\dd t^2+\dd x_{\rm P}^2=-\dd u\dd v,
 \label{eq:nullcoords}
\end{equation}
we restrict the wavefunction \eqref{eq:fullwave} to the reference horizon $u=0$, where $t=v/2$ and $x_{\rm P}=-v/2$. The quadratic denominator simplifies linearly in the null affine coordinate $v$:
\begin{equation}
 (x_{\rm P}-\xi)^2-(t-i\tau)^2
 =\left(-\frac v2-\xi\right)^2-\left(\frac v2-i\tau\right)^2
 =s^2+(\xi+i\tau)v,
 \label{eq:horizonsubstitution}
\end{equation}
so that the horizon wavefunction takes the compact form
\begin{equation}
 \psi_f(v,z)=\frac{(2\tau)^d}{\sqrt{2\pi}}
       \frac{z^d}{[z^2+s^2+(\xi+i\tau)v]^d}.
 \label{eq:wave}
\end{equation}
The horizon wavefunction \eqref{eq:wave} is analytic for all $\tau>0$, and its complex phase dependence in $(\tau,\xi)$ will also govern the off-diagonal interference fluxes studied in Paper II~\cite{CompanionTomography}.

\subsection{Positive null momenta}
Because our subsequent scattering and overlap calculations integrate products of horizon wavefunctions over $v\in\mathbb{R}$, it is advantageous to expand $\psi_f(v,z)$ into positive null-momentum plane waves $e^{-ipv}$ ($p>0$), so that the longitudinal $v$-integral produces a momentum-conserving delta function.

Defining the complex preparation parameters
\begin{equation}
 w=\tau+i\xi=ise^{-i\delta},\qquad b=\frac1{\bar w},\qquad
 n_w=\frac{(2\tau b)^d}{\sqrt{2\pi}\Gamma(d)},
 \label{eq:wbn}
\end{equation}
which satisfy $w\bar w=s^2$ and the algebraic identities
\begin{equation}
 bs^2=w,\qquad \frac{i}{b}=\xi+i\tau,\qquad
 z^2+s^2+(\xi+i\tau)v=b^{-1}(w+bz^2+iv),
 \label{eq:factorization}
\end{equation}
we can apply the standard Schwinger--Laplace representation
\begin{equation}
 \zeta_0^{-d}=\frac1{\Gamma(d)}\int_0^\infty p^{d-1}e^{-p\zeta_0}\dd p,
 \qquad \operatorname{Re}\zeta_0>0,
 \label{eq:Gammaidentity}
\end{equation}
with $\zeta_0=w+bz^2+iv$, whose real part $\operatorname{Re}(w+bz^2)=\tau(1+z^2/s^2)>0$ is strictly positive for all $\tau>0$ and $z>0$. Substituting \eqref{eq:factorization} and \eqref{eq:Gammaidentity} into \eqref{eq:wave} yields the positive null-momentum representation
\begin{equation}
 \psi_f(v,z)=\int_0^\infty\dd p\,
 n_w z^d p^{d-1}e^{-p(w+bz^2)}e^{-ipv},
 \label{eq:momentum}
\end{equation}
where $p>0$ denotes the null momentum along the $v$-generator (distinct from the mean boundary energy $E_f$), and fractional powers for noninteger $d$ are defined by analytic continuation from $\xi=0$ ($b>0$).

\subsection{The null-flux profile}
In the eikonal limit, graviton exchange between the light-scalar pair and the probe couples the shockwave profile on $u=0$ to the probe's integrated null stress tensor $\int\dd v\,T_{vv}^\phi(v,z)$. For a canonically normalized real scalar field in the one-particle state $|f\rangle$, normal-ordered vacuum subtraction on the horizon gives
\begin{equation}
 \Delta\langle T_{vv}(v,z)\rangle_f=2|\partial_v\psi_f(v,z)|^2.
 \label{eq:Tvvoneparticle}
\end{equation}
Because the transverse cross-section of the horizon carries the proper line element $\dd z/z$, we define the radial null-flux density $t_f(z)$ with respect to the coordinate measure $\dd z$ via
\begin{equation}
 t_f(z)\dd z
 =\frac{\dd z}{z}\int_{\mathbb R}\dd v\,
 \Delta\langle T_{vv}(v,z)\rangle_f,
 \label{eq:tfmeaning}
\end{equation}
so that $t_f(z)\dd z$ measures the total $v$-directed null momentum crossing the horizon within the radial interval $[z,z+\dd z]$.

Inserting the momentum representation \eqref{eq:momentum} into \eqref{eq:tfmeaning} (writing $p$ and $q$ for the null momenta in $\partial_v\psi_f$ and $\partial_v\psi_f^*$), the longitudinal $v$-integral produces $2\pi\delta(p-q)$, and the remaining Gamma integral evaluates to
\begin{align}
 t_f(z)&=4\pi|n_w|^2z^{2d-1}\int_0^\infty p^{2d}
 e^{-2p\operatorname{Re}(w+bz^2)}\dd p\nonumber\\
 &=\frac{E_f}{2}q_d(z;s),\qquad
 q_d(z;s)=\frac{2s^{2d+2}}{B(d,d+1)}
             \frac{z^{2d-1}}{(z^2+s^2)^{2d+1}},
 \label{eq:flux}
\end{align}
where $B(a,b)=\Gamma(a)\Gamma(b)/\Gamma(a+b)$ denotes Euler's beta function. Changing variables to $r=z^2/s^2$ immediately confirms the normalization identities (see Appendix~\ref{app:probenorm})
\begin{equation}
 \int_0^\infty q_d(z;s)\dd z=1,
 \qquad
 \int_0^\infty t_f(z)\dd z=\frac{E_f}{2}.
 \label{eq:fluxnormalization}
\end{equation}
Equation~\eqref{eq:fluxnormalization} has a clear physical meaning: the single-insertion wavepacket carries total $v$-directed horizon null momentum $E_f/2$, distributed radially according to the unit-normalized profile $q_d(z;s)\dd z$. Under an active Rindler boost $\psi(v,z)\mapsto\psi(e^\beta v,z)$, the two $\partial_v$ derivatives in $T_{vv}$ supply $e^{2\beta}$ while the $\dd v$ measure supplies $e^{-\beta}$, scaling the flux density linearly as $t_f(z)\mapsto e^\beta t_f(z)$. As we will establish in Section~\ref{sec:dictionary} by matching the tree-level shockwave source against an independent generalized-entropy calculation, the leading $O(e^\beta)$ growing cross-channel response $\PP_\times^{\rm grow}[f]$ pairs this boosted flux directly with the QES displacement $\eta(z)$,
\[
 \PP_\times^{\rm grow}[f]
 =2\pi e^\beta\int_0^\infty t_f(z)\eta(z)\dd z
 =\pi E_fe^\beta\int_0^\infty q_d(z;s)\eta(z)\dd z,
\]
so that $E_f e^\beta$ sets the overall signal strength while $q_d(z;s)$ acts as the radial sampling weight.

\section{The cross response and the QES displacement}
\label{sec:dictionary}

Having characterized the horizon null-flux distribution of our probe wavepackets, we now compute the leading cross-channel modular response $\PP_\times^{\rm grow}[f]$ via tree-level Einstein graviton exchange and compare its transverse source $\Pi_\Delta(z)$ with the generalized-entropy source $p_{\rm ent}(z)$ that governs the QES position. By proving an exact normalization match between these two independently derived bulk sources, we establish a parameter-free dictionary relating the boundary cross response to the bulk QES displacement.

\subsection{The real-support commutator and physical continuation}
Evaluating $\widehat\PP=\widehat K_R-\widehat K_A$ directly between Euclidean-prepared bra and ket states obscures microcausal cancellations between spacelike-separated regions. Following CFL~\cite{CFL}, we therefore begin with an auxiliary Lorentzian correlator whose external probe insertions lie at real, spacelike-separated supports inside the left and right wedges:
\begin{equation}
 \Phi_{\rm L}\in\mathcal A(A),
 \qquad
 \Phi_{\rm R}\in\mathcal A(A^c),
 \label{eq:externalSupports}
\end{equation}
where the subscripts $\mathrm{L}$ and $\mathrm{R}$ label left- and right-wedge operator placements rather than subregions. Within the common factorizing regulator of Section~\ref{sec:setup}, the one-sided modular generator $K_r$ acts solely on the $r=R^c$ tensor factor and commutes with $\Phi_{\rm L}\in\mathcal{A}(A)\subset\mathcal{A}(R)$, while vacuum modular invariance gives $\langle\Omega|(K_R-K_r)=0$. Combining these identities with their counterparts for $A$ and $A^c$ converts the matrix element of $\widehat\PP$ into a single commutator with $\Phi_{\rm L}$:
\begin{align}
 \langle\Phi_{\rm L}(K_R-K_r)\Phi_{\rm R}\rangle_\Omega
 &=-\langle[K_R,\Phi_{\rm L}]\Phi_{\rm R}\rangle_\Omega,\\
 \mathcal M_{\PP}
 &:=\langle\Phi_{\rm L}\widehat\PP\Phi_{\rm R}\rangle_\Omega
 =\langle[K_A-K_R,\Phi_{\rm L}]\Phi_{\rm R}\rangle_\Omega,
 \label{eq:M}
\end{align}
understood in the continuum as an identity between regulated quadratic forms.

To pass from the auxiliary Lorentzian correlator $\mathcal M_{\PP}$ back to the physical Euclidean-prepared expectation value $\PP_\beta[f]$ at preparation parameters $(s,\delta)$, we parameterize the right and left external insertions in Rindler coordinates by real boosts $(r,\ell)$,
\begin{equation}
 (u_R,v_R)=(s e^{r},-s e^{-r}),\qquad
 (u_L,v_L)=(-s e^{-\ell},s e^{\ell}),
\end{equation}
and analytically continue along the Wightman sheet specified in eq.~(4.3) of ref.~\cite{CFL}:
\begin{equation}
 r=\beta+i\delta,\qquad
 \ell=-\beta-i\pi+i\delta.
 \label{eq:explicitACpath}
\end{equation}
At $\beta=0$, this continuation moves the right insertion to the ket position $(i\tau,\xi)$ and uses the $-i\pi$ Rindler half-turn to bring the left insertion across the horizon into the conjugate bra position $(-i\tau,\xi)$ in the lower half-plane; at general $\beta\ge 0$, it prepares the Rindler-boosted probe $U(\beta)|f\rangle$. Denoting this analytic continuation along \eqref{eq:explicitACpath} to the boosted probe state by $\operatorname{AC}_f$ and the unnormalized state-preparation operator by $V_f$ (so $|\widetilde f\rangle=V_f|\Omega\rangle$), we recover the positive physical expectation value
\begin{equation}
 \operatorname{AC}_f[\mathcal M_{\PP}]
 =\mathcal Z_f\PP_\beta[f],
 \qquad \mathcal Z_f=\langle\Omega|V_f^\dagger V_f|\Omega\rangle=Z(\tau)>0,
 \label{eq:AC}
\end{equation}
which reduces to $\PP[f]$ at $\beta=0$ (with $\mathcal Z_f=Z(\tau)$ independent of $\beta$ by unitarity of $U(\beta)$). Crucially, all microcausal commutator simplifications are performed at real Lorentzian supports prior to applying $\operatorname{AC}_f$, whereas the positivity constraint $\PP_\beta[f]\ge 0$ applies only after the analytic continuation \eqref{eq:explicitACpath} has restored a Hermitian bra--ket expectation value.

\subsection{Small-interval channels and why \texorpdfstring{$BB$}{BB} drops out}
We insert the small-interval expansion of $K_R-K_A$ in \eqref{eq:modularOPE} into the commutator \eqref{eq:M}. Because real modular flow generated by $K_B$ preserves the local algebra $\mathcal{A}(B)$, every real-flowed light insertion $\OO_B(\sigma)$ remains supported in $B$ and therefore commutes with $\Phi_{\rm L}\in\mathcal{A}(A)$ by spacelike locality:
\begin{equation}
 [K_B,\Phi_{\rm L}]=[\delta K_{BB},\Phi_{\rm L}]=0,
 \qquad
 [\OO_A\OO_B,\Phi_{\rm L}]=[\OO_A,\Phi_{\rm L}]\OO_B.
 \label{eq:BB}
\end{equation}
Thus, both the local generator $K_B$ and the same-side $BB$ interaction $\delta K_{BB}$ drop out identically from the commutator $\mathcal M_{\PP}$, leaving only the cross channel $\delta K_\times$ and the same-side $AA$ channel $\delta K_{AA}$ at order $\lambda^{2\Delta}$:
\begin{equation}
 \mathcal M_{\PP}
 =-\langle[\delta K_\times,\Phi_{\rm L}]\Phi_{\rm R}\rangle_\Omega
  -\langle[\delta K_{AA},\Phi_{\rm L}]\Phi_{\rm R}\rangle_\Omega+\cdots.
 \label{eq:remainingchannels}
\end{equation}
Each surviving term in \eqref{eq:remainingchannels} is an integral of a connected four-point function of two light scalars $\OO$ and two probe insertions $\OO_\phi$ against a known modular kernel. We now evaluate the cross term and identify its growing $e^\beta$ piece with the bulk QES displacement; Section~\ref{sec:abel} then proves that the second term $\delta K_{AA}$ is suppressed as $O(e^{-\beta})$.

\subsection{The cross-channel scattering source}
At leading order in $\lambda\ll 1$, the cross-channel modular operator in \eqref{eq:remainingchannels} takes the universal form~\cite{CFL}
\begin{equation}
 \delta K_\times=-\lambda^{2\Delta}\int_{\mathbb R}
 \frac{\dd\sigma}{[2\cosh(\sigma/2)]^{2\Delta+2}}
 \OO_A(\sigma)\OO_B,
 \label{eq:crossK}
\end{equation}
where under the modular flow convention \eqref{eq:flow} the two light scalar insertions sit at lightcone coordinates $(u_A,v_A)=(-e^\sigma,e^{-\sigma})$ in $A$ and $(u_B,v_B)=(1,-1)$ at the center of $B$, with their Wightman $i0$ prescriptions inherited from the analytic continuation \eqref{eq:AC}.

In the bulk, the connected four-point function $\langle[\OO_A(\sigma),\Phi_{\rm L}]\OO_B\Phi_{\rm R}\rangle_{\rm conn}$ at order $\GN$ is governed by tree-level graviton exchange between the light scalar $\varphi$ and the probe scalar $\phi$, with interaction vertex $\frac{1}{2}\int\sqrt{-g}\,h^{\mu\nu}T_{\mu\nu}$ yielding an amplitude of the form $\GN\int T^\varphi G^{\rm grav}T^\phi$. Under a large external boost $\beta\gg 1$ of the probe pair, the exchange enters the Regge regime~\cite{EikonalAdS,SS,CFL}, where the dominant interaction couples the probe's $v$-momentum flux $T_{vv}^\phi$ along the horizon $u=0$ to the light pair's $u$-momentum flux $T_{uu}^\varphi$ along the intersecting horizon $v=0$. Integrating over the longitudinal null coordinates reduces the graviton propagator to the one-dimensional transverse radial differential operator $L$ and its Green function $G_L(z,z')$:
\begin{equation}
 L=-\partial_z(z^{-1}\partial_z),\qquad
 G_L(z,z')=\tfrac12\min(z,z')^2,\qquad
 L_zG_L(z,z')=\delta(z-z').
 \label{eq:Green}
\end{equation}
Here the two homogeneous solutions of $Lh=0$ are $1$ and $z^2$; imposing the Dirichlet condition $G_L(0,z')=0$ at the asymptotic boundary and boundedness as $z\to\infty$ selects $z^2$ for $z<z'$ and a constant for $z>z'$, while continuity at $z=z'$ and the jump condition $-[z^{-1}\partial_z G_L]_{z'^-}^{z'^+}=1$ fix the prefactor $1/2$.

While the probe flux $t_f(z)$ was evaluated on the horizon $u=0$ in Section~\ref{sec:probe}, the light-pair stress tensor $T_{uu}^\varphi$ is evaluated along the complementary horizon $v=0$. Restricting the bulk--boundary propagators $z^\Delta/[\sqrt{2\pi}(z^2-(u-u_i)(v-v_i)+i0)^\Delta]$ of the two light insertions to $v=0$ gives, with $A_z=1+z^2$,
\begin{equation}
 F_A^{\rm src}(u,z)
 =\frac{z^\Delta}{\sqrt{2\pi}(A_z+e^{-\sigma}u+i0)^\Delta},
 \qquad
 F_B^{\rm src}(u,z)
 =\frac{z^\Delta}{\sqrt{2\pi}(A_z-u+i0)^\Delta}.
 \label{eq:sourcehorizonlegs}
\end{equation}
Setting $\upsilon=-u$ and relabeling the integration variable $\sigma\mapsto-\sigma$ (which leaves the even modular weight in \eqref{eq:crossK} invariant) puts the two source legs into the symmetric form
\begin{equation}
 F_A(\upsilon,z)
 =\frac{z^\Delta}{\sqrt{2\pi}(A_z-e^\sigma\upsilon+i0)^\Delta},
 \qquad
 F_B(\upsilon,z)
 =\frac{z^\Delta}{\sqrt{2\pi}(A_z+\upsilon+i0)^\Delta},
 \label{eq:lightlegs}
\end{equation}
where $\upsilon$ is an affine coordinate along $v=0$ and reversing its orientation leaves the quadratic derivative integral $\int\dd u\,\partial_u F_A^{\rm src}\partial_u F_B^{\rm src}$ invariant.

We can now evaluate the longitudinal $\upsilon$-integral directly via the regulated Schwinger parametrization
\begin{equation}
 (X+i0)^{-\Delta}
 =\frac{e^{-i\pi\Delta/2}}{\Gamma(\Delta)}
   \int_0^\infty p^{\Delta-1}e^{ip(X+i0)}\dd p.
 \label{eq:crossSchwinger}
\end{equation}
In this representation, $F_A$ and $F_B$ carry affine frequencies $-e^\sigma p$ and $+q$, so their $\partial_\upsilon$ derivatives bring down a factor of $e^\sigma p q$ and the $\upsilon$-integral yields $2\pi\delta(q-e^\sigma p)$. Setting $a=e^\sigma$, the longitudinal stress integral becomes
\begin{align}
 J_\sigma(z)
 &:=\int_{\mathbb R}2\partial_\upsilon F_A
                 \partial_\upsilon F_B\dd\upsilon\nonumber\\
 &=\frac{2e^{-i\pi\Delta}z^{2\Delta}a^{\Delta+1}}
         {\Gamma(\Delta)^2}
   \int_0^\infty p^{2\Delta}
       e^{iA_z(1+a)p-0^+p}\dd p\nonumber\\
 &=\frac{2i\Gamma(2\Delta+1)}{\Gamma(\Delta)^2}
   \frac{z^{2\Delta}}{(1+z^2)^{2\Delta+1}}
   \frac{e^{\sigma/2}}
        {[2\cosh(\sigma/2)]^{2\Delta+1}},
 \label{eq:lightnull}
\end{align}
where the Schwinger integral over $p$ contributes the phase $e^{i\pi(2\Delta+1)/2}=ie^{i\pi\Delta}$, canceling the factor $e^{-i\pi\Delta}$ from the two propagators and leaving a universal overall factor of $i$ for all $\Delta>0$.

Integrating $J_\sigma(z)$ against the modular kernel in \eqref{eq:crossK} then reduces to the elementary beta integral (via $a=e^\sigma$)
\begin{equation}
 \int_{\mathbb R}\frac{e^{\sigma/2}\dd\sigma}
 {[2\cosh(\sigma/2)]^{4\Delta+3}}
 =B(2\Delta+2,2\Delta+1).
 \label{eq:crossModularBeta}
\end{equation}
In the cross commutator $-\langle[\delta K_\times,\Phi_{\rm L}]\Phi_{\rm R}\rangle_\Omega$, the out-of-time-order term that grows as $e^\beta$ enters with a minus sign (cf.\ eq.~(4.7) of ref.~\cite{CFL}), while expanding the eikonal scattering phase $e^{i\delta_{\rm eik}}$ to first order supplies a factor of $i$, whose product is $-i=1/i$. Including this factor and the transverse horizon measure $1/z$ defines the real, positive, densitized scattering source
\begin{align}
 \Pi_\Delta(z)
 &=\frac{\lambda^{2\Delta}}{iz}
   \int_{\mathbb R}\frac{J_\sigma(z)\dd\sigma}
        {[2\cosh(\sigma/2)]^{2\Delta+2}}\nonumber\\
 &=\frac{\lambda^{2\Delta}}2\mathcal N_\Delta
       \frac{z^{2\Delta-1}}{(1+z^2)^{2\Delta+1}},
 \label{eq:crossPiSource}\\
 \mathcal N_\Delta&=
 \frac{\sqrt\pi\,\Gamma(2\Delta+1)^2}
 {4^{2\Delta}\Gamma(\Delta)^2\Gamma(2\Delta+3/2)}.
 \label{eq:N}
\end{align}

To fix the exact gravitational coupling between $\Pi_\Delta(z)$ and the probe flux $t_f(z)$, consider the effective shockwave stress tensor and Aichelburg--Sexl metric ansatz on $u=0$~\cite{AichelburgSexl,DraytHooft},
\begin{equation}
 T^{\rm eff}_{uu}=z\Pi(z)\delta(u),\qquad
 \dd s^2=\frac{-\dd u\dd v+\dd z^2
                  +h(z)\delta(u)\dd u^2}{z^2}.
 \label{eq:crossShockAnsatz}
\end{equation}
The linearized $uu$ Einstein equation in $\text{AdS}_3$ reads
\begin{equation}
 R_{uu}+2g_{uu}
 =-\frac12\left(h''-\frac{h'}z\right)\delta(u)
 =\frac z2 Lh\,\delta(u)
 =8\pi G_NT^{\rm eff}_{uu},
 \label{eq:crossShockEinstein}
\end{equation}
which simplifies to $Lh=16\pi\GN\Pi_\Delta$ and hence $h(z)=16\pi\GN\int_0^\infty G_L(z,z')\Pi_\Delta(z')\dd z'$. Because a probe null ray crossing $u=0$ experiences the Shapiro time shift $\Delta v=h(z)$, each horizon mode $e^{-ipv}$ acquires the tree-level eikonal phase shift $e^{iph(z)}\approx 1+iph(z)$. Weighting the horizon Klein--Gordon mode density by $p h(z)$ reproduces the null-flux density $t_f(z)h(z)$ (with the factor $i$ from $1+iph(z)$ and the out-of-time-order minus sign already absorbed into the real source $\Pi_\Delta$ in \eqref{eq:crossPiSource}), so the boosted probe expectation value yields the growing cross-channel response
\begin{align}
 \PP_\times^{\rm grow}[f]
 &=e^\beta\int_0^\infty t_f(z)h(z)\dd z
  =16\pi\GN e^\beta\int_0^\infty\!\dd z\dd z'\,
        t_f(z)G_L(z,z')\Pi_\Delta(z')\nonumber\\
 &=16\pi\GN\int_0^\infty\!\dd z\dd z'\,
       \frac{E_fe^\beta}{2}q_d(z;s)
       G_L(z,z')\Pi_\Delta(z'),
 \label{eq:scatter}
\end{align}
where the last equality substitutes the single-insertion flux $t_f(z)=(E_f/2)q_d(z;s)$.
In \eqref{eq:scatter}, the three factors play transparent physical roles: $\frac{1}{2}E_f e^\beta q_d(z;s)$ is the boosted null-momentum flux of the probe on $u=0$, $\Pi_\Delta(z')$ is the modular-integrated null stress form factor of the light pair on $v=0$, and $16\pi\GN G_L(z,z')$ propagates the transverse gravitational perturbation between them.

\subsection{The independent generalized-entropy source}
\label{sec:source}
So far, $\Pi_\Delta(z)$ has emerged purely from the Regge limit of the boundary modular commutator. To relate \eqref{eq:scatter} to the bulk geometry without circularly defining the QES through the scattering signal, we now compute the QES displacement $\eta(z)$ independently by extremizing the bulk generalized entropy~\cite{FLM,EW}.

Expanding the classical area functional of the curve $x_{\rm P}=\eta(z)$ to quadratic order and the bulk entanglement entropy $S_{\rm bulk}(A\cup B)$ to linear order in $\eta(z)$ gives the generalized entropy functional
\begin{equation}
 S_{\rm gen}[\eta]=\text{constant}
 +\frac1{8\GN}\int_0^\infty\frac{\eta'(z)^2}{z}\dd z
 -\int_0^\infty p_{\rm ent}(z)\eta(z)\dd z+\cdots,
 \label{eq:Sgen}
\end{equation}
where $p_{\rm ent}(z):=-\delta S_{\rm bulk}/\delta\eta(z)$ is the bulk-entropy shape derivative (generalized-entropy source) evaluated at the unperturbed surface $x_{\rm P}=0$. For the selected variation, the entropy of the isolated $A$ wedge is stationary by vacuum reflection symmetry, and the $B$ wedge is unchanged. Hence $p_{\rm ent}(z)=\delta I_{\rm bulk}(A,B)/\delta\eta(z)$ arises entirely from the shape variation of the bulk mutual information~\cite{Shape,CW}. The exact closed-form expression for $p_{\rm ent}(z)$ was already obtained by Chen and Wang~\cite{Shape,CW}. In Appendix~\ref{app:entropy}, we revisit that entanglement-first-law calculation after excising a regulated tubular neighborhood around $x_{\rm P}=0$ and observe that the resulting double modular endpoint integral is a standard two-variable Selberg integral~\cite{Selberg}, providing a compact first-principles check that reproduces refs.~\cite{Shape,CW} and verifies the exact factor-of-two relation
\begin{equation}
 p_{\rm ent}(z)=\lambda^{2\Delta}p_*(z),\qquad
 p_*(z)=\mathcal N_\Delta\frac{z^{2\Delta-1}}{(1+z^2)^{2\Delta+1}},
 \qquad p_{\rm ent}=2\Pi_\Delta.
 \label{eq:matching}
\end{equation}
(Note that the coordinate-density power $z^{2\Delta-1}$ in \eqref{eq:matching} agrees with the $\text{AdS}_3$ QES solution of ref.~\cite{CW} and our first-principles Selberg derivation in Appendix~\ref{app:entropy}, correcting a typographical power in eqs.~(3.34) and~(4.16) of ref.~\cite{CW}.)
Varying \eqref{eq:Sgen} with respect to $\eta(z)$ yields $\delta S_{\rm gen}/\delta\eta=(4\GN)^{-1}L\eta-p_{\rm ent}=0$, so that the QES profile is governed by the inhomogeneous Jacobi equation
\begin{equation}
 L\eta=4\GN p_{\rm ent},\qquad
 \eta(z)=4\GN\int_0^\infty G_L(z,z')p_{\rm ent}(z')\dd z',
 \label{eq:QES}
\end{equation}
subject to the same boundary conditions ($\eta(0)=0$ and boundedness as $z\to\infty$) that uniquely selected the transverse Green function $G_L(z,z')$ in \eqref{eq:Green}.

\subsection{The source match and the displacement dictionary}
The scattering and shape calculations provide two independent routes to the same displacement coefficient:
\[
 \text{modular scattering}\longrightarrow\Pi_\Delta,
 \qquad
 \text{generalized entropy}\longrightarrow p_{\rm ent}\longrightarrow\eta.
\]
At the retained order, the two sources obey the normalization identity
\begin{equation}
 p_{\rm ent}(z)=2\Pi_\Delta(z).
 \label{eq:sourcematchboxed}
\end{equation}
Substituting $\Pi_\Delta=\frac{1}{2}p_{\rm ent}$ into the scattering response \eqref{eq:scatter} absorbs $\GN$ into the QES displacement $\eta(z)=4\GN\int_0^\infty G_L(z,z')p_{\rm ent}(z')\dd z'$ via $h(z)=16\pi\GN\int_0^\infty G_L(z,z')\Pi_\Delta(z')\dd z'=2\pi\eta(z)$, leaving the purely geometric dictionary
\begin{equation}
 \PP_\times^{\rm grow}[f]
       =2\pi e^\beta\int_0^\infty t_f(z)\eta(z)\dd z
       =\pi E_fe^\beta\int_0^\infty q_d(z;s)\eta(z)\dd z.
 \label{eq:dictionary}
\end{equation}
Equation~\eqref{eq:dictionary} is our fundamental cross-channel displacement dictionary: the growing $e^\beta$ part of the cross-channel modular response equals $\pi E_f e^\beta$ times the $q_d(z;s)$-weighted radial average of the QES coordinate displacement $\eta(z)$.

With the explicit solution \eqref{eq:QES} in hand, we can also verify the asymptotic endpoint behavior of $\eta_*(z)$ and $H(x)$ promised in Section~\ref{sec:setup}. Splitting the Green function integral \eqref{eq:QES} at $z'=z$ gives
\begin{equation}
 \eta_*(z)=2\int_0^z z'^2p_*(z')\dd z'
            +2z^2\int_z^\infty p_*(z')\dd z'.
 \label{eq:splitQES}
\end{equation}
Because $p_*(z)$ in \eqref{eq:matching} has equal zeroth and second moments $\int_0^\infty p_*(z)\dd z=\int_0^\infty z^2 p_*(z)\dd z=\frac{1}{2}\mathcal N_\Delta B(\Delta,\Delta+1)$ for all $\Delta>0$, \eqref{eq:splitQES} implies $\eta_*(z)\sim C_2 z^2$ as $z\to 0$ and $\eta_*(z)\to C_2$ as $z\to\infty$, where $C_2=\mathcal N_\Delta B(\Delta,\Delta+1)$. Consequently, the proper displacement profile $H(x)=e^{-x}\eta_*(e^x)$ decays exponentially at both ends of the reference curve, $H(x)\sim C_2 e^{-|x|}$ as $x\to\pm\infty$.

\section{Complete same-side suppression and the averaged boundary inequality}
\label{sec:abel}

While the cross-channel response \eqref{eq:dictionary} directly samples the QES displacement $\eta(z)$, the exact boundary inequality $\PP[f]\ge 0$ in \eqref{eq:P} constrains the sum of all modular OPE channels. At the leading perturbative order $\GN\lambda^{2\Delta}$, the commutator \eqref{eq:remainingchannels} retains not only $\delta K_\times$ but also the same-side channel $\delta K_{AA}$. We now prove that for every fixed single-insertion Euclidean probe $|f_{s,\delta}\rangle$, both the complete tree-level Einstein graviton exchange in the $AA$ channel---including its entire scalar double-trace completion and keeping both commutator operator orderings on their prescribed Lorentzian sheets---and the boost-rescaled integrated cross-channel remainder are suppressed as $O(e^{-\beta})$ at large boost $\beta$. Applying a positivity-preserving Laplace projection then isolates the maximal $e^\beta$ cross-channel residue and yields an all-scale averaged boundary inequality for the QES displacement.

\subsection{The holographic coefficient}
To extract the leading holographic response for a fixed single-insertion preparation $(s,\delta)$, let $U(\beta)$ denote the physical vacuum Rindler boost and let $|f_c\rangle$ denote the corresponding normalized single-insertion probe state at central charge $c$. Recalling $\PP_\beta[f_c]:=\PP[U(\beta)f_c]\ge 0$, we define the ordered perturbative coefficient at fixed finite boost $\beta$ by taking the small-interval limit $\lambda\to 0^+$ followed by the weak-gravity limit $\GN\to 0^+$:
\begin{equation}
 p_f(\beta)=\lim_{G_N\to0^+}\frac1{G_N}
 \lim_{\lambda\to0^+}\frac{\PP_\beta[f_c]}{\lambda^{2\Delta_c}},
 \label{eq:ordered}
\end{equation}
where $\Delta_c$ denotes the exact scaling dimension of the light scalar primary at central charge $c$ (with $\Delta_c\to\Delta$ as $c\to\infty$). Throughout our boundary-to-bulk implications, we assume that the integrated small-interval expansion of $\PP_\beta[f_c]$ contains no surviving lower power of $\lambda$ after the full-modular subtraction $\widehat K_R-\widehat K_A$ and that the ordered limits in \eqref{eq:ordered} exist for each fixed preparation $(s,\delta)$. In particular, taking $\lambda\to 0^+$ first at fixed finite $c$ (with $\Delta_c<2$) cleanly isolates the $\lambda^{2\Delta_c}$ scalar-pair channel before the $\lambda^4$ stress-tensor channel---which also exhibits spin-2 Regge growth $e^\beta$---can mix with it, while the subsequent $1/\GN$ limit is finite because $\OO$ and $\OO_\phi$ are distinct scalar primaries whose $O(\GN^0)$ connected four-point function vanishes and whose disconnected product cancels identically in the commutator \eqref{eq:remainingchannels}.

Because $\lambda^{2\Delta_c}>0$ and $\GN>0$, dividing the exact inequality $\PP_\beta[f_c]\ge 0$ by these positive factors preserves nonnegativity at every fixed finite $\beta$, giving $p_f(\beta)\ge 0$ without subtracting any lower-order positive terms. With the effective-action matching specified in Section~\ref{sec:setup}, this ordered limit selects the Einstein tree coefficient before the large-boost limit. The corresponding finite-coupling boost regime is described in Section~\ref{sec:finitewindow}.

\subsection{Large-boost suppression of \texorpdfstring{$AA$}{AA} and the cross-channel remainder}
\label{sec:suppression}
Within the selected bulk theory, the ordered coefficient $p_f(\beta)$ receives contributions from the two surviving channels identified in \eqref{eq:remainingchannels}:
\begin{equation}
 p_f(\beta)=p_\times(\beta)+a_{AA}(\beta).
 \label{eq:projected_decomposition}
\end{equation}
Here $p_\times(\beta)$ is the complete cross-channel response and $a_{AA}(\beta)$ is the complete same-side Einstein contribution with both light insertions in $A$, with the perturbative factor $\GN\lambda^{2\Delta}$ removed from both. Stripping the perturbative and boost factors from the growing cross-channel response \eqref{eq:dictionary} defines the reduced cross-channel readout
\begin{equation}
 Q^\times[f]:=\frac{\PP_\times^{\rm grow}[f]}{\GN\lambda^{2\Delta}e^\beta}
 =2\pi\int_0^\infty t_f(z)\eta_*(z)\dd z
 =\pi E_f\int_0^\infty q_d(z;s)\eta_*(z)\dd z,
 \label{eq:Qcross}
\end{equation}
in terms of which we separate the leading large-boost term of $p_\times(\beta)$ from its finite-boost remainder:
\begin{equation}
 p_\times(\beta)=e^\beta Q^\times[f]+r_\times(\beta).
 \label{eq:crossReggeExpansion}
\end{equation}
Thus $r_\times(\beta):=p_\times(\beta)-e^\beta Q^\times[f]$ belongs entirely to the cross channel and collects all subleading finite-boost terms beyond the leading Regge contribution at the same Einstein tree order. We now evaluate the large-boost behavior of both $a_{AA}(\beta)$ and $r_\times(\beta)$ under the minimally coupled bulk effective action \eqref{eq:selectedaction} and the ordered Euclidean-vacuum real-time prescription of Section~\ref{sec:setup}.

Beginning with the same-side $AA$ channel, the bilocal modular correction $\delta K_{AA}$ is an integral over modular times $(j,k)$ of the light-scalar pair $\OO_A(j+k)\OO_A(j)$ along the Sommerfeld--Watson contour $j\in\mathbb R$, $k\in\mathbb R+i\epsilon$ ($0<\epsilon<\pi$) representing the Hermitian operator $\delta K_{AA}$, written explicitly in \eqref{eq:AAkernel} of Appendix~\ref{app:AAfull}. Denoting the two light insertions by $\OO_a:=\OO_A(j)$ and $\OO_g:=\OO_A(j+k)$, the $AA$ commutator $-\langle[\delta K_{AA},\Phi_L]\Phi_R\rangle_\Omega$ yields the difference $W_{\rm sand}-W_{\rm out}$ of the two connected four-point words
\begin{equation}
 W_{\rm out}=\langle\OO_g\OO_a\Phi_L\Phi_R\rangle_{\rm conn},
 \qquad
 W_{\rm sand}=\langle\Phi_L\OO_g\OO_a\Phi_R\rangle_{\rm conn},
 \label{eq:AA_words_main}
\end{equation}
while the disconnected product $\langle\OO_g\OO_a\rangle\langle\Phi_L\Phi_R\rangle$ cancels between the two orderings because $\OO$ and $\OO_\phi$ are distinct species. With the word-dependent external $i0$ in \eqref{eq:AA_joint_i0}, one can shift the $(j,k)$ contours and continue the preparation along \eqref{eq:explicitACpath} with all four cyclic KMS gaps positive. At the physical Euclidean preparation $(s,\delta)$, vacuum boost covariance allows us to transfer the external probe boost $\beta$ onto the common modular flow parameter $j$ by setting $\rho=e^{-\beta}$, $a=e^{j-\beta}$, and $g=e^{j+k-\beta}$, so that the external probe insertions remain fixed while $a$ and $g$ run along fixed, word-dependent rays in the complex plane. Including the Jacobian $\dd j\,\dd k=\dd a\,\dd g/(ag)$, the exact Sommerfeld--Watson $AA$ modular kernel \eqref{eq:AAkernel} becomes
\begin{equation}
 K_\rho(a,g)=e^{i\pi\Delta}
 \frac{\rho(ag)^\Delta}
 {(\rho+a)(\rho+g)(g-a)^{2\Delta+1}},
 \label{eq:AA_boost_kernel}
\end{equation}
on the displaced rays $\Im k=\kappa>0$. The joint boundary value is specified in Appendix~\ref{app:AA-contours}. For either commutator operator ordering ($\mathrm{word}\in\{\mathrm{out},\mathrm{sand}\}$), the connected four-point integrand is $\langle\OO_g\OO_a\rangle F_{\rm word}(a,g)$, where $F_{\rm word}=W_{\rm word}/(\langle\OO_g\OO_a\rangle\langle\Phi_L\Phi_R\rangle)$ denotes the connected four-point function normalized by the product of the light and probe two-point functions.

In Appendix~\ref{app:AAfull}, we bound the full tree-level Einstein graviton exchange graph on the carried Lorentzian sheets across the entire compactified quadrant $(|a|,|g|)\in[0,\infty]^2$---mapping the interior, all four one-leg edges, and both mixed corners (for both outside and sandwich orderings) into compact subsets of the open ordered Wightman tube via an explicit conformal chart, and controlling the common small-$L$ and large-$L$ endpoints by the direct light-pair OPE and chiral inversion---establishing the uniform unboosted graph envelope
\begin{equation}
 |F_{\rm word}(a,g)|\le C G_N f(L),\qquad
 f(L)=\min(L^2,1),\qquad L=|a|+|g|,
 \label{eq:AA_graph_envelope}
\end{equation}
where here $L=|a|+|g|$ denotes the radial coordinate sum (distinct from the Jacobi operator $L=-\partial_z(z^{-1}\partial_z)$). Along the displaced, non-intersecting integration rays for $a$ and $g$ (with moduli $A=|a|$ and $G=|g|$, where $A$ here denotes the modulus $|a|$ rather than the half-line subregion), the modular kernel \eqref{eq:AA_boost_kernel} and the light two-point function obey the pointwise bounds
\begin{equation}
 |K_\rho|\le
 C\frac{\rho(AG)^\Delta}
 {(\rho+A)(\rho+G)L^{2\Delta+1}},\qquad
 |\langle\OO_g\OO_a\rangle|
 \le C\frac{(AG)^\Delta}{L^{2\Delta}}.
 \label{eq:AA_kernel_pair_bounds}
\end{equation}
Changing to polar simplex coordinates $A=Lt$ and $G=L(1-t)$ with $L\in(0,\infty)$, $t\in(0,1)$, and Jacobian $\dd A\,\dd G=L\,\dd L\,\dd t$, the absolute value of the modular integral for each word is bounded by
\begin{align}
 \int|\dd a\dd g\,K_\rho\langle\OO_g\OO_a\rangle F_{\rm word}|
 &\le C G_N\rho\int_0^\infty\!\dd L\int_0^1\!\dd t\,
 \frac{[t(1-t)]^{2\Delta}f(L)}
 {(\rho+Lt)(\rho+L(1-t))}\nonumber\\
 &\le C G_N\rho B(2\Delta,2\Delta)
       \int_0^\infty\frac{f(L)}{L^2}\dd L
 =2C G_N\rho B(2\Delta,2\Delta),
 \label{eq:AA_uniform_integral}
\end{align}
where the second inequality uses $(\rho+Lt)(\rho+L(1-t))\ge L^2 t(1-t)$ to bound the moving endpoint region $L\sim\rho$ and the relative-time tails $t\to 0,1$ simultaneously. As shown in Appendix~\ref{app:AAfull}, removing the auxiliary Gaussian contour regulator (introduced at real supports to justify deforming to the displaced KMS rays) by dominated convergence recovers the original modular pairing and its coincidence contact terms, and subtracting the two commutator words preserves the $O(\rho)=O(e^{-\beta})$ decay. Dividing out $\GN$ gives the complete $AA$ suppression bound
\begin{equation}
 |a_{AA}(\beta)|\le C_{AA} e^{-\beta},\qquad \beta\ge0,
 \label{eq:AAcompletebound}
\end{equation}
where the constant $C_{AA}<\infty$ depends on the chosen preparation $(s,\delta)$ and the thermal contour margins. Although $a_{AA}(\beta)$ is generally nonzero at finite boost $\beta$, \eqref{eq:AAcompletebound} shows that it decays at least as fast as $e^{-\beta}$ and therefore cannot compete with the $e^{+\beta}$ growth of the cross-channel term $e^\beta Q^\times[f]$. Appendix~\ref{app:AA} gives a separate stress-family Ward check of this decay.

For the cross-channel remainder $r_\times(\beta)=p_\times(\beta)-e^\beta Q^\times[f]$, we must control the moving tail of the modular $\sigma$-integral in \eqref{eq:crossK}. Appendix~\ref{app:crossremainder} splits the contour at $(1+|e^\sigma|)e^{-\beta}=\delta_*$ for a fixed small $\delta_*>0$. In the outer region, conformal maps into the ordered Wightman tube bound both commutator words uniformly. In the inner small-cross-ratio region, the time-ordered graph stays on the direct light-pair sheet and is $O_f(\ell^2)$, whereas the out-of-time-ordered graph acquires one positive $+2\pi i$ winding of $\mathsf Y=(1-z_\times)(1-w_\times)$. Summing the complete coefficient of its logarithm yields a leading Barnes integral whose Fourier transform in $\log s$ matches the independent generalized-entropy kernel in \eqref{eq:Qcross} identically; after subtracting this growing contribution, shifting the remaining Barnes contour crosses only a single pole of constant residue, yielding a uniformly $O(1)$ integrated remainder for each fixed preparation and hence
\begin{equation}
 h_\times(\beta):=e^{-\beta}|r_\times(\beta)|
 \le C_\times e^{-\beta},\qquad \beta\ge0,
 \label{eq:crossRemainderNorm}
\end{equation}
where $C_\times<\infty$ depends on $(s,\delta)$ and the thermal contour margins.

\subsection{The positive Laplace projection and the averaged boundary inequality}
\label{sec:Abelresidue}
To extract the coefficient of the maximal $e^\beta$ growing mode while manifestly preserving positivity, we apply a positive Laplace projection (an Abel-type limit in the boost variable $\beta$). For $\epsilon_A>0$, we define
\begin{equation}
 \mathcal R_{\epsilon_A}[f]
 =\epsilon_A\int_0^\infty e^{-(1+\epsilon_A)\beta}p_f(\beta)\dd\beta,
 \qquad \mathfrak R[f]=\lim_{\epsilon_A\downarrow0}
                                  \mathcal R_{\epsilon_A}[f].
 \label{eq:Abel}
\end{equation}
Because the weight $\epsilon_A e^{-(1+\epsilon_A)\beta}$ is strictly positive for all $\beta\ge 0$ and $\epsilon_A>0$, $\mathcal R_{\epsilon_A}[f]\ge 0$ whenever $p_f(\beta)\ge 0$ and the integral converges, so the limiting residue obeys $\mathfrak R[f]\ge 0$. Acting on a pure Regge mode $e^{(J-1)\beta}$ of spin intercept $J$, the Laplace integral evaluates to
\begin{equation}
 \epsilon_A\int_0^\infty e^{-(1+\epsilon_A)\beta}
 e^{(J-1)\beta}\dd\beta
 =\frac{\epsilon_A}{2+\epsilon_A-J},
 \qquad \operatorname{Re}J<2+\epsilon_A,
 \label{eq:spinfilter}
\end{equation}
which acts as a projective filter: as $\epsilon_A\downarrow 0$, the graviton intercept $J=2$ is preserved with unit weight, whereas any mode with $\operatorname{Re}J<2$ is projected to zero. For the complete $AA$ exchange, which involves an infinite sum of conformal families integrated over modular parameters, the uniform bound \eqref{eq:AAcompletebound} allows us to take $\epsilon_A\downarrow 0$ directly:
\begin{equation}
 \left|\epsilon_A\int_0^\infty e^{-(1+\epsilon_A)\beta}
 a_{AA}(\beta)\dd\beta\right|
 \le C_{AA}\frac{\epsilon_A}{2+\epsilon_A}\longrightarrow0.
 \label{eq:AA_zero_residue}
\end{equation}

The complete $AA$ channel therefore has zero maximal Laplace residue. Likewise, the cross-channel remainder bound \eqref{eq:crossRemainderNorm} implies immediately that the integrated cross-channel remainder has zero Laplace residue,
\begin{equation}
 \left|\epsilon_A\int_0^\infty e^{-(1+\epsilon_A)\beta}
 r_\times(\beta)\dd\beta\right|
 \le\epsilon_A\int_0^\infty e^{-\epsilon_A\beta}h_\times(\beta)\dd\beta
 \le C_\times\frac{\epsilon_A}{1+\epsilon_A}\longrightarrow0,
 \label{eq:remainderAbel}
\end{equation}
and more generally for any locally integrable $h_\times(\beta)\to 0$ as $\beta\to\infty$ and any boost cutoff $\beta_0>0$,
\begin{equation}
 \left|\epsilon_A\int_0^\infty e^{-(1+\epsilon_A)\beta}
 r_\times(\beta)\dd\beta\right|
 \le\epsilon_A\int_0^{\beta_0}h_\times(\beta)\dd\beta
       +e^{-\epsilon_A\beta_0}\sup_{\beta\ge\beta_0}h_\times(\beta).
 \label{eq:crossRemainderAbelBound}
\end{equation}
Combining the suppression bounds \eqref{eq:AAcompletebound} and \eqref{eq:crossRemainderNorm} with the ordered positivity \eqref{eq:ordered} yields
\begin{equation}
 \mathfrak R[f]=Q^\times[f]\ge 0,
 \label{eq:positive_cross_residue}
\end{equation}
completing the dynamical bridge from exact boundary relative-entropy monotonicity to positivity of the geometric cross-channel readout within the selected Einstein model.

Substituting the displacement dictionary \eqref{eq:Qcross} into \eqref{eq:positive_cross_residue} and dividing out the strictly positive energy factor $\pi E_f=\pi d/(s\sin\delta)>0$ gives the all-scale averaged boundary inequality for the QES displacement:
\begin{equation}
 \int_0^\infty q_d(z;s)\eta_*(z)\dd z\ge 0
 \qquad\text{for every }s>0,
 \label{eq:geometricinequality}
\end{equation}
or equivalently $\int_0^\infty t_f(z)\eta(z)\dd z\ge 0$ at leading perturbative order $\GN\lambda^{2\Delta}$. At fixed scale $s$, varying the Euclidean preparation angle $\delta$ rescales the signal amplitude $E_f=d/(s\sin\delta)$ without altering the normalized radial sampling profile $q_d(z;s)$; scanning $s>0$ sweeps this fixed-width positive window across the entire reference extremal curve.

\subsection{Perturbative boost regime}
\label{sec:finitewindow}
The Laplace projection \eqref{eq:Abel} acts on the extracted Einstein tree coefficient $p_f(\beta)$. At finite central charge $c=3/(2\GN)$, a weak-scattering calculation instead requires the boosted horizon null momentum $e^\beta\mathcal E_+$ (with $\mathcal E_+=\int_0^\infty t_f(z)\dd z=E_f/2=d/(2\tau)$) to remain below the multi-graviton eikonal threshold $G_N e^\beta\mathcal E_+\ll 1$, giving the Regge boost window
\begin{equation}
 1\ll e^\beta\ll\frac{2\tau}{G_N d}=\frac{2s\sin\delta}{G_N d}.
 \label{eq:singlewindow}
\end{equation}
This window is compatible with arbitrarily large boosts as $G_N\to 0$, and is distinct from taking an infinite boost at fixed gravitational coupling. At finite $\GN$ and finite $\lambda$, isolating the leading scalar-pair response across this boost window also requires $\lambda$ to be small enough that subleading small-interval OPE channels---most notably the $\lambda^4$ stress-tensor channel, which shares the same $e^\beta$ Regge growth, together with $\lambda^{2\Delta+2}$ descendants---remain negligible relative to $\lambda^{2\Delta}$. Moreover, because \eqref{eq:AAcompletebound} and \eqref{eq:crossRemainderNorm} establish exponential decay in $\beta$ for both $e^{-\beta}a_{AA}(\beta)$ and $e^{-\beta}r_\times(\beta)$, averaging $e^{-\beta}p_f(\beta)$ over a shifted finite window $[\beta_{\min},\beta_{\max}]$ inside \eqref{eq:singlewindow} preserves the signal $Q^\times[f]$ with unit weight while suppressing both remainders by powers of $G_N$ (as developed quantitatively in Paper II~\cite{CompanionTomography}).

\section{Discussion}
\label{sec:discussion}

The positive full-modular difference $\PP[f]=\langle f|\widehat K_R-\widehat K_A|f\rangle\ge 0$ provides a boundary answer to the question of how bulk correlations constrain a perturbative quantum causal shadow. In the Einstein regime considered here, its growing cross-channel response measures the horizon-flux average of the QES displacement, with the exact prefactor fixed non-circularly by matching the scattering source $\Pi_\Delta(z)$ against an independent generalized-entropy calculation of $p_{\rm ent}(z)$ (verified via a two-variable Selberg integral). For every fixed single-insertion Euclidean preparation, both the complete $AA$ exchange and the boost-rescaled integrated cross-channel remainder decay as $O(e^{-\beta})$. Their zero Laplace residues leave the all-scale averaged boundary inequality \eqref{eq:geometricinequality}.

The use of the complete exchange diagram is essential. The stress-family response is a useful check, but the graph also contains double-trace contributions and must be followed on the Lorentzian sheets selected by each commutator ordering. Transferring the boost to the modular kernel makes the possible moving endpoint explicit in the $AA$ channel, where the graph envelope controls this endpoint together with the relative-time tails; in the cross channel, summing the logarithmic Mellin coefficient and shifting the subtracted Barnes integral bounds the integrated remainder beyond $e^\beta Q^\times[f]$ by $O(1)$. This separation between the boost-transferred modular kernel and the ordered Wightman-tube and Mellin graph estimates may also be useful for other nonlocal modular interactions.

The bulk action \eqref{eq:selectedaction} and the ordered coefficient expansion \eqref{eq:ordered} delineate the exact holographic scope of our result. Independent matter contact interactions---most notably higher-derivative spin-2 contact terms such as $(\nabla_\mu\varphi\nabla_\nu\varphi)(\nabla^\mu\phi\nabla^\nu\phi)$, which would shift the $O(G_Ne^\beta)$ cross response without changing the QES displacement $\eta(z)$---are excluded by the minimally coupled Einstein action, and in Regge-behaved theories where a finite higher-spin gap $\Delta_{\rm gap}<\infty$ lowers the leading Regge intercept to $J_0<2$~\cite{Regge,SS}, the strict $\epsilon_A\downarrow 0$ Laplace filter projects an $O(e^{(J_0-1)\beta})$ cross response to zero. Extending the construction to broader matter sectors, stringy Regge trajectories, or non-vacuum backgrounds in which both the causal surface and the QES move requires a separate dynamical analysis.

Within the present setting, the single-insertion boundary observable and its averaged geometric meaning are now fixed. What remains is to understand the resolving power of boundary probes: because $q_d(z;s)$ has a fixed radial width set by $d$, positivity of single-insertion averages at all scales $s>0$ need not determine the local pointwise sign of $H(x)$. Paper II~\cite{CompanionTomography} takes up this next step. After analyzing what single-insertion scale scans can and cannot reconstruct, it extends the boundary observable and our tree-level $AA$ and cross-remainder bounds to finite coherent superpositions of the same scalar species, harnesses off-diagonal quantum interference via finite preparation derivatives to establish a complete hierarchy of local positivity criteria, applies this dictionary to derive exact benchmarks and sharp pointwise bounds for the scalar QES, and analyzes the resulting saturation condition and finite-coupling boost window.

\section*{Acknowledgments}
L.C. is supported by the
Shuimu Tsinghua Scholar Program of Tsinghua University and the China
Postdoctoral Science Foundation under Grant No.~2025M783401. In the core part of this work, AI assistance was provided primarily by OpenAI's GPT-5.6 Sol, while OpenAI's subsequent GPT-6 to 6.2 Astra/Sol series and Google's Gemini 4-Neon played a major role in manuscript preparation, cross-validation, and further refinements. The author takes responsibility for the manuscript.

\appendix
\section{Elementary probe derivations and normalization}
\label{app:probenorm}

In this appendix, we collect the fundamental normalization identities underlying the single-insertion probe construction of Section~\ref{sec:probe}. Presenting these derivations together highlights how three distinct normalizations enter our dictionary: the unit Hilbert-space norm of the boundary state, the canonical bulk-to-boundary propagator normalization relative to a unit-normalized CFT primary, and the unit integral of the radial horizon flux distribution.

\subsection{Canonical bulk--boundary coefficient in \texorpdfstring{$\text{AdS}_3$}{AdS3}}
To keep track of dimensional factors transparently, let us temporarily denote the boundary spacetime dimension by $D$ and the scaling dimension of the probe scalar primary by $\Delta_\phi$. For a canonically normalized Euclidean bulk scalar action on unit-radius Poincar\'e $\text{AdS}_{D+1}$,
\begin{equation}
 I[\phi]=\frac12\int\sqrt g\,
 \bigl[(\nabla\phi)^2+m^2\phi^2\bigr],
 \qquad m^2=\Delta_\phi(\Delta_\phi-D),
 \label{eq:canonicalbulkaction}
\end{equation}
the standard Dirichlet bulk--boundary propagator is
\begin{equation}
 K_{\Delta_\phi}(z,x;x')
 =C_{\Delta_\phi,D}
 \left(\frac{z}{z^2+|x-x'|^2}\right)^{\Delta_\phi},
 \qquad
 C_{\Delta_\phi,D}
 =\frac{\Gamma(\Delta_\phi)}
 {\pi^{D/2}\Gamma(\Delta_\phi-D/2)}.
 \label{eq:Dirichletkernel}
\end{equation}
By holographic renormalization, the canonically conjugate boundary operator $\OO_{\rm can}$ dual to $\phi$ satisfies
\begin{equation}
 \langle\OO_{\rm can}(x)\OO_{\rm can}(0)\rangle
 =\frac{(2\Delta_\phi-D)C_{\Delta_\phi,D}}
 {|x|^{2\Delta_\phi}},
 \qquad
 \langle\phi(z,x)\OO_{\rm can}(x')\rangle=K_{\Delta_\phi}(z,x;x').
 \label{eq:canonicalboundarynorm}
\end{equation}
Rescaling $\OO_{\rm can}$ to define the unit-normalized CFT primary $\OO_\phi$,
\begin{equation}
 \OO_\phi=\frac{\OO_{\rm can}}
 {\sqrt{(2\Delta_\phi-D)C_{\Delta_\phi,D}}},
\end{equation}
ensures that $\langle\OO_\phi(x)\OO_\phi(0)\rangle=|x|^{-2\Delta_\phi}$ has unit two-point coefficient and fixes the mixed bulk--boundary correlator to be
\begin{equation}
 \langle\phi(z,x)\OO_\phi(x')\rangle
 =\sqrt{\frac{C_{\Delta_\phi,D}}{2\Delta_\phi-D}}
 \left(\frac{z}{z^2+|x-x'|^2}\right)^{\Delta_\phi}.
 \label{eq:mixedcoefficientgeneral}
\end{equation}
Specializing to $D=2$ and $\Delta_\phi=d$ gives
\begin{equation}
 C_{d,2}=\frac{\Gamma(d)}{\pi\Gamma(d-1)}=\frac{d-1}{\pi},
 \qquad
 \sqrt{\frac{C_{d,2}}{2d-2}}=\frac1{\sqrt{2\pi}},
 \label{eq:mixedcoefficientAdS3}
\end{equation}
which establishes the exact prefactor $1/\sqrt{2\pi}$ used in the bulk wavefunction \eqref{eq:fullwave}. (Note that $C_{d,2}$ here denotes the Dirichlet propagator normalization, not the CFT central charge $c$.)

\subsection{Boundary norm and energy check}
For the unnormalized single-insertion preparation $|\widetilde f_{\tau,\xi}\rangle=\OO_\phi(i\tau,\xi)|\Omega\rangle$ defined in \eqref{eq:unnormalizedstate}, evaluating the unit-normalized boundary two-point function \eqref{eq:boundary2pt} at equal spatial positions $x_1=x_2=\xi$ and time separation $t_1-t_2=-2i\tau$ immediately yields $Z(\tau)=\langle\widetilde f_{\tau,\xi}|\widetilde f_{\tau,\xi}\rangle=(2\tau)^{-2d}$. Differentiating $Z(\tau)$ with respect to $\tau$ gives
\begin{equation}
 \partial_\tau Z(\tau)
 =-2\langle\widetilde f|H_{\rm CFT}|\widetilde f\rangle,
 \qquad
 -\frac12\partial_\tau\log Z(\tau)=\frac d\tau,
\end{equation}
confirming the mean energy formula $E_f=d/\tau$ in \eqref{eq:Ef}.

\subsection{Flux profile and its measure}
Integrating $2|\partial_v\psi_f(v,z)|^2$ over $v\in\mathbb{R}$ via the positive null-momentum representation \eqref{eq:momentum} produces
\begin{equation}
 t_f(z)=4\pi|n_w|^2z^{2d-1}
 \frac{\Gamma(2d+1)}{[2\operatorname{Re}(w+bz^2)]^{2d+1}}.
 \label{eq:tfintermediate}
\end{equation}
Substituting $\operatorname{Re}(w+bz^2)=\tau(1+z^2/s^2)$, $|n_w|^2=(2\tau/s)^{2d}/[2\pi\Gamma(d)^2]$, $\Gamma(2d+1)/\Gamma(d)^2=d/B(d,d+1)$, and $E_f=d/\tau$ simplifies \eqref{eq:tfintermediate} directly to $t_f(z)=\frac{1}{2}E_f q_d(z;s)$ in \eqref{eq:flux}. Changing variables to $r=z^2/s^2$ ($\dd z=\frac{s}{2}r^{-1/2}\dd r$) verifies that $q_d(z;s)$ is unit-normalized on $(0,\infty)$:
\begin{equation}
 \int_0^\infty q_d(z;s)\dd z
 =\frac1{B(d,d+1)}\int_0^\infty
 \frac{r^{d-1}}{(1+r)^{2d+1}}\dd r=1.
 \label{eq:qnormalizationappendix}
\end{equation}
Here the factor of $1/z$ in the definition \eqref{eq:tfmeaning} converts the proper transverse horizon measure $\dd z/z$ into a density with respect to the coordinate measure $\dd z$.

\section{Entropy-source normalization}
\label{app:entropy}

\subsection{The modular endpoint in the shape variation}
In this appendix, we revisit the generalized-entropy source $p_{\rm ent}(z)$ quoted in \eqref{eq:matching}---whose exact closed-form expression was already derived in Chen and Wang~\cite{Shape,CW}---and show that its double modular endpoint integral reduces to a standard two-variable Selberg integral~\cite{Selberg}, providing a compact first-principles check of the exact normalization relation $p_{\rm ent}(z)=2\Pi_\Delta(z)$. We emphasize that this calculation concerns the static shape variation of the vacuum bulk entanglement entropy $S_{\rm bulk}(A\cup B)$ with respect to the QES position $\eta(z)$, rather than the high-boost $AA$ probe scattering response analyzed in Section~\ref{sec:abel}.

Our starting point is the stress-tensor formulation of the bulk entanglement first law under a smooth diffeomorphism vector field $v^\mu$ supported in a neighborhood of the reference surface, as derived in eq.~(2.36) of ref.~\cite{CW} and eq.~(2.6) of ref.~\cite{Shape}:
\begin{equation}
 \delta S_{
 \rm bulk}(a)=\int_{\mathcal B}\!\dd^3X\sqrt g\,
 \nabla^\mu v^\nu\,
 \langle T^{\varphi}_{\mu\nu}(X)K_a\rangle_c ,
 \label{eq:shapeinput}
\end{equation}
where $\mathcal B$ is the Euclidean bulk spacetime, $K_a$ is the bulk vacuum modular Hamiltonian of the spatial region $a$, the subscript $c$ denotes the connected correlator, and $T_{\mu\nu}^\varphi$ is the stress tensor of the canonically normalized bulk scalar $\varphi$ of mass $m_\varphi^2=\Delta(\Delta-2)$ (cf.\ eq.~(3.25) of ref.~\cite{CW}).

\paragraph{Mutual-information reduction and tubular surface flux.}
To fix the overall sign unambiguously, let $a_1$ and $a_2$ denote the classical bulk entanglement wedges of $A$ and $B$, respectively, and define the bulk modular interaction $\Delta K=K_{a_1\cup a_2}-K_{a_1}-K_{a_2}$ as in eq.~(3.8) of ref.~\cite{CW}. Because $S_{\rm bulk}(a_1\cup a_2)=S_{\rm bulk}(a_1)+S_{\rm bulk}(a_2)-I_{\rm bulk}(a_1,a_2)$ and the isolated $a_1$ entropy is stationary by reflection symmetry while $a_2$ is held fixed, the bulk-entropy source $p_{\rm ent}(z)=-\delta S_{\rm bulk}(a_1\cup a_2)/\delta\eta(z)=\delta I_{\rm bulk}(a_1,a_2)/\delta\eta(z)$ is governed by $-\int_{\mathcal B}\sqrt{g}\,\nabla^\mu v^\nu\langle T_{\mu\nu}^\varphi\Delta K\rangle_c$. Excising a thin tubular neighborhood $\mathcal T_\epsilon$ of radius $\epsilon\ll 1$ around the reference surface $x_{\rm P}=t_E=0$ and integrating by parts (using $\nabla^\mu T_{\mu\nu}^\varphi=0$ away from the surface and noting that the outward normal of the punctured domain $\mathcal B\setminus\mathcal T_\epsilon$ is $n_{\rm out}^\mu=-n_{\rm tube}^\mu$, where $n_{\rm tube}^\mu$ points radially away from the tube axis) cancels the minus sign and reduces the variation to the tubular surface flux~\cite{Shape,CW}:
\begin{equation}
 \delta I=\lim_{\epsilon\to0}
 \int_{\mathcal T_\epsilon}\!\dd^2Y\sqrt h\,
 n_{\rm tube}^{\mu}v^\nu
 \langle T^\varphi_{\mu\nu}(Y)\Delta K\rangle_c .
 \label{eq:shapetubeflux}
\end{equation}
Crucially, as emphasized in refs.~\cite{Shape,CW}, the tubular cutoff limit $\epsilon\to 0$ must be taken \emph{after} extracting the modular integral endpoint contribution below.

In complex transverse coordinates $\zeta=x_{\rm P}+it_E=\epsilon e^{i\theta}$ and $\bar\zeta=x_{\rm P}-it_E=\epsilon e^{-i\theta}$, the Euclidean $\text{AdS}_3$ metric reads $\dd s^2=(\dd\zeta\,\dd\bar\zeta+\dd z^2)/z^2$, the displacement vector on $\mathcal T_\epsilon$ is $v^\zeta=v^{\bar\zeta}=\eta(z)$, and the induced area element and outward tube normal are
\begin{equation}
 \sqrt h\,\dd^2Y=\frac{\epsilon}{z^2}\dd\theta\dd z,
 \qquad
 n_{\rm tube}^{\zeta}=ze^{i\theta},\qquad
 n_{\rm tube}^{\bar\zeta}=ze^{-i\theta}.
 \label{eq:shapetubemeasure}
\end{equation}
At leading order $\lambda^{2\Delta}$, the only components of $\langle T_{\mu\nu}^\varphi\Delta K\rangle_c$ with nonvanishing $\epsilon\to 0$ tube flux are the holomorphic and antiholomorphic pole terms $\langle T_{\zeta\zeta}^\varphi\Delta K\rangle_c=\mathcal A_\Delta(z)/\zeta$ and $\langle T_{\bar\zeta\bar\zeta}^\varphi\Delta K\rangle_c=\mathcal A_\Delta(z)/\bar\zeta$ (see the discussion below eq.~(3.32) of ref.~\cite{CW}). Integrating over $\theta\in(0,2\pi)$ gives $\int_0^{2\pi}\epsilon e^{\pm i\theta}(\mathcal A_\Delta/\epsilon e^{\pm i\theta})\dd\theta=2\pi\mathcal A_\Delta$ for each chirality, and hence
\begin{equation}
 p_{\rm ent}(z)=\frac{4\pi}{z}\mathcal A_\Delta(z),
 \label{eq:entropyflux}
\end{equation}
where the factor $1/z$ arises from $\sqrt{h}\,n_{\rm tube}^\zeta=(\epsilon/z)e^{i\theta}$ and converts the tube flux into a coordinate density multiplying $\eta(z)\dd z$ in \eqref{eq:Sgen}.

\paragraph{Channel selection and the $j\to+\infty$ modular endpoint.}
For the free bulk scalar $\varphi$ on fixed $\text{AdS}_3$ (order $\GN^0$), the bulk entanglement-wedge algebra of $a_1\cup a_2$ is isomorphic to the boundary generalized-free-field algebra of $A\cup B$, so the $O(\lambda^{2\Delta})$ bulk modular interaction $\Delta K$ shares the same small-interval decomposition $\delta K_{AA}+\delta K_\times+\delta K_{BB}$ in terms of the boundary primary $\OO$~\cite{Shape,CW}. Remarkably, the roles of the channels here are opposite to those in the high-boost probe response of Section~\ref{sec:abel}: because $\OO_B$ sits at $x_{\rm P}=1$ bounded away from $\mathcal T_\epsilon$, both $\langle T_{\zeta\zeta}^\varphi\delta K_{BB}\rangle_c$ and $\langle T_{\zeta\zeta}^\varphi\delta K_\times\rangle_c$ remain finite at $\zeta=0$ (the $\sigma$-integral of $\delta K_\times$ converges absolutely at $\zeta=0$ since $[2\cosh(\sigma/2)]^{-2\Delta-2}e^\sigma=O(e^{-\Delta\sigma})$ as $\sigma\to+\infty$) and therefore produce zero $\epsilon\to 0$ tube flux. Only the same-side term $\delta K_{AA}$ in \eqref{eq:AAkernel} of Appendix~\ref{app:AAfull} (cf.\ eq.~(3.10) of ref.~\cite{CW}) develops a $1/\zeta$ pole via its asymptotic $j\to+\infty$ modular endpoint.

Under the Wick rotation $t_E=it$, the lightcone coordinates \eqref{eq:nullcoords} map to $(u,v)=(\bar\zeta,-\zeta)$, so for an insertion $\OO_A(j)$ at $(u_A,v_A)=(-e^j,e^{-j})$ the bulk--boundary propagator denominator $z^2-(u-u_A)(v-v_A)$ becomes $a_z+e^j\zeta+e^{-j}\bar\zeta+\zeta\bar\zeta$ with $a_z=1+z^2$. Contracting $T_{\zeta\zeta}^\varphi=(\partial_\zeta\varphi)^2$ with $\OO_A(j+k)\OO_A(j)$ yields two identical Wick contractions, two bulk--boundary normalization factors $(1/\sqrt{2\pi})^2=1/(2\pi)$ from Appendix~\ref{app:probenorm}, and two $\partial_\zeta$ derivatives contributing $\Delta^2 e^{2j+k}$ near $\zeta=0$. In the large-$j$ endpoint regime where $e^j\zeta=O(1)$, the modular kernel factor $[4\cosh^2(j/2)]^{-1}\sim e^{-j}$ in \eqref{eq:AAkernel} reduces $e^{2j+k}$ to $e^{j+k}$, while the second bracket term $(e^{k+j}+1)^{-1}=O(e^{-j})$ drops out, leaving
\begin{equation}
 \int\dd j\,
 \frac{e^{j+k}}
 {(a_z+e^{j+k}\zeta)^{\Delta+1}(a_z+e^j\zeta)^{\Delta+1}}
 \sim\frac{b_k\mathfrak h_\Delta(b_k)}{\zeta a_z^{2\Delta+1}},
 \label{eq:entendpoint}
\end{equation}
where we substituted $y_0=e^j\zeta/a_z$ and rotated the complex ray $y_0\in e^{i\theta}(0,\infty)$ back to $(0,\infty)$ (avoiding the branch points on the negative real axis for $\theta\in(-\pi,\pi)$), with
\begin{equation}
 b_k=e^k,\qquad
 \mathfrak h_\Delta(b_k)=\int_0^\infty
       \frac{\dd y_0}{(1+b_ky_0)^{\Delta+1}(1+y_0)^{\Delta+1}}.
\end{equation}
Changing variables to $\dd k=\dd b_k/b_k$ cancels the factor of $b_k$ in \eqref{eq:entendpoint}. Combining the overall $\frac{i\lambda^{2\Delta}}{2\pi}$ prefactor of \eqref{eq:AAkernel} with the Wick/propagator/derivative factor $2\cdot\frac{1}{2\pi}\cdot\Delta^2$, the thermal light-pair boundary value $c_1(k)=e^{i\pi\Delta}b_k^\Delta/(b_k-1+i0)^{2\Delta}$, and the surviving $(e^k-1)^{-1}=(b_k-1+i0)^{-1}$ kernel factor fixes the residue $\mathcal A_\Delta(z)$ to be
\begin{align}
 \mathcal A_\Delta(z)&=\frac{i\lambda^{2\Delta}\Delta^2}{2\pi^2}
       \frac{z^{2\Delta}}{(1+z^2)^{2\Delta+1}}\mathcal B_\Delta,
 \label{eq:Aresidue}\\
 \mathcal B_\Delta&=e^{i\pi\Delta}\int_0^\infty
       \frac{b_k^\Delta\mathfrak h_\Delta(b_k)}
            {(b_k-1+i0)^{2\Delta+1}}\dd b_k,
 \label{eq:entresidue}
\end{align}
while the opposite endpoint $j\to -\infty$ (using $\mathfrak h_\Delta(1/b_k)=b_k\mathfrak h_\Delta(b_k)$) produces the identical residue $\mathcal A_\Delta(z)$ for the antiholomorphic component $\langle T_{\bar\zeta\bar\zeta}^\varphi\Delta K\rangle_c$.

\subsection{Selberg evaluation and matching}
While \eqref{eq:entresidue} was evaluated directly in refs.~\cite{Shape,CW}, recognizing it as a two-variable Selberg integral~\cite{Selberg} yields the most compact closed-form evaluation. Changing variables to $x_0=b_k y_0$ and compactifying $(x_0,y_0)\in(0,\infty)^2$ onto the unit square $(t_0,v_0)\in(0,1)^2$ via $t_0=x_0/(1+x_0)$ and $v_0=y_0/(1+y_0)$ transforms \eqref{eq:entresidue} into
\begin{equation}
 \mathcal B_\Delta=e^{i\pi\Delta}\int_0^1\!\dd t_0\dd v_0\,
 \frac{(t_0v_0)^\Delta[(1-t_0)(1-v_0)]^{2\Delta}}
      {(t_0-v_0+i0)^{2\Delta+1}}.
 \label{eq:Selberg}
\end{equation}
Working initially in the strip $-1/2<\operatorname{Re}\Delta<0$ where the diagonal singularity at $t_0=v_0$ is integrable and then continuing analytically in $\Delta$ with the $+i0$ prescription fixed, we split the unit square into the two triangles $t_0>v_0$ (phase $1$) and $t_0<v_0$ (phase $-e^{-2\pi i\Delta}$). In terms of the standard two-variable Selberg integral~\cite{Selberg}
\begin{equation}
 S_2(\alpha,\beta_s,\gamma_s)
 =\int_0^1\!\dd t_0\dd v_0\,
 (t_0v_0)^{\alpha-1}[(1-t_0)(1-v_0)]^{\beta_s-1}
 |t_0-v_0|^{2\gamma_s},
\end{equation}
summing the two triangular regions gives
\begin{align}
 \mathcal B_\Delta&=i\sin(\pi\Delta)
                  S_2(\Delta+1,2\Delta+1,-\Delta-1/2),\\
 S_2&=\frac{\sqrt\pi\,\Gamma(2\Delta+1)\Gamma(\Delta+1/2)\Gamma(-2\Delta)}
 {\Gamma(2\Delta+3/2)\Gamma(1/2-\Delta)}.
 \label{eq:Selbergvalue}
\end{align}
Applying Euler's reflection formula $\Gamma(-2\Delta)\Gamma(2\Delta+1)=-\pi/\sin(2\pi\Delta)$ and $\Gamma(1/2-\Delta)=\pi/[\cos(\pi\Delta)\Gamma(\Delta+1/2)]$ cancels the apparent trigonometric poles and zeroes across all $\Delta>0$, leaving the smooth, nonsingular expression
\begin{equation}
 \mathcal B_\Delta=-\frac{i\sqrt\pi}{2}
              \frac{\Gamma(\Delta+1/2)^2}{\Gamma(2\Delta+3/2)}.
 \label{eq:Bvalue}
\end{equation}
(For example, at the half-integer benchmark $\Delta=3/2$, \eqref{eq:Bvalue} immediately gives $\mathcal B_{3/2}=-8i/105$.) Substituting \eqref{eq:Bvalue} and \eqref{eq:Aresidue} into \eqref{eq:entropyflux} yields
\begin{equation}
 p_{\rm ent}(z)=\lambda^{2\Delta}
 \frac{\Delta^2\Gamma(\Delta+1/2)^2}
      {\sqrt\pi\,\Gamma(2\Delta+3/2)}
 \frac{z^{2\Delta-1}}{(1+z^2)^{2\Delta+1}},
\end{equation}
and applying the Legendre duplication formula $\Delta\Gamma(\Delta+1/2)=\sqrt{\pi}\,\Gamma(2\Delta+1)/[2^{2\Delta}\Gamma(\Delta)]$ reproduces the exact expression of refs.~\cite{Shape,CW} with prefactor $\mathcal N_\Delta$ in \eqref{eq:N}, completing the first-principles verification that $p_{\rm ent}(z)=2\Pi_\Delta(z)$.

\section{The complete \texorpdfstring{$AA$}{AA} exchange: sheets and endpoint estimates}
\label{app:AAfull}

Our goal in this appendix is to prove the $O(e^{-\beta})$ same-side suppression bound \eqref{eq:AAcompletebound} of Section~\ref{sec:suppression} in two clearly separated steps. First, after transferring the external boost $\beta$ into the $AA$ modular kernel on displaced complex rays (Appendix~\ref{app:AA-contours}), we establish the uniform unboosted graph envelope \eqref{eq:AA_graph_envelope}, $|F_{\rm word}(a,g)|\le C\GN\min(L^2,1)$, across the entire compactified radial quadrant $(|a|,|g|)\in[0,\infty]^2$ using a single ordered conformal chart into the Wightman forward tube together with the direct light-pair OPE as $L\to 0$ (Appendix~\ref{app:AA-chart}); substituting \eqref{eq:AA_graph_envelope} into the elementary polar-simplex integral \eqref{eq:AA_uniform_integral} immediately yields the $O(e^{-\beta})$ bound on the displaced contours. Second, because $\delta K_{AA}$ is originally defined at real Lorentzian supports ($h=0$, $\Im k=0^+$), we verify via an auxiliary Gaussian regulator and the tempered-distribution growth bound \eqref{eq:AA_tempered_growth} that deforming the contour to the displaced rays at the physical Euclidean preparation ($h=1$) loses neither boundary terms at infinity nor $k=0$ contact distributions, and then remove the Gaussian cutoff by dominated convergence using \eqref{eq:AA_uniform_integral} (Appendix~\ref{app:AA-contacts}). Table~\ref{tab:AAroadmap} summarizes the covering of $[0,\infty]^2$ in Appendix~\ref{app:AA-chart}.

\begin{table}[t]
\centering
\small
\begingroup
\renewcommand{\arraystretch}{1.22}
\begin{tabularx}{\linewidth}{@{}>{\raggedright\arraybackslash}p{0.27\linewidth}>{\raggedright\arraybackslash}p{0.23\linewidth}Y@{}}
\toprule
\textbf{Region of $(|a|,|g|)\in[0,\infty]^2$} & \textbf{Method} & \textbf{Controlling mechanism \& bound}\\
\midrule
Common endpoints ($L\to 0$; $A,G\to\infty$) & Direct light-pair OPE + inversion & Stress tensor starts at dimension $2$ ($L^{2\Delta}|\log L|\le CL^2$ for $\Delta>1$); $|F_{\rm word}|\le C\GN\min(L^2,1)$ (App.~\ref{app:AA-chart}).\\
\addlinespace[2pt]
Mixed corners ($a\to 0,g\to\infty$ \& swap) & Ordered conformal chart + dilation & Map \eqref{eq:AA_ordered_chart} + small complex dilation sends both outside and sandwich words into a compact subset of the open tube; $O(G_N)$ (App.~\ref{app:AA-chart}).\\
\addlinespace[2pt]
Four one-leg edges (one radius finite) & Ordered conformal chart & Finite light radius retains strict tube gaps in \eqref{eq:AA_ordered_chart} without extra dilation; $O(G_N)$ (App.~\ref{app:AA-chart}).\\
\addlinespace[2pt]
Compact interior of $(0,\infty)^2$ & Ordered conformal chart & Strict imaginary gaps in both null coordinates of \eqref{eq:AA_ordered_chart}; $O(G_N)$ (App.~\ref{app:AA-chart}).\\
\bottomrule
\end{tabularx}
\endgroup
\caption{Covering of the compactified radial quadrant $(|a|,|g|)\in[0,\infty]^2$ establishing the global graph envelope $|F_{\rm word}(a,g)|\le C\GN\min(L^2,1)$ in Appendix~\ref{app:AA-chart}.}
\label{tab:AAroadmap}
\end{table}

\subsection{The regulated observable and its two contours}
\label{app:AA-contours}
\paragraph{Sommerfeld--Watson representation of $\delta K_{AA}$.}
In the small-interval modular OPE, the same-side $AA$ bilinear correction is represented by the Sommerfeld--Watson integral along $j\in\mathbb R$, $k\in\mathbb R+i\epsilon$ ($0<\epsilon<\pi$)~\cite{CFL,CW}
\begin{align}
 \delta K_{AA}&=\frac{i\lambda^{2\Delta}}{2\pi}
 \int_{\mathbb R}\dd j\int_{\mathbb R+i\epsilon}\dd k\,
 \frac{c_1(k)}{4\cosh^2(j/2)}
 \left[\frac1{e^{k}-1}+\frac1{e^{k+j}+1}\right]
 \OO_A(j+k)\OO_A(j),\nonumber\\
 c_1(k)&=[2-2\cosh k]^{-\Delta},\qquad 0<\epsilon<\pi,
 \label{eq:AAkernel}
\end{align}
with the kernel and the adjacent fields $\OO_1=\OO_A(j+k)$ and $\OO_2=\OO_A(j)$ continued together from $\Im k=0^+$. For the left-wedge time $t_g=-\sinh(j+k)$ at real $j,k$, this prescription puts $\OO_g$ at earlier imaginary Minkowski time than $\OO_a$. Changing variables to $a=e^j$ and $g=e^{j+k}$ (with $\arg(g/a)=\Im k=\epsilon\in(0,\pi)$, so that $2-2\cosh k=-(g-a)^2/(ag)$ yields $c_1(k)=e^{i\pi\Delta}(ag)^\Delta/(g-a)^{2\Delta}$), the factor $a+1$ cancels one factor of $1+a$ identically and the measure and modular kernel in \eqref{eq:AAkernel} combine (up to the overall factor $i\lambda^{2\Delta}/(2\pi)$) into the manifestly Hermitian kernel
\begin{equation}
 \dd a\dd g\;e^{i\pi\Delta}
 \frac{(ag)^\Delta}
 {(1+a)(1+g)(g-a)^{2\Delta+1}}.
 \label{eq:AApositive}
\end{equation}
At lightcone coordinates $(u_1,v_1)=(-g,g^{-1})$ and $(u_2,v_2)=(-a,a^{-1})$, the compatible light two-point function is
\begin{equation}
 \langle\OO_1\OO_2\rangle
 =e^{i\pi\Delta}\frac{(ag)^\Delta}{(g-a)^{2\Delta}}.
 \label{eq:AAlightpair}
\end{equation}

The factors in \eqref{eq:AApositive} and \eqref{eq:AAlightpair} therefore carry the same boundary prescription at $g=a$. The joint pairing with the external fields requires their positions in the thermal trace.

\paragraph{Thermal KMS slots and boost transfer.}
In the commutator $-\langle[\delta K_{AA},\Phi_L]\Phi_R\rangle_\Omega$, the two operator orderings define the connected four-point words
\begin{align}
 W_{\rm out}&=\langle\OO_g\OO_a\Phi_L\Phi_R\rangle_{\rm conn},
 \qquad W_{\rm sand}=\langle\Phi_L\OO_g\OO_a\Phi_R\rangle_{\rm conn},\nonumber\\
 F_{\rm word}&=\frac{W_{\rm word}}
 {\langle\OO_g\OO_a\rangle\langle\Phi_L\Phi_R\rangle}.
 \label{eq:AA_words_full}
\end{align}
Because $\OO$ and $\OO_\phi$ are distinct scalar species, the disconnected generalized-free product $\langle\OO_g\OO_a\rangle\langle\Phi_L\Phi_R\rangle$ is identical in both orderings and cancels in the commutator difference.

\begin{figure}[t]
\centering
\includegraphics[width=\linewidth]{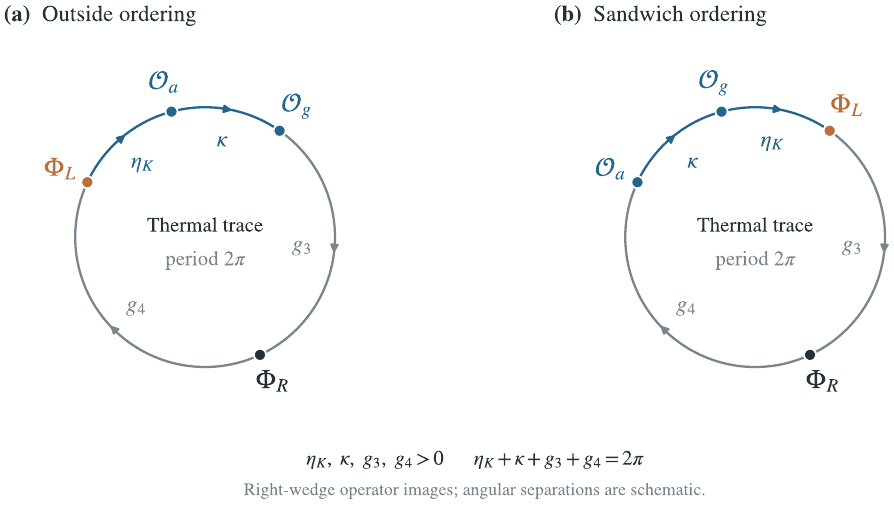}
\caption{Cyclic thermal-slot order of the two $AA$ words on a thermal circle of period $2\pi$. Clockwise arrows indicate increasing imaginary right-Rindler time. The outside word places the transposed probe before the light pair; the sandwich word places it after the pair. Operator labels denote the corresponding right-wedge images of the left insertions, with transpose superscripts suppressed. The fixed gaps $\eta_K,\kappa>0$ exchange positions. The symbols $g_3,g_4$ denote the positive third and fourth gaps listed above; their sum with the first two is $2\pi$. Positions are schematic; the actual slots and displaced integration cycles are given in \eqref{eq:AA_slots_full}--\eqref{eq:AA_cycles_full}.}
\label{fig:AAslots}
\end{figure}

Parameterizing the analytic continuation \eqref{eq:explicitACpath} from real Rindler supports ($h=0$) to the physical single-insertion Euclidean preparation $(s,\delta)$ ($h=1$) by $h\in[0,1]$ (unrelated to the shockwave profile $h(z)$ of Section~\ref{sec:dictionary}), with endpoint angles $(\alpha_L,\theta_R)=(\pi-\delta,\delta)$, we use the regulated thermofield identity $\langle X_L Y_R\rangle=\operatorname{Tr}(\varrho^{1/2}X_L^T\varrho^{1/2}Y_R)$, with normalized $\varrho=Z_R^{-1}e^{-2\pi H_R}$, $Z_R=\operatorname{Tr}e^{-2\pi H_R}$, and right-Rindler Hamiltonian $H_R$. Transposition reverses the left product; moving its thermal factors by conjugation with $\varrho^{-1/2}$ produces the $-i\pi$ time shift (which does not arise from transposition alone). Reading the four operators from left to right inside the right-wedge thermal trace places their imaginary right-Rindler time slots at
\begin{equation}
\begin{array}{c|cccc}
\mathrm{out}&\alpha_L-\pi&\Im j-\pi&\Im(j+k)-\pi&\theta_R\\
\mathrm{sand}&\Im j-\pi&\Im(j+k)-\pi&\alpha_L-\pi&\theta_R
\end{array}.
\label{eq:AA_slots_full}
\end{equation}
At the real-support end ($h=0$), we smear the external insertions and assign each word its own infinitesimal Wightman ordering: for $0<\epsilon\ll\epsilon_p\ll 1$ and any fixed $C_s>2$, we choose
\begin{equation}
 \Im j=0,\quad \Im k=\epsilon,\quad \theta_{R,0}=-C_s\epsilon_p,\qquad
 \alpha_{L,0}=\begin{cases}-2\epsilon_p & (\mathrm{out}),\\
                              +2\epsilon_p & (\mathrm{sand}).
               \end{cases}
 \label{eq:AA_joint_i0}
\end{equation}
Because $\Phi_R$ is spacelike separated from all three left-wedge insertions at $h=0$, the negative shift $\theta_{R,0}=-C_s\epsilon_p$ preserves positivity of all four cyclic thermal gaps in \eqref{eq:AA_slots_full}---the first two gaps are $(2\epsilon_p,\epsilon)$ for $\mathrm{out}$ and $(\epsilon,2\epsilon_p-\epsilon)$ for $\mathrm{sand}$, while the remaining two gaps approach $\pi$---and simultaneously ensures that the unfolded probe span $\pi+\theta_{R,0}-\alpha_{L,0}=\pi-(C_s\mp 2)\epsilon_p$ stays strictly below $\pi$. At these fixed external slots, we linearly interpolate $(\Im j,\Im k)$ from \eqref{eq:AA_joint_i0} to the displaced contours
\begin{equation}
\begin{array}{c|cc}
\mathrm{out}&j=x+i(\alpha_L+\eta_K)&k=y+i\kappa\\
\mathrm{sand}&j=x+i(\alpha_L-\eta_K-\kappa)&k=y+i\kappa,
\end{array}\qquad x,y\in\mathbb R,
\label{eq:AA_cycles_full}
\end{equation}
with small initial margins $\eta_K,\kappa>0$. Because both endpoints lie in the convex open ordering domain, all four cyclic gap inequalities together with $0<\Im k<\pi$, $|\Im j|<\pi$, and $|\Im(j+k)|<\pi$ hold throughout the linear segment, avoiding every modular pole. We then carry \eqref{eq:AA_cycles_full} along the linear interpolation of $(\alpha_L,\theta_R)$ from $(\alpha_{L,0},\theta_{R,0})$ to the physical preparation $(\pi-\delta,\delta)$, adjusting $(\eta_K,\kappa)$ inside the open ordering region to fixed terminal values obeying $\eta_K+\kappa<\delta/3$. Along \eqref{eq:AA_cycles_full}, all four cyclic thermal gaps $g_m=\tau_{m+1}-\tau_m$ on the $2\pi$-periodic thermal circle---which enter the right-wedge thermal trace as damping Boltzmann factors $e^{-g_m H_R}$---remain strictly positive throughout the continuation $h\in[0,1]$ (see Figure~\ref{fig:AAslots}),
\begin{align*}
\mathrm{out}:&\quad\eta_K,\ \kappa,
 \ \pi+\theta_R-\alpha_L-\eta_K-\kappa,
 \ \pi+\alpha_L-\theta_R,\\
\mathrm{sand}:&\quad\kappa,\ \eta_K,
 \ \pi+\theta_R-\alpha_L,
 \ \pi+\alpha_L-\theta_R-\eta_K-\kappa,
\end{align*}
without ever crossing the modular kernel poles at $\Im j=\pm\pi$, $\Im(j+k)=\pm\pi$, or $k=0$. At $h=1$, shifting $j_{\rm old}=J+\beta$ and setting $\rho=e^{-\beta}$, $a=e^J$, and $g=e^{J+k}$ maps the integration contours to fixed rays in the complex $a$- and $g$-planes and factors the boost $\rho=e^{-\beta}$ into the exact Hermitian kernel \eqref{eq:AA_boost_kernel}:
\begin{align*}
\frac1{4\cosh^2(j_{\rm old}/2)}&=\frac{a\rho}{(\rho+a)^2},&
c_1(k)&=e^{i\pi\Delta}
        \frac{(ag)^\Delta}{(g-a)^{2\Delta}},\\
\frac1{e^{k}-1}+\frac1{e^{k+j_{\rm old}}+1}
 &=\frac{g(\rho+a)}
 {(g-a)(\rho+g)},&
\dd j_{\rm old}\dd k&=\frac{\dd a\dd g}{ag}.
\end{align*}

\subsection{Global graph envelope via an ordered conformal chart}
\label{app:AA-chart}
With the boost $\beta$ factored into the kernel $K_\rho$ in \eqref{eq:AA_boost_kernel}, the normalized connected four-point function $F_{\rm word}(a,g)$ is an unboosted, conformally covariant vacuum correlator evaluated along the fixed rays \eqref{eq:AA_cycles_full}. To establish $|F_{\rm word}(a,g)|\le C\GN\min(L^2,1)$ uniformly across the compactified quadrant $(|a|,|g|)\in[0,\infty]^2$, we construct a single real $\mathrm{PSL}(2,\mathbb R)$ conformal chart that maps both the outside and sandwich words simultaneously into the open ordered Wightman forward tube~\cite{CFTDistributions}. Throughout Appendices~\ref{app:AA-chart}--\ref{app:AA-contacts} and~\ref{app:cross-outer}, we take as an analytic regularity input that the fixed tree-level Einstein exchange under the Skenderis--van Rees real-time continuation~\cite{SkenderisVanRees,SkenderisVanReesLong} inherits the standard forward-tube holomorphy, conformal covariance, and tempered distributional boundary values of vacuum CFT four-point functions~\cite{CFTDistributions} (a perturbative graph-level input not deduced here from nonperturbative CFT unitarity). Under this regularity input, the normalized tree graph is holomorphic and uniformly bounded on compact subsets of this open tube.

\paragraph{An ordered conformal chart.}
Passing to the chiral variables $X_+=u$ and $X_-=-1/v$, the two light insertions occupy $-a$ and $-g$ on both chiral lines, while the external probes $\Phi_L,\Phi_R$ sit at radius $s$ on the $X_+$ line and radius $s^{-1}$ on the $X_-$ line. A common imaginary Rindler translation centers the two light angles at $-\kappa/2$ and $+\kappa/2$, placing $\Phi_L$ and $\Phi_R$ at angles $\ell$ and $2\delta+\ell$, where $\ell=-(\eta_K+\kappa/2)$ for the outside word and $\ell=+(\eta_K+\kappa/2)$ for the sandwich word. At fixed $(s,\delta)$, we take $\eta_K$ sufficiently small and then $\kappa$ sufficiently small relative to $\eta_K$; these margins lie inside the open KMS domain \eqref{eq:AA_cycles_full}. Holding the modular kernel on the displaced rays \eqref{eq:AA_cycles_full}, we apply the orientation-preserving $\mathrm{PSL}(2,\mathbb R)$ M\"obius transformation $m(X)=(1+2X)/(1+X)$ to both chiral lines of the normalized four-point function, defining the transformed null coordinates
\begin{equation}
 m(X)=\frac{1+2X}{1+X},\qquad
 Z=(U,V)=\bigl(m(X_+),-1/m(X_-)\bigr).
 \label{eq:AA_ordered_chart}
\end{equation}
For any chiral argument $X=r e^{i\varphi}$, the resulting physical imaginary parts are
\begin{equation}
 \Im m(X)=\frac{r\sin\varphi}{|1+X|^2},\qquad
 \Im[-1/m(X)]=\frac{r\sin\varphi}{|1+2X|^2}.
\end{equation}
For both light insertions, these imaginary parts are uniformly $O(\kappa)$ over all radii $r\in(0,\infty)$ (bounded by $\frac{1}{2}\tan(\kappa/4)$ and $\frac{1}{4}\tan(\kappa/4)$), with negative sign for $a$ and positive sign for $g$. By contrast, both components of $\Phi_L$ have the sign of $\ell$ and magnitude proportional to $\eta_K$, while both components of $\Phi_R$ are bounded below by a strictly positive constant at fixed $(s,\delta)$ since $0<2\delta<\pi$ (and $2\delta+\ell<\pi$ for small $\eta_K,\kappa$). Consequently, the imaginary parts increase strictly in both null coordinates in the right-wedge trace order $L,a,g,R$ (outside) or $a,g,L,R$ (sandwich). This places the configuration inside the open ordered forward tube of the vacuum four-point function~\cite{CFTDistributions}, on which the real-time Einstein tree graph specified in Section~\ref{sec:setup} is holomorphic and conformally covariant under the Skenderis--van Rees continuation.

\paragraph{Sheet identification and mixed-corner compactification.}
To anchor the analytic sheet, set $|a|=|g|=1$. There all four insertions satisfy $v=-u^*$, so the separated Euclidean configuration fixes the analytic germ of the prescribed trace word, and small independent Euclidean perturbations of all four points determine that germ in all four variables. The real $\mathrm{PSL}(2,\mathbb R)$ homotopy $m_b(X)=b+X/(1+bX)$ ($0\le b\le 1$, with $\det=+1$) smoothly connects the identity ($b=0$) to \eqref{eq:AA_ordered_chart} ($b=1$). Along this homotopy, the two imaginary parts replace the denominators $|1+X|^2$ and $|1+2X|^2$ by $|1+bX|^2$ and $|b+(1+b^2)X|^2$, respectively. At the anchor $|a|=|g|=1$, the small-margin choices above keep all four ordered tube inequalities strict uniformly in $b\in[0,1]$, so conformal covariance identifies the normalized correlators in a neighborhood of the anchor. Continuing $(|a|,|g|)$ through all finite positive radii preserves both the thermal ordering and the tube inequalities of \eqref{eq:AA_ordered_chart}, fixing the physical Wightman sheet across the entire open quadrant.

\paragraph{Boundary faces of $[0,\infty]^2$ and global patching.}
The ordered conformal chart \eqref{eq:AA_ordered_chart} controls every boundary face and the interior of the compactified radial quadrant $(|a|,|g|)\in[0,\infty]^2$:
\begin{itemize}
 \item \emph{Both mixed corners ($a\to 0,g\to\infty$ and $a\to\infty,g\to 0$)}: at zero and infinite light radius, the map \eqref{eq:AA_ordered_chart} sends the light insertions to the real timelike-separated limits $P_0=(1,-1)$ and $P_\infty=(2,-1/2)$, whose difference $P_\infty-P_0=(1,1/2)$ has two positive components. Near each mixed corner, we apply the common complex dilation $Z\mapsto e^{i\omega}Z$ with $|\omega|$ small and $\operatorname{sgn}\omega$ chosen so that $\sin\omega\,\Re(Z_g-Z_a)>0$ in both components ($\omega>0$ near $a\to 0,g\to\infty$; $\omega<0$ near $a\to\infty,g\to 0$). Because $\Im(Z_g-Z_a)\ge 0$ already holds throughout the corner neighborhood, this short rotation adds a strictly positive imaginary light--light gap even at the corner itself, while leaving all fixed probe gaps strictly open for sufficiently small $|\omega|$. Real dilation invariance ($Z\mapsto e^t Z$) continues analytically along the short rotation from $0$ to $\omega$ inside the open tube, mapping a closed neighborhood of each mixed corner onto a compact subset of the ordered forward tube where both normalizing two-point functions remain finite and nonzero. Holomorphy therefore bounds each complete normalized tree graph by $|F_{\rm word}|\le C_{s,\delta}G_N$ at both mixed corners (for both outside and sandwich orderings).
 \item \emph{Four one-leg edges}: when one light radius tends to $0$ or $\infty$ while the other stays finite and positive, the finite light insertion retains a strictly nonzero imaginary part in both components of \eqref{eq:AA_ordered_chart}, while the limiting light approaches the real point $P_0$ or $P_\infty$. All four points therefore stay strictly ordered inside a compact subset of the open forward tube, bounding $|F_{\rm word}|\le C_{s,\delta}G_N$ uniformly along every one-leg edge without needing any extra dilation.
 \item \emph{Common small and large endpoints ($L\to 0$; $A,G\to\infty$)}: writing $A=|a|$, $G=|g|$, and $L=A+G$, as $L\to 0$ both light insertions approach $P_0=(1,-1)$ with complex separation $O(L)$, a strictly ordered light--light imaginary gap bounded below by $c_\kappa L$ in each component, and fixed $O(1)$ gaps to the two probes. Linearly interpolating the two lights to a Euclidean separation of size $L$ inside this convex ordered tube keeps both direct light-pair cross ratios inside punctured disks of radius $CL<1/2$ around zero without encircling crossed-channel branch points, uniformly in $A/L\in(0,1)$. Because the identity block cancels in the connected correlator and scalar double-trace dimensions begin at $\min(2\Delta,2d)>2$, the leading small-$L$ behavior in the direct light-pair OPE is governed by the dimension-$2$ stress tensor (with double-trace logarithms obeying $L^{2\Delta}|\log L|\le CL^2$), giving $|F_{\rm word}|\le C_{s,\delta}G_N L^2$ uniformly in $A/L\in(0,1)$. The reciprocal inversion $(u,v)\mapsto(-1/u,v)$ yields the corresponding local bound $|F_{\rm word}|\le C_{s,\delta}G_N$ at the common large-radius corner $A,G\to\infty$ ($P_\infty$).
 \item \emph{Compact interior}: on every compact subset of $(0,\infty)^2$, \eqref{eq:AA_ordered_chart} keeps strict positive gaps in both null coordinates on the physical sheet, so holomorphy bounds $|F_{\rm word}|\le C_{s,\delta}G_N$.
\end{itemize}
Patching these open covers of the compactified quadrant $[0,\infty]^2$ establishes the global uniform graph envelope \eqref{eq:AA_graph_envelope} for both commutator words, and hence the displaced-contour bound \eqref{eq:AA_uniform_integral}. Paper II~\cite{CompanionTomography} extends this conformal-chart construction to off-diagonal preparation pairs.

\subsection{Recovering the original contour and its contacts}
\label{app:AA-contacts}
The estimate \eqref{eq:AA_uniform_integral} bounds the $AA$ integral on the displaced complex rays \eqref{eq:AA_cycles_full} at the physical Euclidean preparation ($h=1$). To complete the proof of \eqref{eq:AAcompletebound}, it remains to verify that deforming the contour from the original real-support definition ($h=0$, $\Im k=0^+$) to those displaced rays loses neither boundary terms at infinity nor $k=0$ coincidence contact distributions.

At the real-support end ($h=0$), fix $\beta\ge 0$, the external regulator $\epsilon_p>0$ in \eqref{eq:AA_joint_i0}, and insert the entire Gaussian damping factor $\exp[-\sigma_G(j^2+k^2)]$ with $\sigma_G>0$. Along the linear $(j,k)$ contour deformation of Appendix~\ref{app:AA-contours}, each word remains strictly inside its ordered thermal domain. On any closed substrip of minimum cyclic gap $m>0$, the tree-level forward-tube temperedness input of Appendix~\ref{app:AA-chart} (analogous to the CFT forward-tube bounds of ref.~\cite{CFTDistributions}) implies polynomial growth of the connected four-point word in the positive radii $e^{\pm\Re j},e^{\pm\Re(j+k)}$, which in logarithmic modular times reads
\begin{equation}
 |W_{\rm word}(j,k)|\le C(1+m^{-1})^N
       \exp\bigl[M(|\Re j|+|\Re k|)\bigr].
 \label{eq:AA_tempered_growth}
\end{equation}
The Gaussian factor $\exp[-\sigma_G(j^2+k^2)]$ therefore suppresses the vertical closing segments at infinity, so Cauchy's theorem equates the original $\Im k=\epsilon$ integral to the displaced-contour integral on \eqref{eq:AA_cycles_full} at fixed $(\epsilon_p,\sigma_G)$, independently of $\epsilon\in(0,\kappa)$. Taking the boundary value $\epsilon\downarrow 0$ at fixed $\epsilon_p$, followed by $\epsilon_p\downarrow 0$ against smeared real external supports, matches the analytically continued functional to the original smeared Wightman pairing and uniquely fixes the coincidence contact distributions supported at $k=0$.

Next, we continue the displaced-contour integral along the linear external path from $(\alpha_{L,0},\theta_{R,0})$ to $(\pi-\delta,\delta)$ at fixed $\sigma_G>0$. Because $\theta_{R,0}=-C_s\epsilon_p$ with $C_s>2$, the unfolded probe span $\pi+\theta_R-\alpha_L$ stays strictly in $(0,\pi)$ throughout the path to $2\delta$; the ordered conformal chart \eqref{eq:AA_ordered_chart} therefore applies on every compact segment of the path (with $\pi+\theta_R-\alpha_L$ in place of $2\delta$ and sufficiently small margins), and at the physical endpoint $h=1$ both the external angles $(\pi-\delta,\delta)$ and the displaced margins $(\eta_K,\kappa)$ are independent of $\epsilon_p$.

Finally, at $h=1$ we apply the real translation $j_{\rm old}=J+\beta$. On the horizontal lines \eqref{eq:AA_cycles_full} with fixed imaginary parts $\Im J$ and $\Im k=\kappa$, we have $|\exp[-\sigma_G((J+\beta)^2+k^2)]|\le\exp[(\Im J)^2+\kappa^2]=:C_0$ uniformly for all $\sigma_G\in(0,1]$ and all $\beta\ge 0$. The uniform integrable majorant \eqref{eq:AA_uniform_integral} is therefore independent of $\sigma_G\in(0,1]$, allowing us to remove the Gaussian regulator by taking $\sigma_G\downarrow 0$ via dominated convergence. Equation~\eqref{eq:AA_uniform_integral} bounds each complete word by $O(G_N e^{-\beta})$, confirming that the $k=0$ contact distributions do not contribute to the $e^\beta$ coefficient in $a_{AA}(\beta)$ and completing the proof of \eqref{eq:AAcompletebound}.

\section{Cross-channel Regge remainder and dominated convergence}
\label{app:crossremainder}

In Section~\ref{sec:dictionary}, the growing cross-channel signal $e^\beta Q^\times[f]$ was extracted by taking the Regge (eikonal) limit $\beta\to\infty$ pointwise at fixed modular time $\sigma$ and then integrating over $\sigma\in\mathbb R$. Because the modular integral extends to arbitrarily large $\sigma$, however, the boosted cross-ratio scale $\ell:=(1+|e^\sigma|)e^{-\beta}$ eventually becomes $O(1)$ in the moving tail $\sigma\gtrsim\beta$, where the eikonal approximation breaks down. The goal of this appendix is to justify interchanging the large-boost limit with the $\sigma$-integral and prove the sharp remainder bound \eqref{eq:crossRemainderNorm}, $|r_\times(\beta)|=|p_\times(\beta)-e^\beta Q^\times[f]|\le C_\times$ uniformly for all $\beta\ge 0$ (hence $h_\times(\beta)=e^{-\beta}|r_\times(\beta)|\le C_\times e^{-\beta}$).

After setting up the $\mathrm{TOC}/\mathrm{OTOC}$ thermal contours and boost-transferred cross ratios (Appendix~\ref{app:cross-contours}), we split the $\sigma$-contour at $\ell=\delta_*$ into two complementary regions (Table~\ref{tab:crossroadmap}):
\begin{enumerate}
 \item \emph{Outer non-Regge tail ($\ell\ge\delta_*$, Appendix~\ref{app:cross-outer})}: we only need a uniform $O(G_N)$ bound on $|F_\times^{\rm word}|$ via ordered conformal maps into the Wightman tube, after which the exponentially decaying modular weight $\mathcal W_\times(\sigma)\sim e^{-(2\Delta+1)|\sigma|}$ suppresses the entire tail exponentially in $\beta$.
 \item \emph{Inner small-cross-ratio region ($\ell<\delta_*$, Appendices~\ref{app:cross-mellin}--\ref{app:cross-small})}: after establishing the complete $\text{AdS}_3$ tree-level Mellin double-trace sum and its single-valued normal-convergence domain (Appendix~\ref{app:cross-mellin}), we show in Appendix~\ref{app:cross-small} that $\mathrm{TOC}$ is $O(G_N\ell^2)$, isolate the exact $O(e^\beta)$ $\mathrm{OTOC}$ discontinuity as a Barnes integral, prove its difference from the full graph is uniformly $O(1)$, and verify via Fourier inversion that its $\sigma$-integral equals $e^\beta Q^\times[f]$ identically so that the growing terms cancel with zero mismatch.
\end{enumerate}
Combining the inner $O(1)$ remainder with the exponentially small outer tail via dominated convergence (Appendix~\ref{app:cross-dom}) establishes \eqref{eq:crossRemainderNorm}.

\begin{table}[!htb]
\centering
\small
\begingroup
\renewcommand{\arraystretch}{1.22}
\begin{tabularx}{\linewidth}{@{}>{\raggedright\arraybackslash}p{0.27\linewidth}>{\raggedright\arraybackslash}p{0.16\linewidth}Y@{}}
\toprule
\textbf{Region of $(\Re\sigma,\beta)$} & \textbf{Word} & \textbf{Method \& bound on $|F_\times^{\rm word}|/G_N$}\\
\midrule
Outer $(1+|a|)\rho\ge\delta_*$: interior, one-leg edges and mixed corner & $\mathrm{TOC},\ \mathrm{OTOC}$ & Ordered compactification of the two light points and fixed probe gaps; $O(1)$ (App.~\ref{app:cross-outer}).\\
\addlinespace[2pt]
Small cross ratio $(1+|a|)\rho<\delta_*$ & $\mathrm{TOC}$ & Direct light-pair OPE of the connected graph; $O([(1+|a|)\rho]^2)$ (App.~\ref{app:cross-small}).\\
\addlinespace[2pt]
Small cross ratio $(1+|a|)\rho<\delta_*$ & $\mathrm{OTOC}$ & One winding of $\mathsf Y$, summed logarithmic coefficient, and subtracted Barnes integral; $-e^\beta q_{\times,f}(\sigma)/\langle\OO_A\OO_B\rangle+O(1)$ (App.~\ref{app:cross-small}).\\
\bottomrule
\end{tabularx}
\endgroup
\caption{Decomposition of the cross-channel integrand across $(1+|a|)\rho\ge\delta_*$ and $(1+|a|)\rho<\delta_*$ in Appendix~\ref{app:crossremainder}.}
\label{tab:crossroadmap}
\end{table}

\subsection{Thermal contours, boost transfer, and exact cross ratios}
\label{app:cross-contours}
\paragraph{Commutator words and KMS thermal contours.}
From \eqref{eq:remainingchannels} and \eqref{eq:crossK} (with the minus sign in \eqref{eq:crossK} canceling the overall minus sign in \eqref{eq:remainingchannels}), the cross-channel commutator $-\langle[\delta K_\times,\Phi_L]\Phi_R\rangle_\Omega$ reads
\begin{equation}
 -\langle[\delta K_\times,\Phi_L]\Phi_R\rangle_\Omega
 =\lambda^{2\Delta}\int_{\mathbb R}\frac{\dd\sigma}{[2\cosh(\sigma/2)]^{2\Delta+2}}
 \bigl(W_\times^{\rm TOC}(\sigma)-W_\times^{\rm OTOC}(\sigma)\bigr),
 \label{eq:cross_commutator_words}
\end{equation}
where $W_\times^{\rm TOC}=\langle\OO_A(\sigma)\Phi_L\OO_B\Phi_R\rangle_{\rm conn}$ is the time-ordered ($\mathrm{TOC}$) word (using $[\OO_B,\Phi_L]=0$ at real Lorentzian supports $h=0$) and $W_\times^{\rm OTOC}=\langle\Phi_L\OO_A(\sigma)\OO_B\Phi_R\rangle_{\rm conn}$ is the out-of-time-ordered ($\mathrm{OTOC}$) word. Writing $F_\times^{\rm word}=W_\times^{\rm word}/(\langle\OO_A(\sigma)\OO_B\rangle\langle\Phi_L\Phi_R\rangle)$ with $\langle\OO_A(\sigma)\OO_B\rangle=[2\cosh(\sigma/2)]^{-2\Delta}$ (and dividing by the diagonal probe norm $\mathcal Z_f=\operatorname{AC}_f[\langle\Phi_L\Phi_R\rangle]=Z(\tau)$), the combined modular weight multiplying $F_\times^{\rm word}(\sigma,\beta)$ is
\begin{equation}
 \mathcal W_\times(\sigma):=[2\cosh(\sigma/2)]^{-(4\Delta+2)},
 \label{eq:cross_weight}
\end{equation}
which is holomorphic throughout the strip $|\Im\sigma|<\pi$ and obeys
\[
 |\mathcal W_\times(\sigma)|
 \le C_\delta[2\cosh(\Re\sigma/2)]^{-(4\Delta+2)}
 \le C_\delta' e^{-(2\Delta+1)|\Re\sigma|}
\]
on every closed substrip $|\Im\sigma|\le\pi-\delta_0$.

With left and right probe angles $\alpha_L,\theta_R$ in the continuation \eqref{eq:explicitACpath}, the right-wedge thermofield trace $\operatorname{Tr}(\varrho^{1/2}X_L^T\varrho^{1/2}Y_R)$ places the four operators at the imaginary right-Rindler time slots
\begin{equation}
\begin{array}{c|cccc}
\mathrm{TOC}&\alpha_L-\pi&\Im\sigma-\pi&0&\theta_R\\
\mathrm{OTOC}&\Im\sigma-\pi&\alpha_L-\pi&0&\theta_R
\end{array}.
\label{eq:cross_slots}
\end{equation}
At the real-support end ($h=0$), we give $\Phi_R$ a small positive angle $\theta_R=\epsilon_p$ and assign $\alpha_L=-2\epsilon_p$ for $\mathrm{TOC}$ and $\alpha_L=+2\epsilon_p$ for $\mathrm{OTOC}$, with $\epsilon_p>0$ infinitesimal (matching Appendix~\ref{app:AA-contours}). The initial line $\Im\sigma=0$ then lies strictly inside each ordered KMS domain. For a fixed small margin $0<\gamma<\delta$ (chosen sufficiently small below), we take the terminal contours
\begin{equation}
 \Im\sigma=\pi-\gamma\quad(\mathrm{TOC}),\qquad
 \Im\sigma=0\quad(\mathrm{OTOC}).
 \label{eq:cross_physical_lines}
\end{equation}
At the physical endpoint $(\alpha_L,\theta_R)=(\pi-\delta,\delta)$, the four successive cyclic gaps are $(\delta-\gamma,\gamma,\delta,2\pi-2\delta)$ for $\mathrm{TOC}$ and $(\pi-\delta,\delta,\delta,\pi-\delta)$ for $\mathrm{OTOC}$. Linearly interpolating all three slots from the regulated start to these terminal values keeps every cyclic gap strictly positive and maintains $|\Im\sigma|<\pi$ throughout, avoiding all poles of the modular weight \eqref{eq:cross_weight}. Smearing the initial real supports and retaining the entire Gaussian factor $e^{-\sigma_G\sigma^2}$ ($\sigma_G>0$) during the contour deformation reproduces the ordered boundary value at $h=0$ just as in Appendix~\ref{app:AA-contacts}, while the uniform terminal bounds below remove $\sigma_G\downarrow 0$ on \eqref{eq:cross_physical_lines} at fixed $\beta$ by dominated convergence.

\paragraph{Boost transfer and exact cross ratios.}
At $h=1$, transferring the external boost $\beta\ge 0$ onto the light pair via $\rho=e^{-\beta}\in(0,1]$ and $a=e^\sigma$ places the four insertions at $(u_1,v_1)=(-a\rho,(a\rho)^{-1})$, $(u_2,v_2)=(\rho,-\rho^{-1})$, $(u_3,v_3)=(L_p,-\bar L_p)$, and $(u_4,v_4)=(R_p,-\bar R_p)$, with $L_p=s e^{-i\delta}$ and $R_p=s e^{i\delta}$. At $(\Re\sigma,\beta)=(0,0)$, all four points satisfy $v_k=-u_k^*$ and lie on the strictly Euclidean configuration. The exact $t$-channel cross ratios $z_\times=u_{12}u_{34}/(u_{13}u_{24})$ and $w_\times=v_{12}v_{34}/(v_{13}v_{24})$ are
\begin{equation}
 z_\times(\sigma,\beta)=\frac{(1+a)\rho(R_p-L_p)}{(L_p+a\rho)(R_p-\rho)},
 \qquad
 w_\times(\sigma,\beta)=\frac{(1+a)\rho(\bar L_p-\bar R_p)}
 {(1+a\rho\bar L_p)(1-\rho\bar R_p)},
 \label{eq:cross_zw}
\end{equation}
with $\mathsf X=z_\times w_\times$ and $\mathsf Y=(1-z_\times)(1-w_\times)$.

\subsection{Outer region \texorpdfstring{$(1+|a|)\rho\ge\delta_*$}{(1+|a|) rho >= delta*}}
\label{app:cross-outer}
In the outer non-Regge region, we only need a uniform $O(G_N)$ bound on $|F_\times^{\rm word}|$ so that the modular weight $\mathcal W_\times(\sigma)$ suppresses the moving tail exponentially at large $\beta$. Fix the preparation $(s,\delta)$ and denote the boosted left-light radius by $k:=|a|\rho$ (distinct from the $AA$ modular time $k$ of Appendix~\ref{app:AAfull}). The outer region is $\rho+k\ge\delta_*$ with $0\le\rho\le 1$ and $k\in[0,\infty]$ compactified at infinity. As in Appendix~\ref{app:AA-chart}, we use conformal covariance of the normalized vacuum tree graph on the two Wightman orderings: in the chiral variables $(X_+,X_-)=(u,-1/v)$, both light insertions carry identical coordinates on the two chiral lines, and any real $\mathrm{PSL}(2,\mathbb R)$ map $f$ acts on physical null coordinates as $(U,V)=\bigl(f(X_+),-1/f(X_-)\bigr)$.

\paragraph{Time-ordered outer bound.}
On the $\mathrm{TOC}$ contour $\Im\sigma=\pi-\gamma$, the chiral light coordinates are $k e^{-i\gamma}$ and $\rho$. A common imaginary Rindler translation by $\gamma/2$ shifts the four angles in right-wedge trace order $\Phi_L,\OO_A,\OO_B,\Phi_R$ to
\begin{equation}
 -\delta+\gamma/2,\qquad -\gamma/2,\qquad +\gamma/2,\qquad +\delta+\gamma/2.
\end{equation}
Applying $f(X)=(1+2X)/(1+X)$ on both chiral lines yields physical imaginary parts $r\sin\phi/|1+X|^2$ and $r\sin\phi/|1+2X|^2$ for $X=re^{i\phi}$. Because $\delta\in(0,\pi/2)$ and $0<\gamma<\delta$, the two probes have fixed negative and positive imaginary parts in both null components, while the two light insertions carry $O(\gamma)$ imaginary parts of opposite sign uniformly over all $k,\rho$. Choosing $\gamma$ sufficiently small therefore keeps fixed positive gaps between the probes and the light pair throughout the outer region. The light--light imaginary gap can close on $\rho+k\ge\delta_*$ only at the mixed corner $(k,\rho)=(\infty,0)$, where the two light points tend in trace order to $P_\infty=(2,-1/2)$ and $P_0=(1,-1)$. Applying the common complex dilation $(U,V)\mapsto e^{-i\omega}(U,V)$ with $0<\omega\ll 1$ adds $(+\sin\omega,+\tfrac{1}{2}\sin\omega)>0$ to $\Im(Z_B-Z_A)$, opening a strictly positive light--light gap across a closed neighborhood of $(\infty,0)$ while preserving the fixed probe gaps. Just as in Appendix~\ref{app:AA-chart}, the ordered real homotopy $m_b$ from the Euclidean anchor $k=\rho=1$ and the short dilation continuation carry the physical $\mathrm{TOC}$ sheet into a compact subset of the open ordered forward tube.

\paragraph{Out-of-time-ordered outer bound.}
On the real $\mathrm{OTOC}$ contour $\Im\sigma=0$, the two light insertions lie on opposite real wedges at $E=-k\le 0$ and $P=\rho\in[0,1]$, while $\Phi_L$ and $\Phi_R$ already sit in the lower and upper imaginary half-planes in both null coordinates throughout the continuation $h\in[0,1]$. Fix any $K>1$ and introduce the orientation-preserving $\mathrm{PSL}(2,\mathbb R)$ map
\begin{equation}
 f(X)=C+\frac{X}{1-X/K}.
\end{equation}
Cover the outer region by two overlapping closed patches: $k\ge\delta_*/2$ (where we take $C>0$ sufficiently small) and $\rho\ge\delta_*/2$ (where we take $C<0$ sufficiently small in magnitude). On either patch, $f$ is finite on $[-\infty,1]$ (its pole at $X=K>1$ lies outside the light range) and separates the two light coordinates strictly across the origin, $f(E)<0<f(P)$, with uniform margins up to the compactified endpoints. At finite positive radii, the real $\mathrm{PSL}(2,\mathbb R)$ homotopy $f_b(X)=bC+X/(1-bX/K)$ ($0\le b\le 1$, $\det=+1$) preserves $f_b(E)<0<f_b(P)$, avoids all poles on the light paths, and keeps $\Phi_L,\Phi_R$ strictly in their respective imaginary half-planes. Consequently, the transformed real light points $\bigl(f(X),-1/f(X)\bigr)$ remain spacelike separated with fixed orientation ($U_A<0<U_B$, $V_B<0<V_A$). A small common imaginary boost $(U,V)\mapsto(e^{i\omega}U,e^{-i\omega}V)$ ($0<\omega\ll 1$) shifts $\Im U_A,\Im V_A<0<\Im U_B,\Im V_B$, opening both light--light imaginary gaps in the physical Wightman order $\Phi_L,\OO_A,\OO_B,\Phi_R$ (matching the spacelike boundary value) while preserving both probe gaps. Every patch thus maps onto a compact subset of the open ordered tube, and the tree-level forward-tube holomorphy input of Appendix~\ref{app:AA-chart} yields $|F_\times^{\rm word}|\le C_f G_N$ for both $\mathrm{TOC}$ and $\mathrm{OTOC}$ across the entire outer region.

\subsection{Complete \texorpdfstring{$\text{AdS}_3$}{AdS3} Mellin representation and convergence domain}
\label{app:cross-mellin}
To extract the exact $\mathrm{OTOC}$ monodromy in the small-cross-ratio region without dropping any double-trace family, we first record the complete $\text{AdS}_3$ tree-level graviton exchange Mellin representation~\cite{PenedonesMellin,FKPRvR,SpinningPropagators} and prove that both its regular and logarithmic double-trace sums converge normally (and hence define single-valued holomorphic functions) throughout the slit domain \eqref{eq:AA_normal_domain}.

\paragraph{Mellin amplitude and double-trace layers.}
Writing $\mathsf X=z_\times w_\times$, $\mathsf Y=(1-z_\times)(1-w_\times)$ from \eqref{eq:cross_zw} and $s_\Sigma=\Delta+d$, the complete tree-level graviton exchange Mellin amplitude---including both its pole series and regular contact polynomial---simplifies to
\begin{align}
 \mathcal M(s_M,t_M):=M/G_N
 &=-2\Delta d C_\Sigma
 -2\Delta d\,s_M(s_M+t_M)\mathcal S(t_M),\nonumber\\
 C_\Sigma&=\frac{\Gamma(s_\Sigma-1)}{\Gamma(\Delta)^2\Gamma(d)^2},
 \qquad
 \mathcal S(t_M)=\sum_{m\ge0}\frac{a_m}{t_M-2m},\nonumber\\
 a_m&=\frac1{m!(m+1)!\Gamma(\Delta-m)\Gamma(d-m)},
 \label{eq:AA_complete_Mellin}
\end{align}
with the crossed-channel inverse Mellin representation
\begin{align}
 F/G_N=\int\frac{\dd s_M\dd t_M}{(4\pi i)^2}\,
 &\mathcal M(s_M,t_M)\mathsf X^{t_M/2}\mathsf Y^{s_M/2}
 \Gamma(-s_M/2)^2\Gamma((s_M+t_M)/2)^2\nonumber\\
 &\times\Gamma(\Delta-t_M/2)\Gamma(d-t_M/2).
 \label{eq:AA_Mellin_inverse}
\end{align}
Closing the $s_M$ contour to the right across the double poles of $\Gamma(-s_M/2)^2$ at $s_M=2n$,
\begin{equation}
 \Gamma(-s_M/2)^2=
 \frac4{n!^2(s_M-2n)^2}
 -\frac{4\psi(n+1)}{n!^2(s_M-2n)}+O(1),
\end{equation}
where $\psi(z)=\Gamma'(z)/\Gamma(z)$ is the digamma function (distinct from the bulk wavefunction $\psi_f$), generates the crossed-channel double-trace expansion
\begin{equation}
 F/G_N=\mathcal A(\mathsf X,\mathsf Y)
       +\tfrac12\operatorname{Log}\mathsf Y\,
                              \mathcal B(\mathsf X,\mathsf Y),
 \qquad
 (\mathcal A,\mathcal B)=\sum_{n\ge0}\mathsf Y^n(A_n,b_n),
 \label{eq:AA_full_normal_sum}
\end{equation}
where $\operatorname{Log}\mathsf Y$ denotes the complex logarithm carried continuously along the analytic-continuation path (rather than a fixed principal branch). For $n\ge 1$ ($0<a_0<2n$), the logarithmic and regular coefficient functions are
\begin{equation}
 \begin{aligned}
  b_n(\mathsf X)=-\frac1{n!^2}
  \int_{\Re t_M=-a_0}\frac{\dd t_M}{2\pi i}\,
  &\mathcal M(2n,t_M)\Gamma(n+t_M/2)^2\\
  &\times\Gamma(\Delta-t_M/2)\Gamma(d-t_M/2)\mathsf X^{t_M/2},
 \end{aligned}
 \label{eq:AA_bn}
\end{equation}
and
\begin{align}
 A_n(\mathsf X)=-\frac1{n!^2}
 \int_{\Re t_M=-a_0}\frac{\dd t_M}{2\pi i}
 &\left[\partial_{s_M}\mathcal M(2n,t_M)
 +\mathcal M(2n,t_M)
 \{\psi(n+t_M/2)-\psi(n+1)\}\right]\nonumber\\
 &\times\Gamma(n+t_M/2)^2
 \Gamma(\Delta-t_M/2)\Gamma(d-t_M/2)\mathsf X^{t_M/2}.
 \label{eq:AA_An}
\end{align}

For the $n=0$ layer, the $m=0$ stress-tensor twist pole of $\mathcal S(t_M)$ at $t_M=0$ collides in $D=2$ with the left double pole of $\Gamma((s_M+t_M)/2)^2$ at $s_M=t_M=0$. Regulating the boundary half-dimension to $h_b=D/2=1+\epsilon_b$ in the $D$-dimensional graviton exchange Mellin amplitude~\cite{SpinningPropagators} separates the $m=0$ twist pole to $t_M=2(h_b-1)=2\epsilon_b$ (allowing the initial contour $-2\epsilon_b<\Re s_M<0$, $0<\Re t_M<2\epsilon_b$ with $\Re(s_M+t_M)>0$). Pushing the $t_M$-contour across $t_M=2\epsilon_b$ onto $\Re t_M=a_+\in(0,2\min(1,\Delta,d))$ extracts the $t_M=2\epsilon_b$ residue entering the regular layer $A_0(\mathsf X)$,
\begin{equation}
 -4\pi^{-\epsilon_b}\Delta d\,\Gamma(h_b)\mathsf X^{\epsilon_b}
 \left[\frac1{h_b\epsilon_b}
 +\frac{\psi(\epsilon_b)-\psi(1)}{2h_b-1}\right]
 \xrightarrow{\;\epsilon_b\to 0\;}-4\Delta d,
\end{equation}
whose $1/\epsilon_b$ poles cancel identically, while the corresponding $t_M=2\epsilon_b$ residue in the logarithmic layer likewise contributes the constant $-4\Delta d$ to $b_0(\mathsf X)$. Taking $\epsilon_b\to 0$ therefore yields the finite lowest-layer coefficients
\begin{align}
 b_0(\mathsf X)&=-4\Delta d+
 \frac{4\Delta d}{s_\Sigma-1}\mathsf X^d
 {}_2F_1(d,d;s_\Sigma;1-\mathsf X),\nonumber\\
 A_0(\mathsf X)&=-4\Delta d+
 2\Delta d\int_{\Re t_M=a_+}\frac{\dd t_M}{2\pi i}
 \left[t_M\mathcal S(t_M)
 +C_\Sigma\{\psi(t_M/2)-\psi(1)\}\right]\nonumber\\
 &\hspace{24mm}\times\Gamma(t_M/2)^2
 \Gamma(\Delta-t_M/2)\Gamma(d-t_M/2)\mathsf X^{t_M/2}.
 \label{eq:AA_lowest_layers}
\end{align}

\paragraph{Normal-convergence radius via Euler integrals.}
To bound the growth of $b_n(\mathsf X)$ and $A_n(\mathsf X)$ uniformly in $n$, write the contact weight $J_n(z)$ as
\begin{align}
 J_n(z)&=\int\frac{\dd t_M}{2\pi i}\Gamma(n+t_M/2)^2
 \Gamma(\Delta-t_M/2)\Gamma(d-t_M/2)z^{t_M/2}\nonumber\\
 &=\frac{2\Gamma(n+\Delta)^2\Gamma(n+d)^2}{\Gamma(s_\Sigma+2n)}
 z^d\,{}_2F_1(n+d,n+d;s_\Sigma+2n;1-z),
\end{align}
which admits the Euler $v$-integral representation given by \eqref{eq:AA_shifted_J} below at $\alpha=0$. On the slit plane $z=w^2\notin(-\infty,0]$ ($\Re w>0$), rotating the integration variable $v\mapsto v/w$ transforms the $n$-th power kernel $[v/((1+v)(1+zv))]^n$ into $[v/((w+v)(1+wv))]^n$; since
\begin{equation}
 \frac{(w+v)(1+wv)}v=(1+w)^2+w(v+v^{-1}-2),\qquad
 \Re[(1+w)^2\bar w]>0,
\end{equation}
and $v+v^{-1}-2\ge 0$ for all $v>0$, taking the squared modulus gives $|(w+v)(1+wv)/v|^2\ge |1+w|^4$, bounding the $n$-th power kernel uniformly by $|1+w|^{-2n}$. Expressing $(t_M-2m)^{-1}=-\frac{1}{2}\int_0^1 y^{m-t_M/2-1}\dd y$ (convergent on the contour $\Re t_M=-a_0<0$) to sum $\mathcal S(t_M)$ into ${}_2F_1(1-\Delta,1-d;2;y)$, we obtain the integral representation
\begin{align}
 b_n(\mathsf X)={}&\frac{2\Delta d C_\Sigma}{n!^2}J_n(\mathsf X)
 \nonumber\\
 &-\frac{4\Delta d\,n}{n!^2\Gamma(\Delta)\Gamma(d)}
 \int_0^1\frac{\dd y}{y}\,{}_2F_1(1-\Delta,1-d;2;y)
 (n+D_z)J_n(z)\big|_{z=\mathsf X/y},
 \label{eq:AA_bn_Euler}
\end{align}
where $D_z=z\partial_z$.

To bound the regular coefficients $A_n(\mathsf X)$ in \eqref{eq:AA_An} with the same exponential base $R=|1+\sqrt{\mathsf X}|^2$, note that the first term $\partial_{s_M}\mathcal M(2n,t_M)=-4\Delta d(2n+t_M/2)\mathcal S(t_M)$ yields the identical $y$-integral as \eqref{eq:AA_bn_Euler} with $n(n+D_z)J_n$ replaced by $(2n+D_z)J_n$. For the digamma term $\psi(n+t_M/2)-\psi(n+1)$, we introduce the shifted contact weight
\begin{align}
 J_n^{(\alpha)}(z)
 &=\int\frac{\dd t_M}{2\pi i}\Gamma(n+\alpha+t_M/2)\Gamma(n+t_M/2)
   \Gamma(\Delta-t_M/2)\Gamma(d-t_M/2)z^{t_M/2}\nonumber\\
 &=2\Gamma(n+\Delta+\alpha)\Gamma(n+d)z^d
   \int_0^\infty\dd v\,v^{n+d+\alpha-1}(1+v)^{-(n+\Delta+\alpha)}(1+zv)^{-(n+d)},
 \label{eq:AA_shifted_J}
\end{align}
whose $\alpha$-derivative at $\alpha=0$ produces the factor $\psi(n+t_M/2)$ inside the Barnes integral:
\begin{align}
 \partial_\alpha J_n^{(\alpha)}(z)\big|_{\alpha=0}
 &=\psi(n+\Delta)J_n(z)+2\Gamma(n+\Delta)\Gamma(n+d)z^d\nonumber\\
 &\quad\times\int_0^\infty\dd v\,\log\!\left(\frac{v}{1+v}\right)
   v^{n+d-1}(1+v)^{-(n+\Delta)}(1+zv)^{-(n+d)}.
 \label{eq:AA_dalpha_J}
\end{align}
Since $|\psi(n+\Delta)-\psi(n+1)|\le C$, the logarithm $\log[v/(1+v)]$ is integrable at $v=0$ (for $d>2$) and vanishes as $v\to\infty$, and the rotated $n$-th power kernel in \eqref{eq:AA_dalpha_J} is identical to that of $J_n(z)$. Moreover, at $z=\mathsf X/y$ ($y\in(0,1]$), $\Re(\sqrt{y}+w)>\Re w>0$ implies $y/|\sqrt{y}+w|^2\le|1+w|^{-2}=R^{-1}$ uniformly for all $y\in(0,1]$. Therefore, on every compact subset $\mathcal K$ of the slit plane,
\begin{equation}
 |D_{\mathsf X}^k b_n|+|D_{\mathsf X}^k A_n|
 \le C_{\mathcal K,k}(n+1)^{p_k}R^{-n},
 \label{eq:AA_normal_coefficient_bound}
\end{equation}
with $D_{\mathsf X}=\mathsf X\partial_{\mathsf X}$, so that the double-trace sums \eqref{eq:AA_full_normal_sum} converge normally throughout the domain
\begin{equation}
 \mathsf X\notin(-\infty,0],\qquad
 |\mathsf Y|<|1+\sqrt{\mathsf X}|^2,\qquad
 \Re\sqrt{\mathsf X}>0.
 \label{eq:AA_normal_domain}
\end{equation}

\subsection{Small-cross-ratio region \texorpdfstring{$(1+|a|)\rho<\delta_*$}{(1+|a|) rho < delta*} and exact Regge match}
\label{app:cross-small}
We now evaluate the cross-channel integrand uniformly for $\ell:=(1+|a|)\rho<\delta_*$ in three steps: (i)~using the single-valued convergence domain \eqref{eq:AA_normal_domain}, we show that $\mathrm{TOC}$ is $O_f(G_N\ell^2)$ while $\mathrm{OTOC}$ winds once around $\mathsf Y=0$, yielding $\Delta F/G_N=i\pi\mathcal B(\mathsf X,\mathsf Y)$; (ii)~we isolate the leading $O(\ell^{-1})=O(e^\beta)$ Barnes integral from $\mathcal B$ and shift the subtracted contour from $\Re u=-1/4$ to $\Re u=+1/4$ to bound the remainder uniformly by $O_f(1)$; and (iii)~we match the resulting Barnes coefficient analytically against $Q^\times[f]$ via Fourier inversion so that the growing $e^\beta$ terms cancel identically.

\paragraph{The logarithm on the ordered contours.}
Set $\mathsf X=z_\times w_\times$, $\mathsf Y=(1-z_\times)(1-w_\times)$, $h_Y=1-\mathsf Y$, and $\ell:=(1+|a|)\rho$ (local shorthand distinct from the left Rindler boost $\ell$ of Section~\ref{sec:dictionary}). On the real $\mathrm{OTOC}$ line, abbreviate the positive preparation scales by $A:=\bar w+w=2s\sin\delta$ and $B:=\bar b+b=2\sin\delta/s$ (distinct from the regions $A,B$). At fixed preparation $(s,\delta)$, expanding \eqref{eq:cross_zw} uniformly for $\ell<\delta_*$ yields
\begin{equation}
 z_\times=iB\ell[1+O_f(\ell)],\qquad
 w_\times=iA\ell[1+O_f(\ell)],\qquad
 h_Y=i(A+B)\ell[1+O_f(\ell)].
 \label{eq:cross_small_zw}
\end{equation}
On the $\mathrm{TOC}$ contour, the adjacent light pair in the ordered chart of Appendix~\ref{app:cross-outer} has complex separation $O(\ell)$, an imaginary gap bounded below by $c_\gamma\ell$ in both null components, and fixed $O(1)$ gaps to the two probes. Its direct light-pair OPE is therefore uniform in $k/(\rho+k)$: the stress tensor starts at dimension two and the double-trace families at dimensions strictly above two, just as in Appendix~\ref{app:AA-chart}, giving $F_\times^{\rm TOC}=O_f(G_N\ell^2)$.

For the $\mathrm{OTOC}$, we fix the analytic sheet at the symmetric point $s=1$, $a=1$ by tracking $\rho$ continuously from $1$ down to $0$. At the Euclidean anchor $\rho=1$ (where $v_i=-u_i^*$ for all four points), we have $z_\times=w_\times=2$, $\mathsf X=4$, $\mathsf Y=1$, and $\operatorname{Log}\mathsf Y=0$. Along $\rho\in(0,1]$, \eqref{eq:cross_zw} simplifies to
\begin{equation}
 1-z_\times=1-w_\times=
 \frac{1-\rho^2-2i\rho\sin\delta}{1-\rho^2+2i\rho\sin\delta}.
\end{equation}
As $\rho$ decreases from $1$ to $0$, each factor moves counterclockwise along the unit circle from argument $-\pi$ to $0$, so their product $\mathsf Y=(1-z_\times)(1-w_\times)$ acquires $+2\pi i$ in its continuously carried logarithm. Throughout the path, $\Re\sqrt{\mathsf X}=\Re z_\times>0$ and $|\mathsf Y|=1<|1+z_\times|^2=|1+\sqrt{\mathsf X}|^2$, keeping the Mellin expansion \eqref{eq:AA_full_normal_sum} inside its normal-convergence domain \eqref{eq:AA_normal_domain} on its continued branch; small independent Euclidean perturbations of all four points at $\rho=1$ fix the same analytic germ in both cross ratios. Deforming the positive ratio $k/\rho$ and the preparation $(s,\delta)$ along a compact path inside the small-cross-ratio region keeps $z_\times,w_\times$ inside small punctured disks and fixes the lifted phases $\arg\mathsf X=\pi+O_f(\ell)$ and $\arg(\mathsf X/h_Y^2)=O_f(\ell)$ via \eqref{eq:cross_small_zw}. The explicit Barnes continuation below therefore transports the same $+2\pi i$ winding to every fixed preparation. Writing $\log_{\rm loc}$ for the local branch with $\log_{\rm loc}(1)=0$, we obtain
\begin{equation}
 \operatorname{Log}_{\rm OTOC}\mathsf Y=2\pi i+\log_{\rm loc}\mathsf Y,\qquad
 -\log_{\rm loc}\mathsf Y=h_Y+O(h_Y^2).
 \label{eq:cross_phase_margins}
\end{equation}
By \eqref{eq:AA_full_normal_sum}, the difference between the $\mathrm{OTOC}$ graph and the direct-sheet connected graph is precisely $\Delta F/G_N=i\pi\mathcal B(\mathsf X,\mathsf Y)$, while the direct-sheet term itself is $O_f(G_N\ell^2)$ by the light-pair OPE.

\paragraph{Subtracting the growing term.}
Abbreviating $X=\mathsf X$, $Y=\mathsf Y$, $D=Y\partial_Y$, and $\int_{(b)}=\int_{\Re u=b}\frac{\dd u}{2\pi i}$ (oriented upwards), we insert \eqref{eq:AA_complete_Mellin} into \eqref{eq:AA_bn}, change variables to $t_M=2u$, and sum over $n\ge 1$ inside the Euclidean convergence domain (where $D$ multiplies $Y^n$ by $n$):
\begin{align}
 \mathcal B(X,Y)
 &=b_0(X)+4\Delta d\int_{(-1/4)} M(u)X^u
   \bigl[C_\Sigma+4\mathcal S(2u)D(D+u)\bigr]K(u,Y),
 \label{eq:cross_summed_log}\\
 M(u)&=\Gamma(\Delta-u)\Gamma(d-u),\qquad
 K(u,Y)=\Gamma(u)^2\bigl[{}_2F_1(u,u;1;Y)-1\bigr].\nonumber
\end{align}
The subtracted constant in $K(u,Y)$ omits the $n=0$ layer, whose dimensionally regulated contribution $b_0(X)$ is given in \eqref{eq:AA_lowest_layers}. Near $Y=1$ ($h_Y=1-Y\to 0$), Gauss' connection formula on the local branch reads
\begin{align}
 K(u,Y)+\Gamma(u)^2={}&
 \frac{\Gamma(u)^2\Gamma(1-2u)}{\Gamma(1-u)^2}
 {}_2F_1(u,u;2u;h_Y) \nonumber\\
 &+\Gamma(2u-1)h_Y^{1-2u}
 {}_2F_1(1-u,1-u;2-2u;h_Y).
 \label{eq:cross_gauss}
\end{align}
Acting with $D(D+u)$ on the leading power of the second term produces $2u(2u-1)\Gamma(2u-1)h_Y^{-1-2u}=\Gamma(1+2u)h_Y^{-1-2u}$. We isolate this leading $O(h_Y^{-1})$ contribution by defining the Barnes integral $I(v)$ and the subtracted kernel $\mathcal T(u,Y)$,
\begin{align}
 I(v)&=\int_{(-1/4)} M(u)\Gamma(1+2u)\mathcal S(2u)v^u,\qquad v=X/h_Y^2,
 \label{eq:cross_Barnes_I}\\
 \mathcal T(u,Y)&=D(D+u)K(u,Y)-\Gamma(1+2u)h_Y^{-1-2u}.
 \label{eq:cross_subtracted_kernel}
\end{align}
Then $\mathcal B(X,Y)-16\Delta d\,I(v)/h_Y$ equals $b_0(X)$ plus the integral of $4\Delta d\,M(u)X^u\bigl[C_\Sigma K(u,Y)+4\mathcal S(2u)\mathcal T(u,Y)\bigr]$ along $\Re u=-1/4$. We now shift this subtracted contour to $\Re u=+1/4$. At fixed $Y$, the defining power series $K(u,Y)=\sum_{n\ge 1}\Gamma(n+u)^2 Y^n/n!^2$ is holomorphic across $u=0$ and continues analytically to $K(0,Y)=\operatorname{Li}_2(Y)$, giving
\begin{equation}
 \mathcal T(0,Y)=D^2\operatorname{Li}_2(Y)-h_Y^{-1}
       =\frac{Y}{1-Y}-\frac{1}{1-Y}=-1.
\end{equation}
Within the strip $-1/4\le\Re u\le 1/4$, only $\mathcal S(2u)$ has a pole, located at $u=0$ with residue $\operatorname{Res}_{u=0}\mathcal S(2u)=[2\Gamma(\Delta)\Gamma(d)]^{-1}$. The residue of $M(u)X^u[C_\Sigma K+4\mathcal S\mathcal T]$ at $u=0$ is therefore the $Y$-independent constant $-2$, and Cauchy's theorem equates the $\Re u=-1/4$ integral to the $\Re u=+1/4$ integral minus this residue.

To bound the shifted integral on $\Re u=+1/4$ uniformly as $\ell\to 0$, let $t=\Im u$. By Stirling's formula, $M(u)$ together with the connection prefactors in \eqref{eq:cross_gauss} (as well as $-\Gamma(u)^2$) supplies an overall exponential decay $e^{-2\pi|t|}$ up to powers of $1+|t|$. The phase factors of the regular and singular branches multiply this by $e^{-t\arg X}$ and $e^{-t\arg v}$, respectively; since $\arg X=\pi+O_f(\ell)$ and $\arg v=O_f(\ell)$, both directions retain a strict exponential damping margin (at least $e^{-(\pi-o(1))|t|}$). In the Gauss hypergeometric series of \eqref{eq:cross_gauss}, on $\Re u=1/4$ the Pochhammer ratios $(u)_j/(2u)_j$ and $(1-u)_j/(2-2u)_j$ have modulus at most unity for every $j\ge 0$ (where $(u)_j=\Gamma(u+j)/\Gamma(u)$), so summing the remaining factor $|(u)_j||h_Y|^j/j!$ (and up to two $D$-derivatives) is bounded by a polynomial in $1+|t|$ times $(1-|h_Y|)^{-|u|}\le e^{C|h_Y|(1+|t|)}$; taking $\delta_*$ sufficiently small absorbs this exponential completely into the Stirling margin. The same large-$|t|$ bounds hold throughout the shift strip $-1/4\le\Re u\le 1/4$ (where $|(u)_j/(2u)_j|$ grows at most polynomially in $j$), justifying both the contour deformation and the term-by-term estimates on $\Re u=+1/4$: the analytic terms in $h_Y$ (including $-\Gamma(u)^2$) integrate to $O_f(|X|^{1/4})$, while the subtracted singular branch left in $\mathcal T(u,Y)$ begins at $h_Y^{-2u}$ and integrates to $O_f(|X|^{1/4}|h_Y|^{-1/2})$. Finally, \eqref{eq:AA_lowest_layers} gives $b_0(X)=-4\Delta d+O_f(|X|^\Delta|\log|X||)=O_f(1)$ on this sheet. Because $|X|\asymp_f |h_Y|^2$, we obtain the uniform small-cross-ratio expansion
\begin{equation}
 \mathcal B(X,Y)=\frac{16\Delta d}{h_Y}I(X/h_Y^2)+O_f(1),
 \label{eq:cross_strong_B}
\end{equation}
which automatically absorbs the regular Mellin contact polynomial into the $O_f(1)$ remainder. Moreover, $I(v)$ is holomorphic near $v_0=AB/(A+B)^2$, so replacing $v=v_0+O_f(\ell)$ and $h_Y=i(A+B)\ell[1+O_f(\ell)]$ via \eqref{eq:cross_small_zw} shifts $\Delta F/G_N=i\pi\mathcal B$ by $O_f(1)$, isolating the leading $\mathrm{OTOC}$ growing piece:
\begin{equation}
 \frac{\Delta F^{\rm lead}}{G_N}
 =\frac{16\pi\Delta d}{(1+a)\rho(A+B)}
 I\!\left(\frac{AB}{(A+B)^2}\right).
 \label{eq:cross_exact_eikonal}
\end{equation}

\paragraph{Analytic Fourier--Barnes match to the radial QES kernel.}
In the cross commutator \eqref{eq:cross_commutator_words}, the $\mathrm{OTOC}$ enters with a minus sign. Setting $m=2\Delta+1$ and $J_m=B(m,m+1)$, the modular measure \eqref{eq:cross_weight} takes the form $\dd\mu=a^{m-1}(1+a)^{-2m}\dd a$, with $\int_0^\infty\dd\mu/(1+a)=J_m$. Integrating $-\Delta F^{\rm lead}/G_N$ in \eqref{eq:cross_exact_eikonal} against $\dd\mu$ therefore yields the $e^\beta$-coefficient
\begin{equation}
 Q_M(A,B)=-\frac{16\pi\Delta d\,J_m}{A+B}
 I\!\left(\frac{AB}{(A+B)^2}\right).
\end{equation}
To match $Q_M(A,B)$ analytically against the independently computed radial QES overlap $Q^\times[f]$ at all probe scales $s>0$, we pass to the logarithmic scale variable $y$ at fixed $AB$ and take the Fourier transform of $\sqrt{AB}\,Q_M$ with the convention
\begin{equation}
 \widehat g(\nu)=\int_{\mathbb R}e^{i\nu y}g(y)\,\dd y,
 \qquad
 y=\frac12\log\frac{A}{B}=\log s,
 \qquad
 AB=4\sin^2\delta.
 \label{eq:cross_fourier_conv}
\end{equation}
Writing $b_\pm=(1\pm i\nu)/2$, noting that $\sqrt{AB}/(A+B)=\tfrac{1}{2}\operatorname{sech}y$ and $v_0=AB/(A+B)^2=\tfrac{1}{4}\operatorname{sech}^2y$, and applying the beta-integral transform
\begin{equation}
 \widehat{\operatorname{sech}^{2j+1}}(\nu)
 =\frac{2^{2j}\Gamma(j+b_+)\Gamma(j+b_-)}{\Gamma(2j+1)}
\end{equation}
with $j=u$ cancels $2^{-2u}\Gamma(1+2u)$ identically and produces
\begin{equation}
 \widehat{\sqrt{AB}\,Q_M}(\nu)
 =-8\pi\Delta d\,J_m\int_{(-1/4)}P(u)\mathcal S(2u),
 \quad P(u):=M(u)\Gamma(u+b_+)\Gamma(u+b_-).
\end{equation}
The Mellin pole residues in \eqref{eq:AA_complete_Mellin} satisfy the exact recurrence $j(j+1)a_j=(\Delta-j)(d-j)a_{j-1}$ for all $j\ge 1$. Substituting this into the partial-fraction expansion of $\mathcal S(2u)$ eliminates all poles simultaneously:
\begin{equation}
 (\Delta-u)(d-u)\mathcal S(2u-2)-u(u+1)\mathcal S(2u)
 =-\frac{\Delta+d}{2}\sum_{j\ge 0}a_j.
\end{equation}
We now multiply $\widehat{\sqrt{AB}\,Q_M}(\nu)$ by $b_+b_-=(1+\nu^2)/4$ and insert the identity $b_+b_-=(u+b_+)(u+b_-)-u(u+1)$ under the integral sign. In the first term, $(u+b_+)(u+b_-)\Gamma(u+b_+)\Gamma(u+b_-)=\Gamma(u+1+b_+)\Gamma(u+1+b_-)$; changing variables to $u'=u+1$ (on $\Re u'=3/4$) yields $M(u'-1)=(\Delta-u')(d-u')M(u')$, and shifting $\Re u'=3/4$ back to $\Re u'=-1/4$ crosses no poles (since $\mathcal S(2u'-2)$ has its leftmost pole at $u'=1$). Combining the two terms invokes the recurrence above and removes the entire pole series $\mathcal S$ in one stroke. Evaluating $\sum_{j\ge 0}a_j=\Gamma(\Delta+d)/[\Gamma(\Delta)\Gamma(d)\Gamma(1+\Delta)\Gamma(1+d)]$ by Gauss' ${}_2F_1(1)$ summation theorem and $\int_{(-1/4)}P(u)=\Gamma(\Delta+b_+)\Gamma(\Delta+b_-)\Gamma(d+b_+)\Gamma(d+b_-)/\Gamma(\Delta+d+1)$ by Barnes' first lemma yields
\begin{equation}
 \widehat{\sqrt{AB}\,Q_M}(\nu)=
 \frac{16\pi J_m}{\Gamma(\Delta)^2\Gamma(d)^2}
 \frac{\Gamma(\Delta+b_+)\Gamma(\Delta+b_-)
       \Gamma(d+b_+)\Gamma(d+b_-)}{1+\nu^2}.
 \label{eq:cross_barnes_geometry}
\end{equation}

On the independent generalized-entropy side, changing variables to $x=\log z$ and $H(x)=e^{-x}\eta_*(e^x)$ casts the QES shape equation \eqref{eq:QES} (with \eqref{eq:matching}) and the cross readout \eqref{eq:Qcross} (with \eqref{eq:flux}) into the exact convolution form
\begin{align}
 (1-\partial_x^2)H(x)&=
 \frac{4\mathcal N_\Delta}{2^{2\Delta+1}}
 \operatorname{sech}^{2\Delta+1}x, \nonumber\\
 \sqrt{AB}\,Q^\times[f]&=
 \frac{4\pi d}{2^{2d+1}B(d,d+1)}
 \int_{\mathbb R}\operatorname{sech}^{2d+1}(x-y)H(x)\,\dd x.
\end{align}
Because the anchored Green solution decays exponentially at both ends by \eqref{eq:splitQES}, taking the Fourier transform divides by $1+\nu^2$ and multiplies the two $\operatorname{sech}$-power transforms; inserting $\mathcal N_\Delta=4J_m\Gamma(2\Delta+1)/\Gamma(\Delta)^2$ from \eqref{eq:N} reproduces \eqref{eq:cross_barnes_geometry} identically. Fourier inversion therefore establishes $Q_M(A,B)=Q^\times[f]$ for every scale $s>0$, fixing both the sign and the absolute normalization non-circularly from the two independent integrals.

Returning to the original modular variable $\sigma$ (where the horizon source in \eqref{eq:lightnull} is $J_{-\sigma}$), define $q_{\times,f}(\sigma):=16\pi\int_0^\infty\!\dd z\,\dd z'\,t_f(z)G_L(z,z')J_{-\sigma}(z')/(iz')$. From the $\sigma$-dependence of \eqref{eq:lightnull} and $\langle\OO_A(\sigma)\OO_B\rangle=[2\cosh(\sigma/2)]^{-2\Delta}$, we have $q_{\times,f}(\sigma)/\langle\OO_A(\sigma)\OO_B\rangle=Q^\times[f]/[J_m(1+a)]$. Substituting this identity into \eqref{eq:cross_exact_eikonal} and \eqref{eq:cross_strong_B} yields the uniform small-cross-ratio remainder bound
\begin{equation}
 \left|-\frac{F_\times^{\rm OTOC}(\sigma,\beta)}{G_N}
 -e^\beta\frac{q_{\times,f}(\sigma)}{\langle\OO_A(\sigma)\OO_B\rangle}\right|\le C_f,
 \qquad (1+a)\rho<\delta_*.
 \label{eq:cross_Regge_uniform}
\end{equation}

\subsection{Dominated convergence and remainder decay}
\label{app:cross-dom}
Combining the outer bound $|F_\times^{\rm word}|\le C G_N$ of Appendix~\ref{app:cross-outer} on $(1+|a|)\rho\ge\delta_*$ with the small-cross-ratio remainder bound \eqref{eq:cross_Regge_uniform} of Appendix~\ref{app:cross-small} on $(1+|a|)\rho<\delta_*$ gives the global pointwise bound
\begin{equation}
 e^{-\beta}\bigl(|F_\times^{\rm TOC}(\sigma,\beta)|+|F_\times^{\rm OTOC}(\sigma,\beta)|\bigr)
 \le C G_N\left(e^{-\beta}+\frac{1}{1+|e^\sigma|}\right)
 \le 2C G_N,
\end{equation}
so $2C|\mathcal W_\times(\sigma)|\in L^1(\mathbb R)$ serves as a $\beta$-independent dominating function for $e^{-\beta}p_\times(\beta)$. Splitting the real $\mathrm{OTOC}$ integral at $\sigma_*(\beta)$ defined by $(1+e^{\sigma_*})e^{-\beta}=\delta_*$ (for $\beta$ sufficiently large that $\delta_* e^\beta>2$; for bounded $\beta$, both $p_\times(\beta)$ and $e^\beta Q^\times[f]$ are continuous and bounded):
\begin{itemize}
 \item The entire $\mathrm{TOC}$ contribution satisfies $|p_\times^{\rm TOC}(\beta)|\le C\int_{\mathbb R}|\mathcal W_\times(\sigma)|\,\dd\Re\sigma\le C_1$.
 \item On $\Re\sigma<\sigma_*(\beta)$, \eqref{eq:cross_Regge_uniform} bounds the $\mathrm{OTOC}$ remainder by $C\int_{\mathbb R}|\mathcal W_\times(\sigma)|\,\dd\Re\sigma\le C_2$.
 \item On $\Re\sigma\ge\sigma_*(\beta)$, the actual $\mathrm{OTOC}$ tail is bounded by $C\int_{\sigma_*(\beta)}^\infty e^{-(2\Delta+1)\Re\sigma}\,\dd\Re\sigma=O(e^{-(2\Delta+1)\beta})$, while the subtracted eikonal tail satisfies
 \[
  e^\beta\int_{\sigma_*(\beta)}^\infty
  \frac{|q_{\times,f}(\sigma)|}{[2\cosh(\sigma/2)]^{2\Delta+2}}\,\dd\Re\sigma
  =e^\beta\,O\!\left(e^{-(2\Delta+2)\beta}\right)
  =O\!\left(e^{-(2\Delta+1)\beta}\right).
 \]
\end{itemize}
The full real-axis integral of the leading term is $e^\beta Q^\times[f]$ by \eqref{eq:cross_barnes_geometry}. Therefore $|r_\times(\beta)|=|p_\times(\beta)-e^\beta Q^\times[f]|\le C_\times$ uniformly for all $\beta\ge 0$, which yields $h_\times(\beta)=e^{-\beta}|r_\times(\beta)|\le C_\times e^{-\beta}$ and completes the derivation of \eqref{eq:crossRemainderNorm}. (The off-diagonal generalization to finite superpositions is proved in Paper II~\cite{CompanionTomography}.)

\section{A stress-family check}
\label{app:AA}

The complete $AA$ calculation can be checked against the two chiral global stress families. With $\langle T(u)T(0)\rangle=c/(2u^4)$, their light-pair OPE block is
\begin{align}
 [\OO_1\OO_2]_{[T]}&=\langle\OO_1\OO_2\rangle\frac{6\Delta}{c}
   [\mathcal I_T(u_1,u_2)+\mathcal I_{\bar T}(v_1,v_2)],\\
 \mathcal I_T(x_1,x_2)&=(x_1-x_2)^2
       \int_0^1 t(1-t)T(x_2+t(x_1-x_2))\dd t.
 \label{eq:stressline}
\end{align}

Here $\mathcal I_T$ smears the stress tensor along the segment between its two arguments, and $\mathcal I_{\bar T}$ is its antiholomorphic counterpart. Inserting this block into \eqref{eq:AAkernel} and writing $\ell=(1-t)a+tg$, $r=g/a$ reduces the modular integral to
\begin{equation}
 \delta K_{AA}^{[T]}=-\frac{\lambda^{2\Delta}}c
       \int_0^\infty W_\Delta(\ell)[T(-\ell)+\bar T(\ell)]\dd\ell,
 \label{eq:AAweightoperator}
\end{equation}
\begin{equation}
 W_\Delta(\ell)=-\frac{3i\Delta e^{2\pi i\Delta}}\pi\ell^2
 \int_{\Gamma_r}\!\dd r\,r^{2\Delta}(r-1)^{1-4\Delta}
 \int_0^1\frac{t(1-t)\dd t}{D_t(D_t+\ell)(D_t+\ell r)},
 \label{eq:weighttwo}
\end{equation}
where $D_t=1-t+tr$ and $\Gamma_r=e^{i\theta}\mathbb R_+$ with $0<\theta<\pi/2$ preserves the light-coincidence bypass. At small $\ell$, the $t$ integral is $[(r-r^{-1})/2-\log r]/(r-1)^3$; the resulting beta integral gives $\kappa_\Delta=3\Delta B(2\Delta+1,2\Delta+1)$. The substitution $t'=(1-t)/(1-t+tr)$ gives $W_\Delta(\ell)=\ell^2W_\Delta(1/\ell)$, so $W_\Delta(\ell)\sim\kappa_\Delta\ell^2$ at zero and $W_\Delta(\ell)\to\kappa_\Delta$ at infinity. At $\Delta=3/2$, residues give the explicit check
\begin{equation}
 W_{3/2}(\ell)=\frac9{280}
 \frac{\ell^2(\ell^4+6\ell^3+15\ell^2+6\ell+1)}{(1+\ell)^6},
 \qquad \kappa_{3/2}=\frac9{280}.
 \label{eq:AAhalf}
\end{equation}

For a diagonal Euclidean preparation, the two chiral Ward identities then give
\begin{equation}
 \PP_{AA}^{[T]}[f;\beta]
 =\frac{\pi d\kappa_\Delta\lambda^{2\Delta}}{c\tau}
 (1+s^2)e^{-\beta}+\frac{\lambda^{2\Delta}}c\,o(e^{-\beta}).
 \label{eq:AAleading}
\end{equation}

The full graph estimate of Appendix~\ref{app:AAfull} includes the remaining conformal families and establishes the single-insertion bound \eqref{eq:AAcompletebound} used in the main text.

\bibliographystyle{JHEP}
\bibliography{ref}

\end{document}

%% file: modular_tomography_preamble.tex
\usepackage{jheppub}
\usepackage[T1]{fontenc}
\usepackage{lmodern,amsthm,mathtools,bm,mathrsfs,booktabs,tabularx,array,microtype}
\usepackage{enumitem,needspace,xcolor,placeins}

\setlist{nosep,leftmargin=1.8em}
\renewcommand{\Re}{\operatorname{Re}}
\renewcommand{\Im}{\operatorname{Im}}
\newcolumntype{Y}{>{\raggedright\arraybackslash}X}
\numberwithin{equation}{section}

\newcommand{\dd}{\mathrm d}
\newcommand{\GN}{G_N}
\newcommand{\OO}{\mathcal O}
\newcommand{\PP}{\mathcal P}

%% file: modular_tomography_I.bbl
\providecommand{\href}[2]{#2}\begingroup\raggedright\begin{thebibliography}{10}

\bibitem{DensityMatrixDual}
B.~Czech, J.~L. Karczmarek, F.~Nogueira and M.~Van~Raamsdonk, \emph{{The
  Gravity Dual of a Density Matrix}},
  \href{https://doi.org/10.1088/0264-9381/29/15/155009}{\emph{Class. Quant.
  Grav.} {\bfseries 29} (2012) 155009}
  [\href{https://arxiv.org/abs/1204.1330}{{\ttfamily 1204.1330}}].

\bibitem{CausalInfo}
V.~E. Hubeny and M.~Rangamani, \emph{{Causal Holographic Information}},
  \href{https://doi.org/10.1007/JHEP06(2012)114}{\emph{JHEP} {\bfseries 06}
  (2012) 114} [\href{https://arxiv.org/abs/1204.1698}{{\ttfamily 1204.1698}}].

\bibitem{RT}
S.~Ryu and T.~Takayanagi, \emph{{Holographic derivation of entanglement entropy
  from AdS/CFT}},
  \href{https://doi.org/10.1103/PhysRevLett.96.181602}{\emph{Phys. Rev. Lett.}
  {\bfseries 96} (2006) 181602}
  [\href{https://arxiv.org/abs/hep-th/0603001}{{\ttfamily hep-th/0603001}}].

\bibitem{HRT}
V.~E. Hubeny, M.~Rangamani and T.~Takayanagi, \emph{{A Covariant holographic
  entanglement entropy proposal}},
  \href{https://doi.org/10.1088/1126-6708/2007/07/062}{\emph{JHEP} {\bfseries
  07} (2007) 062} [\href{https://arxiv.org/abs/0705.0016}{{\ttfamily
  0705.0016}}].

\bibitem{FLM}
T.~Faulkner, A.~Lewkowycz and J.~Maldacena, \emph{{Quantum corrections to
  holographic entanglement entropy}},
  \href{https://doi.org/10.1007/JHEP11(2013)074}{\emph{JHEP} {\bfseries 11}
  (2013) 074} [\href{https://arxiv.org/abs/1307.2892}{{\ttfamily 1307.2892}}].

\bibitem{EW}
N.~Engelhardt and A.~C. Wall, \emph{{Quantum Extremal Surfaces: Holographic
  Entanglement Entropy beyond the Classical Regime}},
  \href{https://doi.org/10.1007/JHEP01(2015)073}{\emph{JHEP} {\bfseries 01}
  (2015) 073} [\href{https://arxiv.org/abs/1408.3203}{{\ttfamily 1408.3203}}].

\bibitem{HKLL}
A.~Hamilton, D.~N. Kabat, G.~Lifschytz and D.~A. Lowe, \emph{{Holographic
  representation of local bulk operators}},
  \href{https://doi.org/10.1103/PhysRevD.74.066009}{\emph{Phys. Rev. D}
  {\bfseries 74} (2006) 066009}
  [\href{https://arxiv.org/abs/hep-th/0606141}{{\ttfamily hep-th/0606141}}].

\bibitem{ADH}
A.~Almheiri, X.~Dong and D.~Harlow, \emph{{Bulk Locality and Quantum Error
  Correction in AdS/CFT}},
  \href{https://doi.org/10.1007/JHEP04(2015)163}{\emph{JHEP} {\bfseries 04}
  (2015) 163} [\href{https://arxiv.org/abs/1411.7041}{{\ttfamily 1411.7041}}].

\bibitem{JLMS}
D.~L. Jafferis, A.~Lewkowycz, J.~Maldacena and S.~J. Suh, \emph{{Relative
  entropy equals bulk relative entropy}},
  \href{https://doi.org/10.1007/JHEP06(2016)004}{\emph{JHEP} {\bfseries 06}
  (2016) 004} [\href{https://arxiv.org/abs/1512.06431}{{\ttfamily
  1512.06431}}].

\bibitem{DHW}
X.~Dong, D.~Harlow and A.~C. Wall, \emph{{Reconstruction of Bulk Operators
  within the Entanglement Wedge in Gauge-Gravity Duality}},
  \href{https://doi.org/10.1103/PhysRevLett.117.021601}{\emph{Phys. Rev. Lett.}
  {\bfseries 117} (2016) 021601}
  [\href{https://arxiv.org/abs/1601.05416}{{\ttfamily 1601.05416}}].

\bibitem{GaoWald}
S.~Gao and R.~M. Wald, \emph{{Theorems on gravitational time delay and related
  issues}}, \href{https://doi.org/10.1088/0264-9381/17/24/305}{\emph{Class.
  Quant. Grav.} {\bfseries 17} (2000) 4999}
  [\href{https://arxiv.org/abs/gr-qc/0007021}{{\ttfamily gr-qc/0007021}}].

\bibitem{Maximin}
A.~C. Wall, \emph{{Maximin Surfaces, and the Strong Subadditivity of the
  Covariant Holographic Entanglement Entropy}},
  \href{https://doi.org/10.1088/0264-9381/31/22/225007}{\emph{Class. Quant.
  Grav.} {\bfseries 31} (2014) 225007}
  [\href{https://arxiv.org/abs/1211.3494}{{\ttfamily 1211.3494}}].

\bibitem{CausalityHEE}
M.~Headrick, V.~E. Hubeny, A.~Lawrence and M.~Rangamani, \emph{{Causality \&
  holographic entanglement entropy}},
  \href{https://doi.org/10.1007/JHEP12(2014)162}{\emph{JHEP} {\bfseries 12}
  (2014) 162} [\href{https://arxiv.org/abs/1408.6300}{{\ttfamily 1408.6300}}].

\bibitem{QFC}
R.~Bousso, Z.~Fisher, S.~Leichenauer and A.~C. Wall, \emph{{Quantum focusing
  conjecture}}, \href{https://doi.org/10.1103/PhysRevD.93.064044}{\emph{Phys.
  Rev. D} {\bfseries 93} (2016) 064044}
  [\href{https://arxiv.org/abs/1506.02669}{{\ttfamily 1506.02669}}].

\bibitem{SeeingEW}
A.~Levine, A.~Shahbazi-Moghaddam and R.~M. Soni, \emph{{Seeing the entanglement
  wedge}}, \href{https://doi.org/10.1007/JHEP06(2021)134}{\emph{JHEP}
  {\bfseries 06} (2021) 134}
  [\href{https://arxiv.org/abs/2009.11305}{{\ttfamily 2009.11305}}].

\bibitem{Shape}
L.~Chen and H.~Wang, \emph{{Shape dependence of mutual information in the OPE
  limit: linear responses}},
  \href{https://doi.org/10.1007/JHEP10(2022)101}{\emph{JHEP} {\bfseries 10}
  (2022) 101} [\href{https://arxiv.org/abs/2207.05268}{{\ttfamily
  2207.05268}}].

\bibitem{CW}
L.~Chen and H.~Wang, \emph{{Causal shadow and non-local modular flow: from
  degeneracy to perturbative genesis by correlation}},
  \href{https://doi.org/10.1007/JHEP02(2023)052}{\emph{JHEP} {\bfseries 02}
  (2023) 052} [\href{https://arxiv.org/abs/2211.12064}{{\ttfamily
  2211.12064}}].

\bibitem{AgonFaulkner}
C.~A. Ag{\'o}n and T.~Faulkner, \emph{{Quantum Corrections to Holographic
  Mutual Information}},
  \href{https://doi.org/10.1007/JHEP08(2016)118}{\emph{JHEP} {\bfseries 08}
  (2016) 118} [\href{https://arxiv.org/abs/1511.07462}{{\ttfamily
  1511.07462}}].

\bibitem{BW}
J.~J. Bisognano and E.~H. Wichmann, \emph{{On the Duality Condition for Quantum
  Fields}}, \href{https://doi.org/10.1063/1.522898}{\emph{J. Math. Phys.}
  {\bfseries 17} (1976) 303}.

\bibitem{CHM}
H.~Casini, M.~Huerta and R.~C. Myers, \emph{{Towards a derivation of
  holographic entanglement entropy}},
  \href{https://doi.org/10.1007/JHEP05(2011)036}{\emph{JHEP} {\bfseries 05}
  (2011) 036} [\href{https://arxiv.org/abs/1102.0440}{{\ttfamily 1102.0440}}].

\bibitem{CTTNull}
H.~Casini, E.~Test{\'e} and G.~Torroba, \emph{{Modular Hamiltonians on the null
  plane and the Markov property of the vacuum state}},
  \href{https://doi.org/10.1088/1751-8121/aa7eaa}{\emph{J. Phys. A} {\bfseries
  50} (2017) 364001} [\href{https://arxiv.org/abs/1703.10656}{{\ttfamily
  1703.10656}}].

\bibitem{CHNonlocal}
H.~Casini and M.~Huerta, \emph{{Reduced density matrix and internal dynamics
  for multicomponent regions}},
  \href{https://doi.org/10.1088/0264-9381/26/18/185005}{\emph{Class. Quant.
  Grav.} {\bfseries 26} (2009) 185005}
  [\href{https://arxiv.org/abs/0903.5284}{{\ttfamily 0903.5284}}].

\bibitem{NullCuts}
G.-C. Lu and H.~Wang, \emph{{Non-local modular flows across deformed
  null-cuts}}, \href{https://doi.org/10.1007/JHEP06(2025)101}{\emph{JHEP}
  {\bfseries 06} (2025) 101}
  [\href{https://arxiv.org/abs/2501.02998}{{\ttfamily 2501.02998}}].

\bibitem{ArakiRE}
H.~Araki, \emph{{Relative Entropy of States of von Neumann Algebras}},
  \href{https://doi.org/10.2977/prims/1195191148}{\emph{Publ. Res. Inst. Math.
  Sci.} {\bfseries 11} (1976) 809}.

\bibitem{CasiniRE}
H.~Casini, \emph{{Relative entropy and the Bekenstein bound}},
  \href{https://doi.org/10.1088/0264-9381/25/20/205021}{\emph{Class. Quant.
  Grav.} {\bfseries 25} (2008) 205021}
  [\href{https://arxiv.org/abs/0804.2182}{{\ttfamily 0804.2182}}].

\bibitem{BCFull}
D.~D. Blanco and H.~Casini, \emph{{Localization of Negative Energy and the
  Bekenstein Bound}},
  \href{https://doi.org/10.1103/PhysRevLett.111.221601}{\emph{Phys. Rev. Lett.}
  {\bfseries 111} (2013) 221601}
  [\href{https://arxiv.org/abs/1309.1121}{{\ttfamily 1309.1121}}].

\bibitem{ModEnergy}
D.~Blanco, H.~Casini, M.~Leston and F.~Rosso, \emph{{Modular energy
  inequalities from relative entropy}},
  \href{https://doi.org/10.1007/JHEP01(2018)154}{\emph{JHEP} {\bfseries 01}
  (2018) 154} [\href{https://arxiv.org/abs/1711.04816}{{\ttfamily
  1711.04816}}].

\bibitem{BCHM}
D.~D. Blanco, H.~Casini, L.-Y. Hung and R.~C. Myers, \emph{{Relative Entropy
  and Holography}}, \href{https://doi.org/10.1007/JHEP08(2013)060}{\emph{JHEP}
  {\bfseries 08} (2013) 060} [\href{https://arxiv.org/abs/1305.3182}{{\ttfamily
  1305.3182}}].

\bibitem{LVFisher}
N.~Lashkari and M.~Van~Raamsdonk, \emph{{Canonical Energy is Quantum Fisher
  Information}}, \href{https://doi.org/10.1007/JHEP04(2016)153}{\emph{JHEP}
  {\bfseries 04} (2016) 153}
  [\href{https://arxiv.org/abs/1508.00897}{{\ttfamily 1508.00897}}].

\bibitem{FLPW}
T.~Faulkner, R.~G. Leigh, O.~Parrikar and H.~Wang, \emph{{Modular Hamiltonians
  for Deformed Half-Spaces and the Averaged Null Energy Condition}},
  \href{https://doi.org/10.1007/JHEP09(2016)038}{\emph{JHEP} {\bfseries 09}
  (2016) 038} [\href{https://arxiv.org/abs/1605.08072}{{\ttfamily
  1605.08072}}].

\bibitem{HKT}
T.~Hartman, S.~Kundu and A.~Tajdini, \emph{{Averaged Null Energy Condition from
  Causality}}, \href{https://doi.org/10.1007/JHEP07(2017)066}{\emph{JHEP}
  {\bfseries 07} (2017) 066}
  [\href{https://arxiv.org/abs/1610.05308}{{\ttfamily 1610.05308}}].

\bibitem{BFKLW}
R.~Bousso, Z.~Fisher, J.~Koeller, S.~Leichenauer and A.~C. Wall, \emph{{Proof
  of the Quantum Null Energy Condition}},
  \href{https://doi.org/10.1103/PhysRevD.93.024017}{\emph{Phys. Rev. D}
  {\bfseries 93} (2016) 024017}
  [\href{https://arxiv.org/abs/1509.02542}{{\ttfamily 1509.02542}}].

\bibitem{QNEC}
S.~Balakrishnan, T.~Faulkner, Z.~U. Khandker and H.~Wang, \emph{{A General
  Proof of the Quantum Null Energy Condition}},
  \href{https://doi.org/10.1007/JHEP09(2019)020}{\emph{JHEP} {\bfseries 09}
  (2019) 020} [\href{https://arxiv.org/abs/1706.09432}{{\ttfamily
  1706.09432}}].

\bibitem{CFQNEC}
F.~Ceyhan and T.~Faulkner, \emph{{Recovering the QNEC from the ANEC}},
  \href{https://doi.org/10.1007/s00220-020-03751-y}{\emph{Commun. Math. Phys.}
  {\bfseries 377} (2020) 999}
  [\href{https://arxiv.org/abs/1812.04683}{{\ttfamily 1812.04683}}].

\bibitem{FL}
T.~Faulkner and A.~Lewkowycz, \emph{{Bulk locality from modular flow}},
  \href{https://doi.org/10.1007/JHEP07(2017)151}{\emph{JHEP} {\bfseries 07}
  (2017) 151} [\href{https://arxiv.org/abs/1704.05464}{{\ttfamily
  1704.05464}}].

\bibitem{KLIntersect}
D.~Kabat and G.~Lifschytz, \emph{{Local bulk physics from intersecting modular
  Hamiltonians}}, \href{https://doi.org/10.1007/JHEP06(2017)120}{\emph{JHEP}
  {\bfseries 06} (2017) 120}
  [\href{https://arxiv.org/abs/1703.06523}{{\ttfamily 1703.06523}}].

\bibitem{FLWToolkit}
T.~Faulkner, M.~Li and H.~Wang, \emph{{A modular toolkit for bulk
  reconstruction}}, \href{https://doi.org/10.1007/JHEP04(2019)119}{\emph{JHEP}
  {\bfseries 04} (2019) 119}
  [\href{https://arxiv.org/abs/1806.10560}{{\ttfamily 1806.10560}}].

\bibitem{InfoMetricEW}
Y.~Suzuki, T.~Takayanagi and K.~Umemoto, \emph{{Entanglement Wedges from
  Information Metric in Conformal Field Theories}},
  \href{https://doi.org/10.1103/PhysRevLett.123.221601}{\emph{Phys. Rev. Lett.}
  {\bfseries 123} (2019) 221601}
  [\href{https://arxiv.org/abs/1908.09939}{{\ttfamily 1908.09939}}].

\bibitem{ShadowsEW}
Y.~Kusuki, Y.~Suzuki, T.~Takayanagi and K.~Umemoto, \emph{{Looking at Shadows
  of Entanglement Wedges}},
  \href{https://doi.org/10.1093/ptep/ptaa152}{\emph{PTEP} {\bfseries 2020}
  (2020) 11B105} [\href{https://arxiv.org/abs/1912.08423}{{\ttfamily
  1912.08423}}].

\bibitem{MES}
L.~Chen, \emph{{Inequalities and Positivity in Modular Flowed Entanglement
  Entropy}},  \href{https://arxiv.org/abs/2608.29329}{{\ttfamily 2608.29329}}.

\bibitem{CFL}
V.~Chandrasekaran, T.~Faulkner and A.~Levine, \emph{{Scattering strings off
  quantum extremal surfaces}},
  \href{https://doi.org/10.1007/JHEP08(2022)143}{\emph{JHEP} {\bfseries 08}
  (2022) 143} [\href{https://arxiv.org/abs/2108.01093}{{\ttfamily
  2108.01093}}].

\bibitem{EikonalAdS}
L.~Cornalba, M.~S. Costa, J.~Penedones and R.~Schiappa, \emph{{Eikonal
  Approximation in AdS/CFT: From Shock Waves to Four-Point Functions}},
  \href{https://doi.org/10.1088/1126-6708/2007/08/019}{\emph{JHEP} {\bfseries
  08} (2007) 019} [\href{https://arxiv.org/abs/hep-th/0611122}{{\ttfamily
  hep-th/0611122}}].

\bibitem{Regge}
M.~S. Costa, V.~Gon{\c{c}}alves and J.~Penedones, \emph{{Conformal Regge
  theory}}, \href{https://doi.org/10.1007/JHEP12(2012)091}{\emph{JHEP}
  {\bfseries 12} (2012) 091} [\href{https://arxiv.org/abs/1209.4355}{{\ttfamily
  1209.4355}}].

\bibitem{SS}
S.~H. Shenker and D.~Stanford, \emph{{Stringy effects in scrambling}},
  \href{https://doi.org/10.1007/JHEP05(2015)132}{\emph{JHEP} {\bfseries 05}
  (2015) 132} [\href{https://arxiv.org/abs/1412.6087}{{\ttfamily 1412.6087}}].

\bibitem{RS}
D.~A. Roberts and D.~Stanford, \emph{{Two-dimensional conformal field theory
  and the butterfly effect}},
  \href{https://doi.org/10.1103/PhysRevLett.115.131603}{\emph{Phys. Rev. Lett.}
  {\bfseries 115} (2015) 131603}
  [\href{https://arxiv.org/abs/1412.5123}{{\ttfamily 1412.5123}}].

\bibitem{MSS}
J.~Maldacena, S.~H. Shenker and D.~Stanford, \emph{{A bound on chaos}},
  \href{https://doi.org/10.1007/JHEP08(2016)106}{\emph{JHEP} {\bfseries 08}
  (2016) 106} [\href{https://arxiv.org/abs/1503.01409}{{\ttfamily
  1503.01409}}].

\bibitem{ModChaos}
J.~de~Boer and L.~Lamprou, \emph{{Holographic Order from Modular Chaos}},
  \href{https://doi.org/10.1007/JHEP06(2020)024}{\emph{JHEP} {\bfseries 06}
  (2020) 024} [\href{https://arxiv.org/abs/1912.02810}{{\ttfamily
  1912.02810}}].

\bibitem{WittenBlocks}
E.~Hijano, P.~Kraus, E.~Perlmutter and R.~Snively, \emph{{Witten Diagrams
  Revisited: The AdS Geometry of Conformal Blocks}},
  \href{https://doi.org/10.1007/JHEP01(2016)146}{\emph{JHEP} {\bfseries 01}
  (2016) 146} [\href{https://arxiv.org/abs/1508.00501}{{\ttfamily
  1508.00501}}].

\bibitem{SpinningPropagators}
M.~S. Costa, V.~Gon{\c{c}}alves and J.~Penedones, \emph{{Spinning AdS
  Propagators}}, \href{https://doi.org/10.1007/JHEP09(2014)064}{\emph{JHEP}
  {\bfseries 09} (2014) 064} [\href{https://arxiv.org/abs/1404.5625}{{\ttfamily
  1404.5625}}].

\bibitem{SkenderisVanRees}
K.~Skenderis and B.~C. van Rees, \emph{{Real-time gauge/gravity duality}},
  \href{https://doi.org/10.1103/PhysRevLett.101.081601}{\emph{Phys. Rev. Lett.}
  {\bfseries 101} (2008) 081601}
  [\href{https://arxiv.org/abs/0805.0150}{{\ttfamily 0805.0150}}].

\bibitem{SkenderisVanReesLong}
K.~Skenderis and B.~C. van Rees, \emph{{Real-time gauge/gravity duality:
  Prescription, Renormalization and Examples}},
  \href{https://doi.org/10.1088/1126-6708/2009/05/085}{\emph{JHEP} {\bfseries
  05} (2009) 085} [\href{https://arxiv.org/abs/0812.2909}{{\ttfamily
  0812.2909}}].

\bibitem{CompanionTomography}
L.~Chen, \emph{{Modular tomography of perturbative quantum causal shadows II:
  Local positivity}},  Paper II of this series (2026).

\bibitem{BrownHenneaux}
J.~D. Brown and M.~Henneaux, \emph{{Central Charges in the Canonical
  Realization of Asymptotic Symmetries: An Example from Three-Dimensional
  Gravity}}, \href{https://doi.org/10.1007/BF01211590}{\emph{Commun. Math.
  Phys.} {\bfseries 104} (1986) 207}.

\bibitem{BreitenlohnerFreedman}
P.~Breitenlohner and D.~Z. Freedman, \emph{{Stability in gauged extended
  supergravity}},
  \href{https://doi.org/10.1016/0003-4916(82)90116-6}{\emph{Annals Phys.}
  {\bfseries 144} (1982) 249}.

\bibitem{HPPS}
I.~Heemskerk, J.~Penedones, J.~Polchinski and J.~Sully, \emph{{Holography from
  Conformal Field Theory}},
  \href{https://doi.org/10.1088/1126-6708/2009/10/079}{\emph{JHEP} {\bfseries
  10} (2009) 079} [\href{https://arxiv.org/abs/0907.0151}{{\ttfamily
  0907.0151}}].

\bibitem{AdSEFT}
A.~L. Fitzpatrick and J.~Kaplan, \emph{{AdS Field Theory from Conformal Field
  Theory}}, \href{https://doi.org/10.1007/JHEP02(2013)054}{\emph{JHEP}
  {\bfseries 02} (2013) 054} [\href{https://arxiv.org/abs/1208.0337}{{\ttfamily
  1208.0337}}].

\bibitem{CEMZ}
X.~O. Camanho, J.~D. Edelstein, J.~Maldacena and A.~Zhiboedov, \emph{{Causality
  Constraints on Corrections to the Graviton Three-Point Coupling}},
  \href{https://doi.org/10.1007/JHEP02(2016)020}{\emph{JHEP} {\bfseries 02}
  (2016) 020} [\href{https://arxiv.org/abs/1407.5597}{{\ttfamily 1407.5597}}].

\bibitem{WittenEntanglement}
E.~Witten, \emph{{APS Medal for Exceptional Achievement in Research: Invited
  article on entanglement properties of quantum field theory}},
  \href{https://doi.org/10.1103/RevModPhys.90.045003}{\emph{Rev. Mod. Phys.}
  {\bfseries 90} (2018) 045003}
  [\href{https://arxiv.org/abs/1803.04993}{{\ttfamily 1803.04993}}].

\bibitem{AichelburgSexl}
P.~C. Aichelburg and R.~U. Sexl, \emph{{On the gravitational field of a
  massless particle}}, \href{https://doi.org/10.1007/BF00758149}{\emph{Gen.
  Rel. Grav.} {\bfseries 2} (1971) 303}.

\bibitem{DraytHooft}
T.~Dray and G.~'t~Hooft, \emph{{The gravitational shock wave of a massless
  particle}}, \href{https://doi.org/10.1016/0550-3213(85)90525-5}{\emph{Nucl.
  Phys. B} {\bfseries 253} (1985) 173}.

\bibitem{Selberg}
P.~J. Forrester and S.~O. Warnaar, \emph{{The importance of the Selberg
  integral}}, \href{https://doi.org/10.1090/S0273-0979-08-01221-4}{\emph{Bull.
  Amer. Math. Soc.} {\bfseries 45} (2008) 489}
  [\href{https://arxiv.org/abs/0710.3981}{{\ttfamily 0710.3981}}].

\bibitem{CFTDistributions}
P.~Kravchuk, J.~Qiao and S.~Rychkov, \emph{{Distributions in CFT. Part II.
  Minkowski space}}, \href{https://doi.org/10.1007/JHEP08(2021)094}{\emph{JHEP}
  {\bfseries 08} (2021) 094}
  [\href{https://arxiv.org/abs/2104.02090}{{\ttfamily 2104.02090}}].

\bibitem{PenedonesMellin}
J.~Penedones, \emph{{Writing CFT correlation functions as AdS scattering
  amplitudes}}, \href{https://doi.org/10.1007/JHEP03(2011)025}{\emph{JHEP}
  {\bfseries 03} (2011) 025} [\href{https://arxiv.org/abs/1011.1485}{{\ttfamily
  1011.1485}}].

\bibitem{FKPRvR}
A.~L. Fitzpatrick, J.~Kaplan, J.~Penedones, S.~Raju and B.~C. van Rees,
  \emph{{A Natural Language for AdS/CFT Correlators}},
  \href{https://doi.org/10.1007/JHEP11(2011)095}{\emph{JHEP} {\bfseries 11}
  (2011) 095} [\href{https://arxiv.org/abs/1107.1499}{{\ttfamily 1107.1499}}].

\end{thebibliography}\endgroup
